\documentclass[1p]{elsarticle}

    \makeatletter
    \def\ps@pprintTitle{%
       \let\@oddhead\@empty
       \let\@evenhead\@empty
       \let\@oddfoot\@empty
       \let\@evenfoot\@oddfoot
    }
    \makeatother

\usepackage{textcomp}
\usepackage{amsmath}
\usepackage{amssymb}
\usepackage{array} 
\usepackage{multirow}
\usepackage{subcaption}
\usepackage{verbatim}
\newcommand{\bfx}{\mathbf{x}}
\newcommand{\bfp}{\mathbf{p}}
\newcommand{\bfa}{\mathbf{a}}
\newcommand{\bfb}{\mathbf{b}}
\newcommand{\bfc}{\mathbf{c}}

\newcommand{\bfe}{\mathbf{e}}
\newcommand{\bfI}{\mathbf{I}}
\newcommand{\bfP}{\mathbf{P}}
\newcommand{\bfu}{\mathbf{u}}
\newcommand{\bfv}{\mathbf{v}}

\newcommand{\bfz}{\mathbf{z}}

\newcommand{\bmalpha}{\boldsymbol{\alpha}}
\newcommand{\bfR}{\mathbf{R}}

\newcommand{\bfK}{\mathbf{K}}
\newcommand{\bff}{\mathbf{f}}
\newcommand{\bfzero}{\mathbf{0}}

\usepackage{bm}
\usepackage{color}
\usepackage{amsmath}
\usepackage{amssymb}
\usepackage{mathtools}
\usepackage{graphicx}
\usepackage{comment}
\usepackage{lscape}
\usepackage{subcaption}
\usepackage{algorithm, algpseudocode, caption}
\algnewcommand{\IIf}[1]{\State\algorithmicif\ #1\ \algorithmicthen} 
\algnewcommand{\EndIIf}{\unskip\ \algorithmicend\ \algorithmicif}  
\newcommand*\diff{\mathop{}\!\mathrm{d}}

\begin{document}

\begin{frontmatter}

\title{Multi-material Topology Optimization of Lattice Structures using Geometry Projection}


\author[mymainaddress]{Hesaneh Kazemi}

\author[mysecondaryaddress]{Ashkan Vaziri}

\author[mymainaddress]{Juli{\'a}n A. Norato\corref{mycorrespondingauthor}}
\cortext[mycorrespondingauthor]{Corresponding author}
\ead{norato@engr.uconn.edu}

\address[mymainaddress]{Department of Mechanical Engineering, The University of Connecticut, 191 Auditorium Road, U-3139, Storrs, CT 06269, USA}
\address[mysecondaryaddress]{Department of Mechanical and Industrial Engineering, Northeastern University, 254 Richards Hall, 360 Huntington Ave., Boston, Massachusetts 02115, USA}

\begin{abstract}
This work presents a computational method for the design of architected truss lattice materials where each strut can be made of one of a set of available materials.  We design the lattices to extremize effective properties. As customary in topology optimization, we design a periodic unit cell of the lattice and obtain the effective properties via numerical homogenization.
Each bar is represented as a cylindrical offset surface of a medial axis parameterized by the positions of the endpoints of the medial axis. These parameters are smoothly mapped onto a continuous density field for the primal and sensitivity analysis via the geometry projection method. A size variable per material is ascribed to each bar and penalized as in density-based topology optimization to facilitate the entire removal of bars from the design. During the optimization, we allow bars to be made of a mixture of the available materials.  However, to ensure each bar is either exclusively made of one material or removed altogether from the optimal design, we impose optimization constraints that ensure each size variable is 0 or 1, and that at most one material size variable is 1.  The proposed material interpolation scheme readily accommodates any number of materials.  To obtain lattices with desired material symmetries, we design only a reference region of the unit cell and reflect its geometry projection with respect to the appropriate planes of symmetry. Also, to ensure  bars remain whole upon reflection inside the unit cell or with respect to the periodic boundaries, we impose a no-cut constraint on the bars. We demonstrate the efficacy of our method via numerical examples of bulk and shear moduli maximization and Poisson's ratio minimization for two- and three-material  lattices with cubic symmetry.
\end{abstract}

\begin{keyword}
\texttt{Topology Optimization}\sep \texttt{Multi-material} \sep \texttt{Lattice Structures} 
\end{keyword}

\end{frontmatter}


\section{Introduction}
\label{sec:introduction}

Topology optimization techniques generate novel structural designs by optimizing the material layout within a prescribed design region. One of their applications is the design of architected materials with desired effective properties. This work focuses on the design of multi-material, periodic lattice structures via topology optimization. Open-cell lattice structures are generally easier to manufacture than most organic designs produced by topology optimization methods, as these may exhibit fully enclosed cavities. As all methods to design periodic structures, the goal of the proposed method is to design the unit cell, which we refer to in this work as the microstructure. 

%
Methods to design microstructures were first introduced in ground-structure approaches. Sigmund presented an inverse homogenization method to design the microstructures by modeling the unit cell as a truss or thin frame structure \cite{sigmund1994materials, sigmund1995tailoring}. These methods benefit from efficient computation and naturally enforce a uniform cross-sectional area for the struts that simplifies their fabrication. However, they do not capture 3-dimensional stress states at strut intersections, and the optimal design may be suboptimal as it is a `subset' of the ground structure. 

The use of topology optimization of continua to design microstructures was first presented in density-based methods to design materials with extreme thermal expansion \cite{sigmund1997design}, and later employed in \cite{gibiansky2000multiphase} to optimize the effective bulk modulus. This and similar approaches were also employed in \cite{sigmund1998design, neves2000optimal, cox2000band, torquato2002multifunctional, guest2006optimizing, guest2007design,  de2007topological}; cf.\ the recent review \cite{osanov2016topology}. Compared to ground-structure approaches, these methods produce more efficient structures and are less dependent on the initial design. However,  they can be more difficult to manufacture, specially closed-cell designs.
Other topology optimization approaches to design microstructures include evolutionary approaches \cite{huang2008optimal, huang2011topological, huang2012evolutionary, radman2013topological} and level-set methods \cite{challis2008design, zhou2010level, otomori2014level, picelli2017stress}. Some methods concurrently design the topology of the micro and macrostructures  \cite{rodrigues2002hierarchical, guedes2006hierarchical, coelho2008hierarchical, liu2008optimum, niu2009optimum, yan2008uniform, deng2013multi, huang2013topology, yan2014concurrent}. This work, however, focuses only on the design of the microstructure. 
The multi-material topology optimization of microstructures was first demonstrated in the aforementioned works \cite{sigmund1997design} and \cite{gibiansky2000multiphase}. Other interpolation schemes (i.e., \cite{bendsoe1999material,stolpe2001alternative}) have also been used to design multi-phase materials with extreme thermal conductivity  \cite{zhou2007relation,zhou2008computational}.  The reader is referred to \cite{kazemi2018topology} for a more detailed description of these approaches.

Although there exist multi-material additive manufacturing methods, the organic designs produced by the preceding methods are not necessarily easy to manufacture. They may possess closed cavities that require supports which are difficult to remove. Therefore, multi-material lattices made of struts may simplify their manufacturing. The geometry projection method \cite{bell2012geometry, norato2015geometry, zhang2016geometry} can be employed to design structures exclusively made of geometric components, whereby an analytic parameterization of the geometry is smoothly mapped onto a density field over a fixed finite element grid. By assigning a size variable to each geometric component that is penalized in the spirit of density-based methods, the geometry projection enables the entire removal of a component from the design. A similar family of methods to design structures made of discrete geometric components is the method of moving morphable components \cite{guo2014doing}, in which geometric components and their union are represented using level set functions.  Unlike the geometry projection method, this method does not employ penalized size variables for the components and therefore can only remove components by engulfing them inside other components or by making their size small enough that they do not affect the analysis. 

The geometry projection method was employed in \cite{watts2017geometric} in conjunction with an adaptation of the multi-material interpolation scheme of \cite{sigmund1997design} to design two-material, 3-dimensional lattice structures.  Its effectiveness is demonstrated via the design of two-material lattices for maximal bulk modulus and for minimal Poisson's ratio. While this interpolation scheme is very effective, extending its application to more than two materials is not straightforward and requires changes to the material interpolation formulation, as noted in \cite{stegmann2005discrete}. 
In the context of topology optimization with discrete geometric components using geometry projection, a new interpolation scheme was presented in \cite{kazemi2018topology} to accommodate the design of multi-material structures. A size variable per material is ascribed to each geometric component. This interpolation is an adaptation of the discrete material optimization (DMO) method  \cite{stegmann2005discrete}.  However, unlike DMO, this method employs optimization constraints to ensure these size variables are 0 or 1, and to guarantee that each component has at most one material with a size variable of 1. The moving morphable component method has been applied to the design of multi-material structures \cite{zhang2017topology} with geometric components made of different materials; however, unlike all of the aforementioned multi-material methods, the choice of material for each component is fixed and therefore not part of the optimization.

This work focuses on the topology optimization of multi-material lattice structures using the geometry projection method. By employing the multi-material interpolation and optimization constraints introduced in \cite{kazemi2018topology}, the proposed method can readily accommodate any number of materials. It employs inverse homogenization to design the unit cell for extremizing the effective properties of the macrostructure. To enforce desired symmetries in the lattice, the geometry projection of a reference region is reflected onto other regions with respect to the symmetry planes corresponding to the desired symmetry. To ensure bars remain whole upon reflection inside the unit cell or with respect to the periodic boundaries and thus facilitate fabrication, we impose a volume difference constraint on the bars.  This work builds on the preliminary results presented in \cite{kazemi2019topology} for maximal bulk modulus design by adding a no-cut constraint to ensure that bars remain whole upon reflection, which results in significantly different designs. We demonstrate the efficacy of our method by designing two- and three-material lattices with maximal bulk modulus, maximal shear modulus and minimal Poisson's ratio subject to a weight constraint.  

The rest of the paper is organized as follows. Section \ref{sec:geom-proj} discusses the projection of bnto analysis grid and the aggregation function and multi-material interpolation scheme. In Section \ref{sec:homogenization} we describe the homogenization method that we use to calculate effective properties of the macrostructure by considering a unit cell. In Section \ref{sec:symmetry} we employ reflection matrices to enforce symmetries on the lattice.  Section \ref{sec:no-cut} presents a new constraint to enforce the bars to stay in the symmetry reference region. The optimization problem in described in Section \ref{sec:opt-problem}.  We present numerical examples to demonstrate our method in Section \ref{sec:examples} , and we draw conclusions in Section \ref{sec:conclusions}.

\section{Geometry Projection}
\label{sec:geom-proj}

\begin{figure}[h]
\centering
\begin{subfigure}[c]{0.4\textwidth}
\centering
\includegraphics[width=\columnwidth]{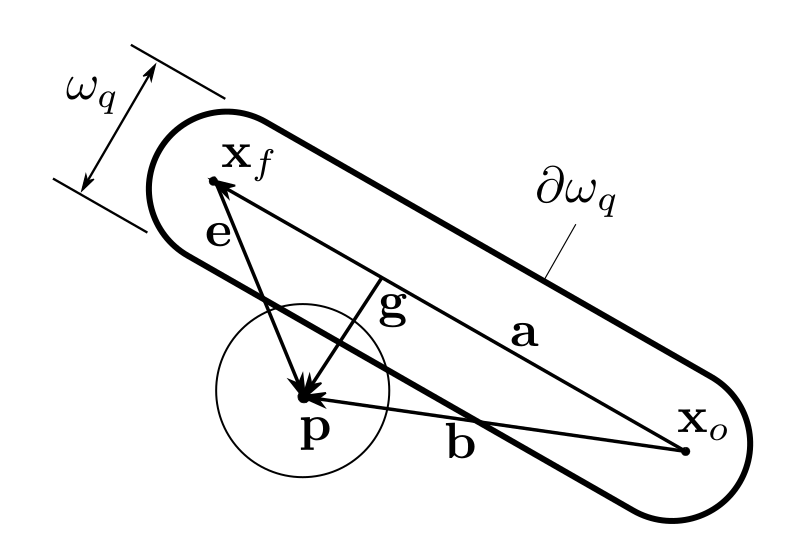}
\caption{Bar geometry}
\label{fig:bar-geom}
\end{subfigure}
\begin{subfigure}[c]{0.4\textwidth}
\centering
\includegraphics[width=0.7\columnwidth]{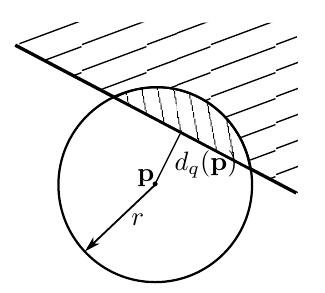}
\caption{Sample window}
\label{fig:sample-window}
\end{subfigure}
\caption{Geometry projection}
\label{fig:geometry-projection}
\end{figure}

We employ the geometry projection method to create a differentiable map between a high-level parametric description of geometry and a density field defined over a fixed design region. For the lattice design, we consider structures made of cylindrical bars of the same diameter. We model a bar $q$ of diameter $w$ as an offset surface of a line segment (the bar's medial axis), resulting in a cylinder with semi-spherical ends. We parameterize this geometric representation with the positions of the endpoints of the medial axis, $\bfx_{q_o}$ and $\bfx_{q_f}$ (cf.~Fig.~\ref{fig:bar-geom}). We calculate the projected density for point $\bfp$ as the volume fraction of the intersection of a sample window $\mathbf{B}_\mathbf{p}^r:=\{ \bfx | \; \| \bfp-\bfx \| \leq r \}$ and the solid structure $\omega$:
\begin{equation}
\rho(\bfx,r) := \frac{|\mathbf{B}_\mathbf{p}^r \cap \omega|}{\mathbf{B}_\mathbf{p}^r}
\label{eq:geom-proj}
\end{equation}
%
This projected density is approximated as the volume fraction of the circular cap of height $r-\phi_q$  (cf.~Fig.~\ref{fig:sample-window}), i.e.:
\begin{equation}
\rho_q(\phi_q, r) =
\left\{
	\begin{array}{ll}
		0  & \mbox{if } \phi_q > r \\
		\frac{1}{2} + \frac{\phi_q^3}{4r^3} - \frac{3\phi_q}{4r}  & \mbox{if } -r \leq \phi_q \leq r \\
		1  & \mbox{if }\phi_q < -r \\
	\end{array}
\right.
\label{eq:rhoq-3d}
\end{equation}
%
The signed distance $\phi_q(\bfp)$ from $\bfp$ to bar $q$ is obtained from
%
\begin{equation}
\phi_q (d_q, w) := d_q (\bfx_{q_o}, \bfx_{q_f}, \mathbf{p}) - \frac{w}{2}
\label{eq:signed-dist}
\end{equation}
%
In the expression above, $d_q$ is the distance from $\bfp$ to the medial axis of bar $q$, given by:
%
\begin{equation}
d_q (\bfx_{q_o}, \bfx_{q_f}, \mathbf{p}) =
\left\{
	\begin{array}{ll}
		\Vert\mathbf{b}\Vert  & \quad \mbox{if } \mathbf{a}\cdot\mathbf{b} \leq 0 \\
		\Vert\mathbf{g}\Vert  & \quad \mbox{if } 0 < \mathbf{a}\cdot\mathbf{b} < \mathbf{a}\cdot\mathbf{a} \\
		\Vert\mathbf{e}\Vert  & \quad \mbox{if } \mathbf{a}\cdot\mathbf{b} > \mathbf{a}\cdot\mathbf{a} \\
	\end{array}
\right.
\end{equation}
where
\[
\bfa := \bfx_{q_f} - \bfx_{q_o} \qquad \bfb := \bfp - \bfx_{q_o} \qquad \bfe := \bfp - \bfx_{q_f} \qquad \mathbf{g} := \bfP_{\bfa}^\perp \bfb
\]
%
and $\bfP_a^\perp = \bfI - (\bfa \otimes \bfa) / \| \bfa \|^2$ is the perpendicular projection matrix on $\bfa$.

To calculate the projected density considering the contribution of multiple bars, a $p$-norm approximation of the maximum function is employed in \cite{bell2012geometry, norato2015geometry}. Unlike these works, in which all bars are made of the same isotropic material, \cite{kazemi2018topology} presented a new aggregation scheme to account for the intersection of bars made of different materials by defining an effective density for material $i$ at point $\bfp$, i.e.,
%
\begin{equation}
\label{eq:rho-eff-mult-mat}
\rho_{eff}^{i}(\bfz, \bfp) = \frac{\sum_{q=1}^{N_b}\tilde{H}_{\epsilon}(-\phi_q(\bfz, \bfp)) \rho_q w_i^q(\bfz)}{A + B}
\end{equation}
where
\begin{align}
A &= \sum_{q=1}^{N_b} \left( \tilde{H}_{\epsilon}(-\phi_q(\bfz, \bfp))\sum_{i=1}^{Nm} \alpha_i^q \right) \\
B &= 1 - \underset{q}{KS}\left(\tilde{H}_{\epsilon}(-\phi_q(\bfz, \bfp))\sum_{i=1}^{Nm} \alpha_i^q \right) \\
\underset{i}{KS}(\bfx) &:= \frac{1}{k} \ln \left(\sum_i e^{kx_i} \right) \label{eq:KS}
\end{align}
%
In the above expressions $N_b$ denotes the number of bars and $N_m$ the number of available materials; the term $A$ is an approximation of the number of solid bars containing $\bfp$, and the term $B$ equals unity to avoid a division by zero if point $\bfp$ lies in the void region; $KS$ is the Kreisselmeier-Steinhauser function, which smoothly approximates the maximum; and $\tilde{H}_{\epsilon}$ is a smooth relaxation of the Heaviside given by
\begin{equation}
\label{eq:smooth-H}
\tilde{H}_{\epsilon}(x) =
\left\{
	\begin{array}{ll}
		0  & \quad \mbox{if } x < -\epsilon \\
		\left[ \frac{1}{2} + \frac{x}{2\epsilon} + \frac{1}{2\pi} \sin(\frac{\pi x}{\epsilon}) \right]^p & \quad \mbox{if }  -\epsilon \leq x \leq \epsilon \\
		1  & \quad \mbox{if } x > \epsilon
	\end{array}
\right.
\end{equation}
%
with the parameter $p$ indicating the sharpness of the approximation. The weight fractions $w_i^q$ are defined in a the spirit of the DMO method, but defined using the size variables instead of the element densities as:
%
\begin{equation}
\label{eq:weight}
w_i^q = (\alpha_i^q) \prod_{j=1}^{N_m} (1- (\alpha^q_{j\neq i}))
\end{equation}
%
In the geometry projection method, as in density-based topology optimization techniques (and some level set optimization techniques), we employ an ersatz material to analyze the structure using a fixed mesh, whereby the effective elastic tensor at $\bfp$ is computed as:
%
\begin{equation}
\label{eq:effC}
\mathbb{C}(\bfz, \bfp) = \mathbb{C}_{min} + \sum_{i=1}^{N_m} \left( \mathbb{C}_i - \mathbb{C}_{min} \right) \, \rho_{eff}^{i}(\bfz, \bfp)
\end{equation}
%
In this expression, $\mathbb{C}_{min}$ is the elasticity tensor of a weak isotropic material to preclude an ill-posed analysis, and $\mathbb{C}_i$ is the elasticity tensor for material $i$. The main difference between this material interpolation and the one used by multi-material density-based methods is that the material interpolation is uncoupled from the penalization (to ensure 0-1 size variables in the optimal design) and the mutual material exclusion (which ensures each bar is made of at most one material).  We achieve this by imposing separate constraints in the optimization, which facilitates accommodating any number of materials.
\section{Homogenization}
\label{sec:homogenization}

We approximate the effective properties of the lattice by using homogenization (cf., \cite{guedes1990preprocessing, hassani1998reviewi, hassani1998reviewii, hassani1998reviewiii}).  The components of the effective elastic tensor $\mathbb{C}^H$ are given by
%
\begin{equation}
\label{eq:effective-elasticity2}
C_{ijkl}^H = \frac{1}{|Y|} \int_Y C_{pqrs}(\epsilon_{pq}^{0(ij)}-\epsilon_{pq}^{*(ij)})(\epsilon_{rs}^{0(kl)}-\epsilon_{rd}^{*(kl)})~ \diff y
\end{equation}
%
with $\bm{\epsilon}^{0(kl)} = \bfe_k \otimes \bfe_l$ corresponding to six unit strains applied on the unit cell, $Y$ denoting the domain of the unit cell, $C_{pqrs}$ indicating the components of the elasticity tensor of Eq.\ \ref{eq:effC},  and 
%
\begin{equation}
\epsilon_{pq}^{*(kl)} = \frac{1}{2} \left( \frac{\partial \chi_p ^{(kl)}}{\partial y_q} + \frac{\partial \chi_q ^{(kl)}}{\partial y_p} \right)
\end{equation}
%
The fields $\bm{\chi}^{(kl)} \in \mathcal{U}_{adm}$ are the solutions to the six problems
%
\begin{equation}
\int_Y C_{ijpq} \frac{\partial \chi_p ^{(kl)}}{\partial y_q} \frac{\partial v_i}{\partial y_j}~ \diff y =\int_Y C_{ijkl} \frac{\partial v_i}{\partial y_j}~ \diff y , \forall \bfv \in \mathcal{U}_{adm}
\end{equation}
%
with $\bfv$ denoting the test function and $\mathcal{U}_{adm} := \{\bfu|  \bfu \in H^1(Y), \bfu \text{ is } Y\text{--periodic} \}$ being the set of admissible solutions. 

\section{Symmetry}
\label{sec:symmetry}

We impose symmetry with respect to an arbitrary number of planes on the unit cell to obtain desired material symmetries on the lattice. The intersection of these planes defines a number of similar regions, among which we choose one as the reference region, wherein we define the bars. To compute the projected density at a point in any of the other regions, we reflect the point with respect to the appropriate symmetry planes so that the reflected point lies on the reference region, and then we perform the geometry projection as usual. This strategy is employed in \cite{kazemi2018topology} and is similar to the one introduced in \cite{watts2017geometric}. To perform the reflection with respect to the appropriate symmetry planes, we multiply all the corresponding reflection matrices (we assume all symmetry planes pass through the origin of the unit cell coordinate system).  The reflected point is obtained as
\begin{equation}
\hat{\bfp} := \prod_{s=1}^{N_s} \bfR_s \bfp \\
\label{eq:sym}
\end{equation}
%
with $N_s$ denoting the number of symmetry planes and $ \bfR_s$ being the reflection matrix corresponding to symmetry plane $s$.

\section{No-Cut Constraint}
\label{sec:no-cut}

A significant difference of the proposed method with the preliminary results of \cite{kazemi2019topology} is that here we introduce a constraint in the optimization to ensure components remain whole in the design space, that is, that lattice bars are not cut upon reflection or across boundaries, which would make fabrication more difficult. A mechanism to achieve this is introduced in \cite{watts2017geometric} by imposing constraints on the positions of the endpoints of the bar's medial axes so that they entirely lie within the reference region; in the case of cubic symmetry, for instance, this amounts to lower and upper bounds on the endpoint positions (which render orthotropic symmetry) plus four additional constraints per bar to restrict the positions to the reference tetrahedral region. 

Here, we follow a simpler but equally effective approach, whereby we introduce a constraint on the difference between the volume of the bars in the reference region computed using the geometric parameters, and the volume that would be computed using the geometry projection. If these two volumes are different, it means a portion of the bar is lying outside of the reference region and the bar is cut.  Therefore, if we impose a constraint that this difference cannot be larger than a small value, we consequently force bars to be wholly placed within the reference region. To account for multiple bars, we place the constraint on the maximum volume difference violation of all bars, which we smoothly approximate using a lower-bound KS function:
\begin{equation}
\label{eq:no-cut-constraint}
 g_n(\bfz) :=\underset{q}{LKS}(V_{geom}^q - V_{num}^q) \leq \varepsilon_n
 \end{equation}
with
\begin{equation}
\label{eq:LKS}
\underset{i}{LKS}(\bfx) := \frac{1}{k} \ln \left( \frac{1}{n} \sum_i e^{kx_i} \right), \, \bfx \in \mathbb{R}^n
 \end{equation}
where $V_{geom}^q$ is the volume of bar $q$ calculated using its geometric parameters (i.e., endpoint locations and width), and $V_{num}$  is the sum of projected densities for bar $q$ inside the symmetry reference region.  The advantages of this approach are that it does not require formulating different placement constraints on the points for different types of material symmetries, and it renders a single optimization constraint regardless of the number of bars and of symmetry planes. We use the lower-bound KS function instead of the KS function of Eq.~\ref{eq:KS} because it approximates the maximum from below and therefore the approximation does not exceed the desired maximum value of zero. Also, we do not need to use adaptive constraint scaling strategies to compensate for the approximation error similar to those used in, e.g., stress-based topology optimization (cf., \cite{le2010stress}) because the constraint limit is zero and we are approximating from below.  

\section{Optimization and Computer Implementation}
\label{sec:opt-problem}

We consider three problems in this work: maximization of the effective bulk modulus, maximization of the effective shear modulus and minimization of the effective Poisson's ratio of the lattice structure, all subject to a material resource constraint. This constraint is imposed on the weight fraction, which considers the physical densities of the materials, as opposed to the more prevalent volume fraction constraint.  A weight fraction constraint is more meaningful for multi-material design problems than specifying arbitrary volume fraction limits for each material. 

We also impose a discreteness constraint to ensure all size variables $\alpha_q^i$ for bar $q$ and material $i$ attain a value of either 0 or 1 in the optimal design.  We also need to guarantee that each bar is made of at most one material. In other words, the size variable corresponding to that material should equal 1 and all other size variables should be 0; or all size variables should equal 0, signifying the bar is entirely removed from the design. We employ the discreteness and mutual-material exclusion constraints introduced in \cite{kazemi2018topology}, which we describe in the sequel. 

 \subsection{Discreteness Constraint}
\label{sec:discreteness}
To facilitate the lattice fabrication, we desire bars that are made of only one material. Therefore, the size variables have to be either 0 or 1 in the optimal design. Thus we need a penalization scheme to make intermediate values of the size variables disadvantageous. We achieve this by imposing the equality constraint
\begin{equation}
\label{eq:disc-constraint}
 g_d(\bfz) := 4 \underset{i,q}{LKS}(\bmalpha^T(\mathbf{1}-\bmalpha)) = 0 
 \end{equation}
%
with $LKS(\bfx)$ being the function defined in Eq.\ \ref{eq:LKS}, and $\bmalpha = [\bmalpha_1^T \; \bmalpha_2^T \ldots \bmalpha_{N_b}^T]^T$ denoting the vector of all size variables in the lattice, where $\bmalpha_q$ is the $N_m$-vector of size variables for bar $q$ and $N_m$ is the number of materials. Since it is easier to enforce inequality constraints with the optimizer we employ in this work, we replace the constraint of Eq.~\ref{eq:disc-constraint} with the inequality constraint
\begin{equation}
\label{eq:disc-ineq}
 g_d(\bfz) \leq \varepsilon_d \ll 1
\end{equation}
%
This explicit penalization of the size variables is similar to approaches employed in some density-based topology optimization techniques (e.g., \cite{borrvall2001topology}) to penalize intermediate density values.  In those approaches, however, the explicit penalization is added as a penalty term to the objective function. As noted in \cite{sigmund2013topology}, the challenge with this approach is to choose an adequate weighting factor for the penalty term that ensures good convergence. In our case, however, the explicit penalization is incorporated as an optimization constraint.  Moreover, the fact that we are using the $LKS$ smooth maximum approximation allows us to choose a value for $\varepsilon$ that works well regardless of the number of bars. 

Additionally, to prevent the design from quickly selecting a material for a bar that results in a poor local minimum, we employ a continuation strategy. We gradually decrease a relatively large initial value $\varepsilon_d^{(0)}$ by a step $\Delta \varepsilon_d$.  This decrease is performed only after the relative change in the objective function in consecutive iterations is less than a specified value $\Delta f^*$, i.e.:
\begin{equation}
\label{eq:cont-gd}
\text{If } \Delta f^{(I+1)} \leq \Delta f^* \text{ then } \varepsilon_d^{(I+1)} \gets \max(\varepsilon_d^{(I)} - \Delta \varepsilon_d, \varepsilon_d^*)
\end{equation}
%
where $\varepsilon_d^*$ is the final constraint limit and $\Delta f^{(I+1)} := (|f^{(I+1)} - f^{(I)}|)/f^{(I)}$ the relative change in the objective function at iteration $I + 1$.  As observed in the numerical experiments in this work and in \cite{kazemi2019topology}, this continuation strategy is effective in preventing premature convergence to poor local minima and renders good convergence.

\subsection{Mutual Material Exclusion Constraint}
\label{sec:mutual}
To ensure each bar is either void or exclusively made of a single material, we impose the constraint
\begin{equation}
\label{eq:mut-mat-constraint}
 g_m(\bfz) := \underset{q}{LKS}\left( \sum_{i=1}^{N_m}\alpha_i^q \right) -1 \leq 0
 \end{equation}
%
We use the same continuation strategy of Section \ref{sec:discreteness} for this constraint.  As with the case of the discreteness constraint described in the preceding section, the mutual material exclusion constraint is an explicit penalization enforced through a constraint in the optimization, as opposed to the implicit penalization employed by the DMO method.

\subsection{Optimization Problem}
\label{subsec:opt-problem}
In this work, we only consider effective lattices with cubic symmetry.  However, as demonstrated in \cite{kazemi2019topology} for the case of bulk modulus maximization, the application to orthotropic materials is straightforward. The effective bulk ($K$) and shear ($G$) moduli are calculated as: 
\begin{equation}
K(\bfz) = \frac{1}{9}(C_{1111}+ C_{1122}+ C_{1133}+ C_{2211}+ C_{2222} + C_{2233}+ C_{3311}+ C_{3322}+ C_{3333})  \label{eq:bulk} \\
\end{equation}
\begin{equation}
G(\bfz) = \frac{1}{3}(C_{2323}+ C_{1313} + C_{1212})  \label{eq:shear} \\
\end{equation}
The Poisson's ratio is given by:
\begin{equation}
\nu(\bfz) = \frac{C_{1122}}{2(C_{1122} + C_{1212})}  \label{eq:poisson} \\
\end{equation}
The optimization problem is given by:
\begin{align}
&\underset{\bfz}{\min} f(\bfz)   \label{eq:opt-problem} \\
&\text{subject to} \nonumber \\
&w_f := \frac{1}{|\Omega| \gamma_{max} } \sum_{i=1}^{N_m} \gamma_i \int_{\Omega} \rho_{eff}^i(\bfz, \bfp)  \diff v \leq w_f^* \label{eq:wf}\\
&\mathsf{a}(\bfu^{(kl)}(\bfz), \bfv) = \mathsf{l}(\bfv, \bm{\epsilon}^{0(kl)}), \, \forall \bfv \in \mathcal{U}_0, \bfu^{(kl)} \in \mathcal{U} \label{eq:equilibrium}\\
&g_d(\bfz) \leq \varepsilon_d^{(I)} \label{eq:disc-cons}\\
&g_m(\bfz) \leq \varepsilon_m^{(I)} \label{eq:mme-cons} \\
&g_n(\bfz) \leq \varepsilon_n \label{eq:no-cut-cons} \\
&\bfx_{q_0} , \bfx_{q_f} \in \Omega \label{eq:point-bounds} \\
&0.0 \leq \alpha_i^q \leq 1.0 \label{eq:size-bounds}
\end{align}
with $f(\bfz) \equiv -K(\bfz)$ for the bulk modulus maximization, $f(\bfz) \equiv -G(\bfz)$ for the shear modulus maximization, and $f(\bfz) \equiv \nu(\bfz)$ for the Poisson's ratio minimization.  In addition to the foregoing constraints, and as is customary for this problem (cf.~\cite{sigmund1994materials}), for the Poisson's ratio minimization we also impose a lower limit on the effective bulk modulus, $K(\bfz) \leq K_{min}$ to avoid removal of all the material.  In the expressions above, $w_f$ is the weight fraction, $\gamma_i$ the physical density for material $i$, and $\gamma_{max}$ is the largest physical density of all available materials, so that if the heaviest material occupies the entire unit cell, $w_f = 1$. The domains $\Omega$ and $\omega \subseteq \Omega$ correspond to the region occupied by the design envelope and the design, respectively. The test functions are denoted by $\bfv$, and $\bfu^{(kl)}$ are the displacements corresponding to the six applied unit strains $\bm{\epsilon}^{0(kl)}$, $k,l=1,\cdots, 3$. The admissible sets for trial and test functions are $\mathcal{U} := \{ \bfu | \bfu \in H^1(\Omega), \bfu \text{ is } Y\text{-periodic}, \bfu(\bfc) = \bfzero \}$ and $\mathcal{U}_0 := \{ \bfv | \bfv \in H^1(\Omega), \bfv|_{\Gamma} = \bfzero, \bfv(\bfc) = \bfzero \}$, respectively.  We prevent rigid-body motions by imposing zero displacements at the center of the unit cell $\bfc$. The energy bilinear form $\mathsf{a}$ and the load linear form $\mathsf{l}$ in  Eq.~\ref{eq:equilibrium} are computed as:
\begin{align}
\mathsf{a}(\bfu, \bfv) &:= \int_\Omega \nabla \bfv \cdot \mathbb{C}(\bfz, \bfp) \nabla \bfu  \diff v \label{eq:bilinear} \\
\mathsf{l}(\bfv, \bm{\epsilon}) &:= \int_{\Omega} \nabla \bfv \cdot \mathbb{C}(\bfz, \bfp) \bm{\epsilon} \diff v \label{eq:linear}
\end{align}
%

For the design variables $\hat{z}$ to fall within the range $[0,1]$, we scale them as in our previous works \cite{norato2015geometry, zhang2016geometry, kazemi2018topology}, and at each optimization iteration $I$ we impose a move limit $m$ on each design variable as:
%
\begin{equation}
\label{eq:move-limit}
\max(0,\bfz^{(I-1)} -m) \leq \bfz^{(I)} \leq \min(1,\bfz^{(I-1)} + m)
\end{equation} 

\subsection{Computer Implementation}
\label{subsec:computer-implementation}

A flowchart describing the proposed method is shown in Algorithm \ref{alg:mmtols}. Our code is implemented in C++ using the deal.II library \cite{BangerthHartmannKanschat2007,dealII90} as a backbone for the finite element solutions. We parallelize the assembly of the stiffness matrix, the computation of geometry projection, and the solution of the linear system of equations by employing the data structures provided by parallel linear algebra libraries. We employ an element-wise uniform effective density, which we compute at the element centroid $\bfx_e$. The geometry projection is computed using a window radius $r$ equal to $c$ times the radius of the sphere that circumscribes the element, i.e., $r = c \sqrt{3} h/2$, where $h$ is the element size.  As the optimizer, we employ the parallel implementation of the method of moving asymptotes (MMA) of \cite{svanberg2002class,svanberg2007mma}, presented in \cite{aage2017giga}. We use the default MMA parameters presented in \cite{svanberg2007mma}.  We stop the optimization when the relative change $\Delta f$ in compliance between consecutive iterations falls below a specified value $\Delta f^*$.  The optimization parameter values we employed for all examples are listed in Table \ref{table:parameters}. 

 \begin{algorithm}[H]
    \caption{Multi-material Topology Optimization of Lattice Structures}\label{alg:mmtols}
\begin{algorithmic}[1]
\small
 \State $k \gets 0$ \Comment{Iteration counter}
 \State $\bfz^{(0)} \gets \bfz_0$ \Comment{Initial design}
 \State $f_{old} = -objtol$
\Repeat
	\For{$q=1,\ldots,N_b$}
		\For{$e=1,\ldots,N_{el}$}
			\State Compute signed distance $\phi_q$  to bar $q$ \Comment{\S\ref{sec:geom-proj}}
			\If{element $e$ outside of the reference region}
				\State Computed reflected element centroid $\hat{\bfx}_e$ \Comment{Eq.~(\ref{eq:sym})}
			\EndIf
			\State Compute projected density $\rho_q$  \Comment{Eq.~(\ref{eq:signed-dist})}
		\EndFor
	\EndFor
	\For{$k=1,\ldots,6$}
		\For{$e=1,\ldots,N_{el}$}
			\State Compute element stiffness matrix $\bfK_e^k$ using $\mathbb{C}(\bfz,\bfp)$ \Comment{Eq.~(\ref{eq:effC})}
			\State Assemble $\bfK_e^k$ into global stiffness matrix $\bfK_k$
			\State Compute element force $\bff_e^k$ contributions from applied unit strains  \Comment{Eq.~(\ref{eq:equilibrium})}
			\State Assemble $\bff_e^k$ into global force vectors $\bff_k$
		\EndFor
		\State Solve $\bfK_k(\bfz) \bfu_k(\bfz) = \bff_k$ for $\bfu(\bfz)$
	\EndFor
	\State Compute $f(\bfz)$, $\nabla f(\bfz)$ \Comment{Eq.~(\ref{eq:bulk}) or Eq.~(\ref{eq:shear}) or Eq.~(\ref{eq:poisson})}
	\State Compute $\mathbf{g}(\bfz)$, $\nabla \mathbf{g}(\bfz)$ \Comment{$\mathbf{g}$ denotes vector of constraints}
	\State Impose move limits and update $\bfz_{low}$ and $\bfz_{upp}$ \Comment{Eq.~(\ref{eq:move-limit})}
	\State  $\bfz^{(k+1)} \gets \texttt{opt}(\bfz^{(k)}, f, \nabla f, \mathbf{g}, \nabla \mathbf{g},  \bfz_{low}, \bfz_{upp})$ \Comment{Update design}
	\State $k \gets k + 1$
         \State Compute relative change in objective $\Delta f = | (f-f_{old})/f_{old} |$ 
         \State $f_{old} \gets f$
\Until{$\Delta f \leq objtol$ }
\normalsize
\end{algorithmic}
\end{algorithm}

\begin{table*}
\begin{centering}
\begin{tabular}{>{\centering}m{2cm}>{\centering}m{2cm}>{\centering}m{3cm} }
Parameter & Value & Equation/Section\tabularnewline 
\hline
\tabularnewline 
$\varepsilon_d^{0}$ & 1.0 &\ref{eq:disc-ineq} \tabularnewline 
$\varepsilon_d^{*}$ & 0.01 &\ref{eq:cont-gd} \tabularnewline 
$\varepsilon_m^{0}$ & 0.3 & \S\ref{sec:mutual} \tabularnewline 
$\varepsilon_m^{*}$ & 0.01 &\S\ref{sec:mutual} \tabularnewline 
$\varepsilon_n$ & $10^{-5}$ &\ref{eq:no-cut-constraint}\tabularnewline 
$K_{min}$ & 0.001 & \S\ref{subsec:opt-problem} \tabularnewline 
$p$ & 2 & \ref{eq:smooth-H} \tabularnewline 
$k$ & 25 & \ref{eq:KS}  \tabularnewline 
$m$ & 0.1& \ref{eq:move-limit}  \tabularnewline 
$c$ & 1.0 & \S\ref{subsec:computer-implementation} \tabularnewline
\hline 
\end{tabular}
\par\end{centering}
\caption{Optimization parameters}
\label{table:parameters}
\end{table*}

\section{Examples}
\label{sec:examples}

We now present examples to illustrate the effectiveness of the proposed method. We employ hexahedral, trilinear elements to mesh the unit cell. We employ a system with six compute nodes with 24 Intel Haswell cores each, CPU speed of 2.59 GHz and 128 GB of memory per node to perform all the examples.

\subsection{Maximal Bulk Modulus of Two- and Three-material Lattices with Cubic Symmetry}
\label{sec:cubic}
\begin{figure}[h]
\centering
 \includegraphics[width=0.25\columnwidth, bb=300 50 750 550, clip=true]{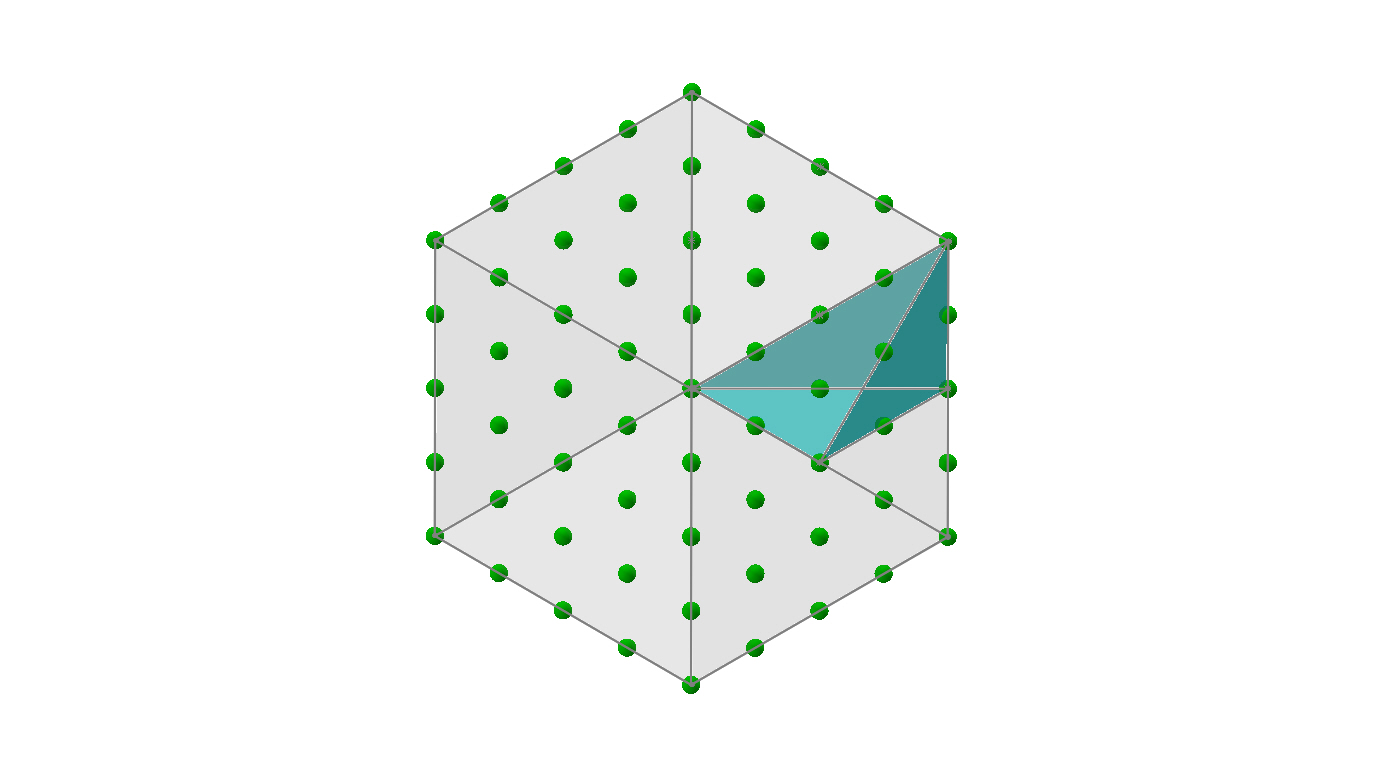}
\caption{Initial design after reflection for cubic symmetry. Blue region indicates reference region. }
\label{fig:initial}
\end{figure}

In this example, we maximize the bulk modulus for lattices with cubic symmetry and made of two and three materials.  To enforce the cubic symmetry, we define nine symmetry planes, including the three orthogonal planes perpendicular to the faces that pass through the center of the unit cell, and the six planes that pass through the origin and two opposite edges that divide the cube into two equal partitions. The initial design consists of 10 bars with near-zero length (which resemble spheres) and width  $w = 0.1$, as shown in Fig.~\ref{fig:initial}. All materials are homogeneous, isotropic, and linearly elastic, with Poisson's ratio $\nu = 0.3$.  The unit cell is meshed with a regular grid of $64 \times 64 \times 64$ elements. 

For the two-material designs, the available materials have Young's moduli $E_1 = 10$ and $E_2 = 5$, and physical densities $\gamma_1 = 0.9$ and $\gamma_2 = 0.45$.  We set the initial size variables corresponding to these two materials to $\alpha_1^q = \alpha_2^q = 0.5$. 

The results of the optimization for different weight fraction limits are presented in Table ~\ref{table:cubic_bulk}. It is worth noting that small changes in the weight fraction limit produce completely different designs. This is expected, since this problem is known to have many local minima \cite{bendsoe2013topology}. The overall trend is that the maximal bulk modulus increases as we increase the weight fraction limit as expected.  We posit the exceptions to a strict monotonic behavior correspond to convergence to local minima.   One possibility to obtain better minima would be to employ the tunneling method proposed in \cite{zhang2018finding}; however, this is outside the scope of this paper.  

An important difference between these results and those previously published in \cite{kazemi2019topology} is that the no-cut constraint is effective in rendering struts that do not have cuts that would be difficult to manufacture.  The no-cut constraint also performs an important task, namely to prevent agglomerations of cut struts on a face of the unit cell to effectively produced a thin wall, which produces a closed-cell design (this situation was observed in  \cite{kazemi2019topology}).

\begin{table*}
\begin{centering}
\begin{tabular}{>{\centering}m{1.4cm}>{\centering}m{0.6cm}>{\centering}m{1.8cm}>{\centering}m{1.8cm}|>{\centering}m{1.4cm}>{\centering}m{.6cm}>{\centering}m{1.8cm}>{\centering}m{1.8cm}}
$K$ & $w_f^*$ & iso & side &$K$ & $w_f^*$ & iso & side 
\tabularnewline 
\hline \\[-1ex]
$ 0.06999 $ & $0.0444$ & \includegraphics[width=1.8cm, bb=250 50 950 650, clip=true]{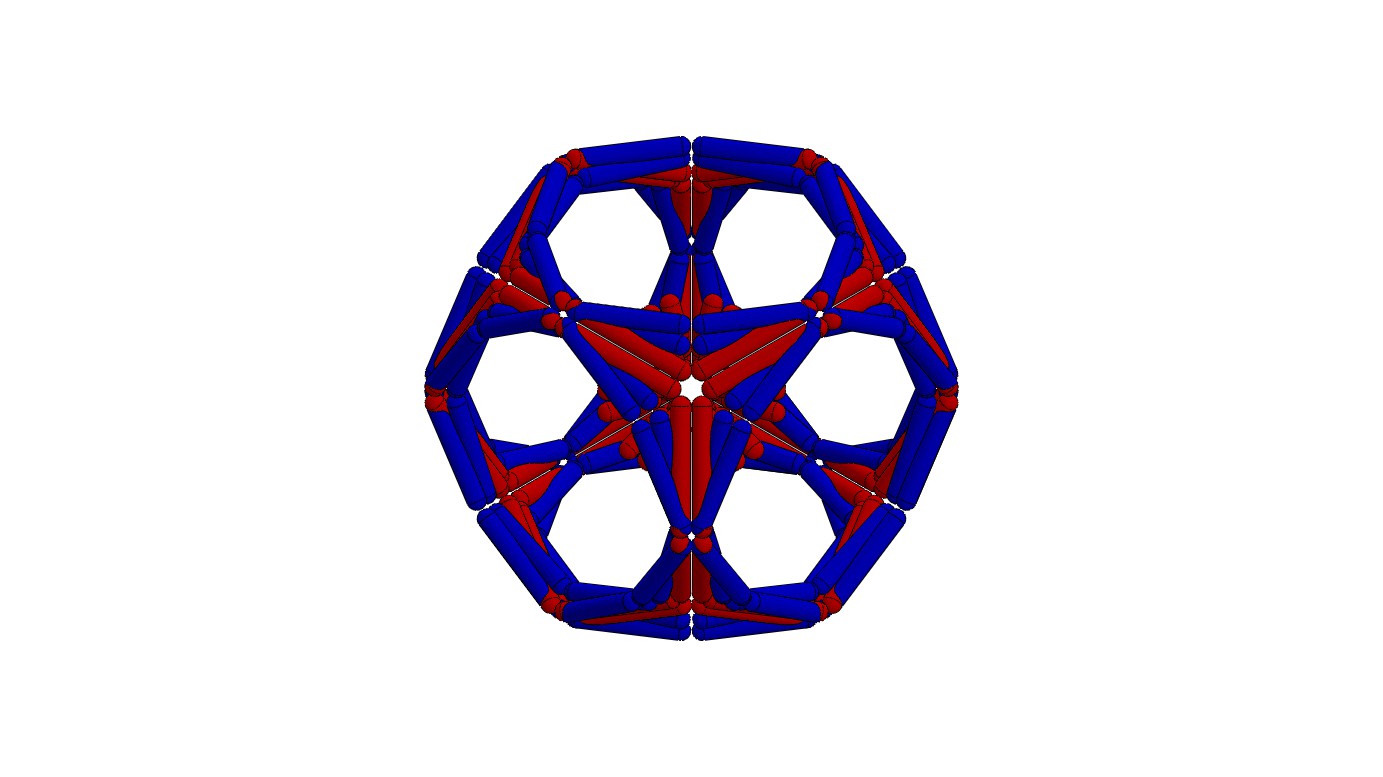} &  \includegraphics[width=1.8cm, bb=250 50 1000 700, clip=true]{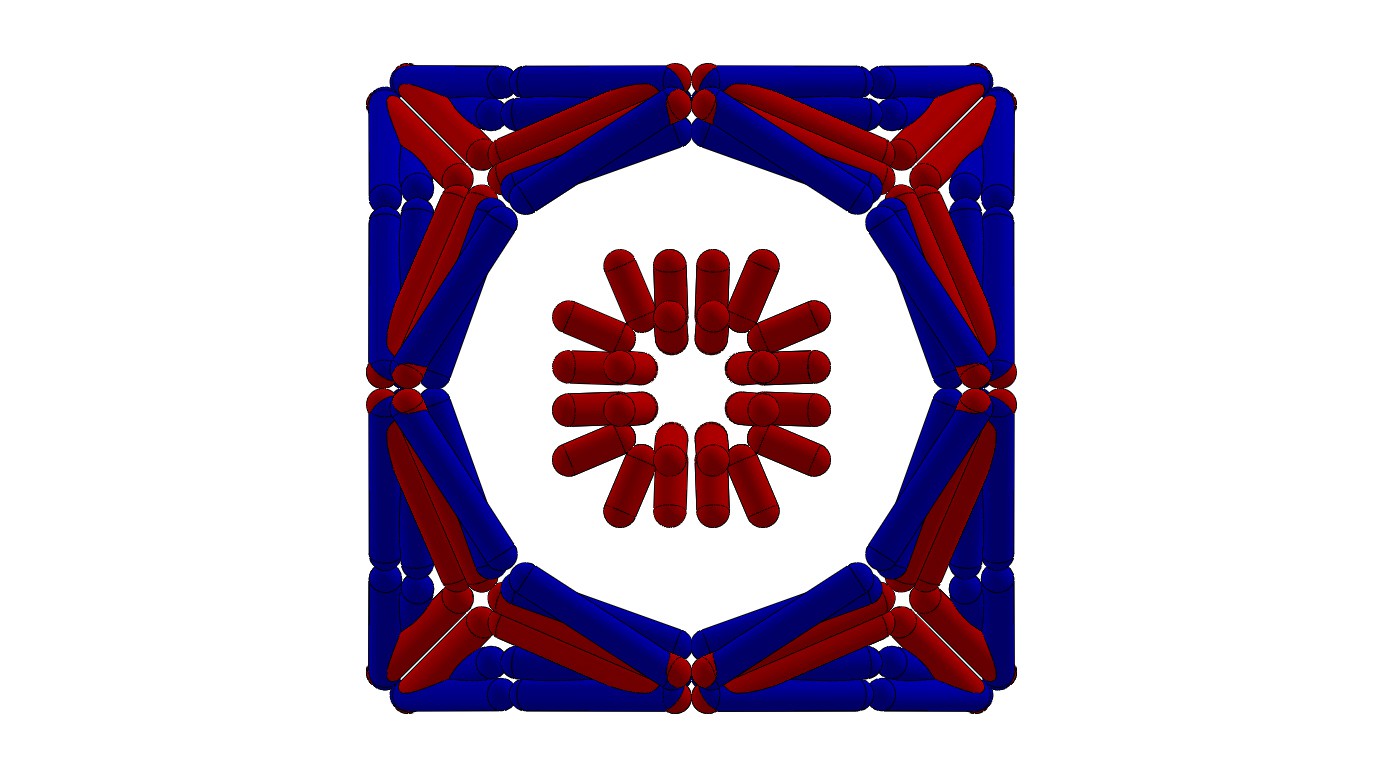}  &$ 0.13853 $ & $0.0778$ & \includegraphics[width=1.8cm, bb=250 50 1000 700, clip=true]{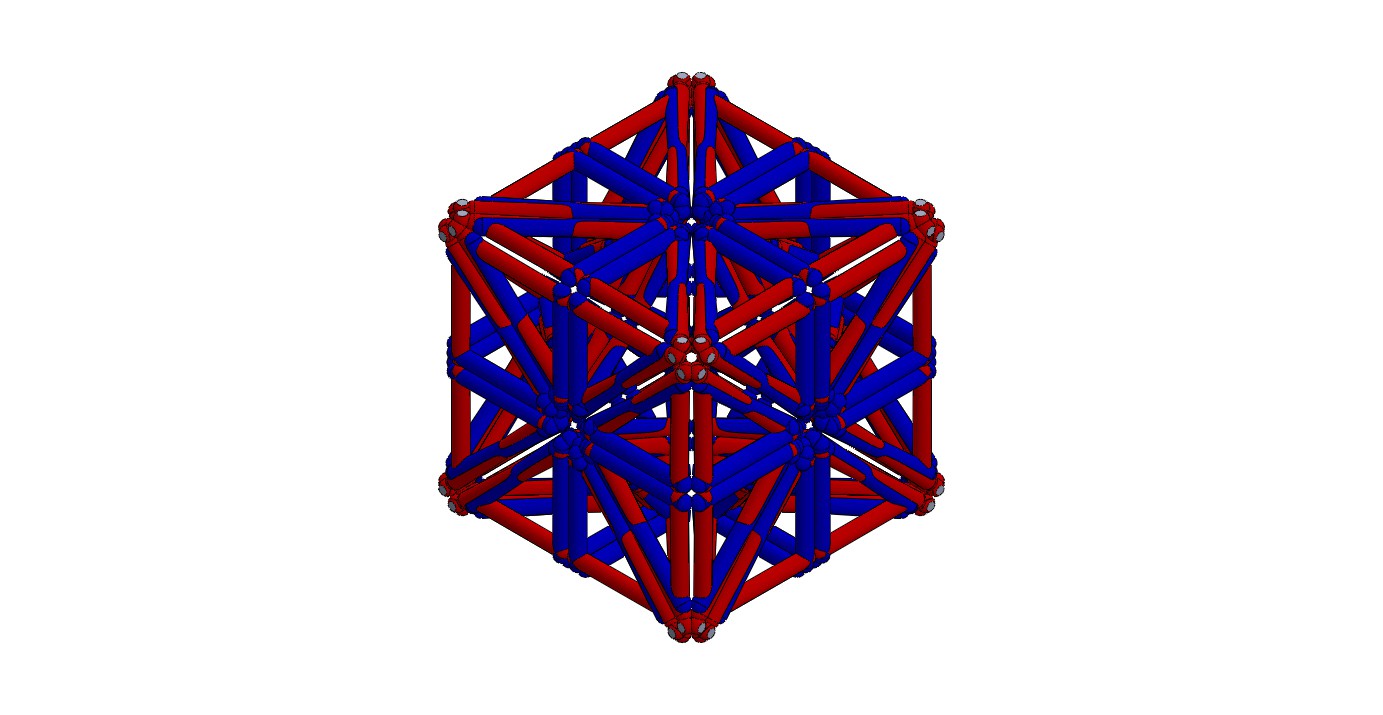} &  \includegraphics[width=1.8cm, bb=250 50 1000 700, clip=true]{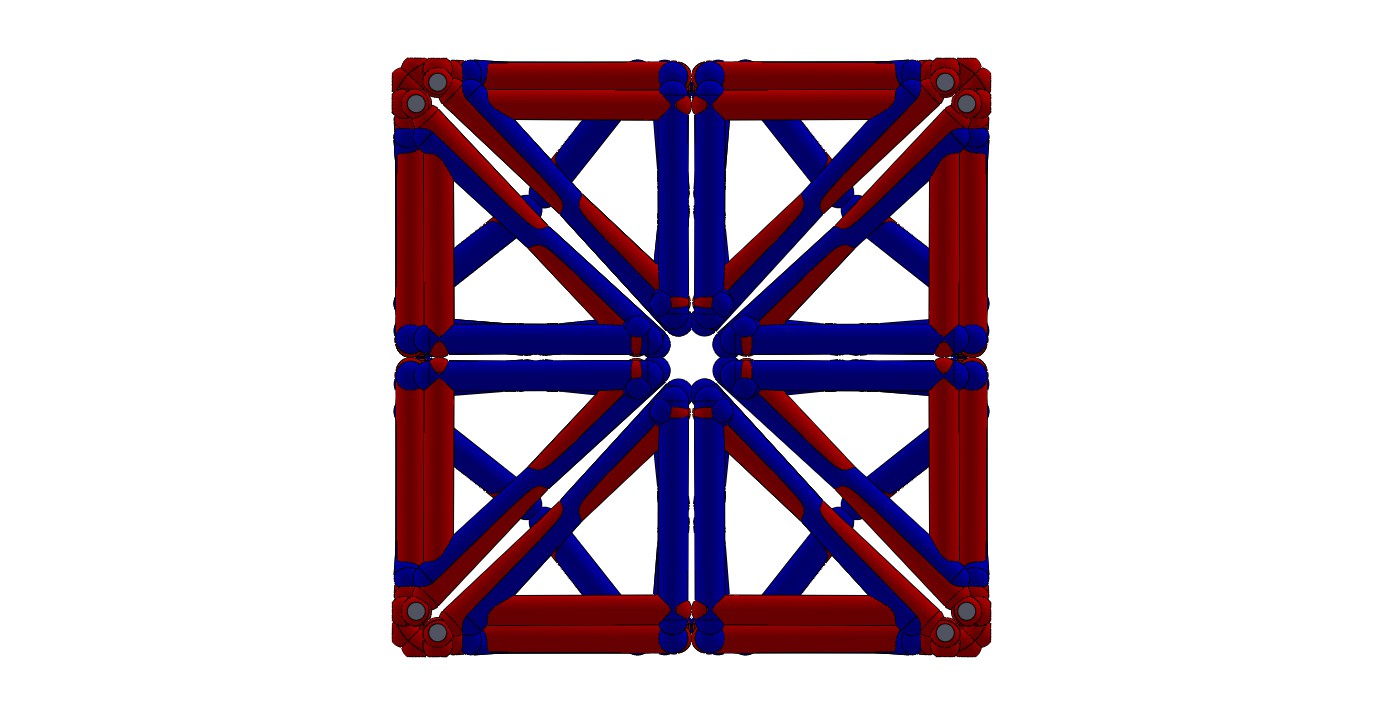} \tabularnewline
$ 0.08031 $ & $0.05$ & \includegraphics[width=1.8cm, bb=250 50 1000 700, clip=true]{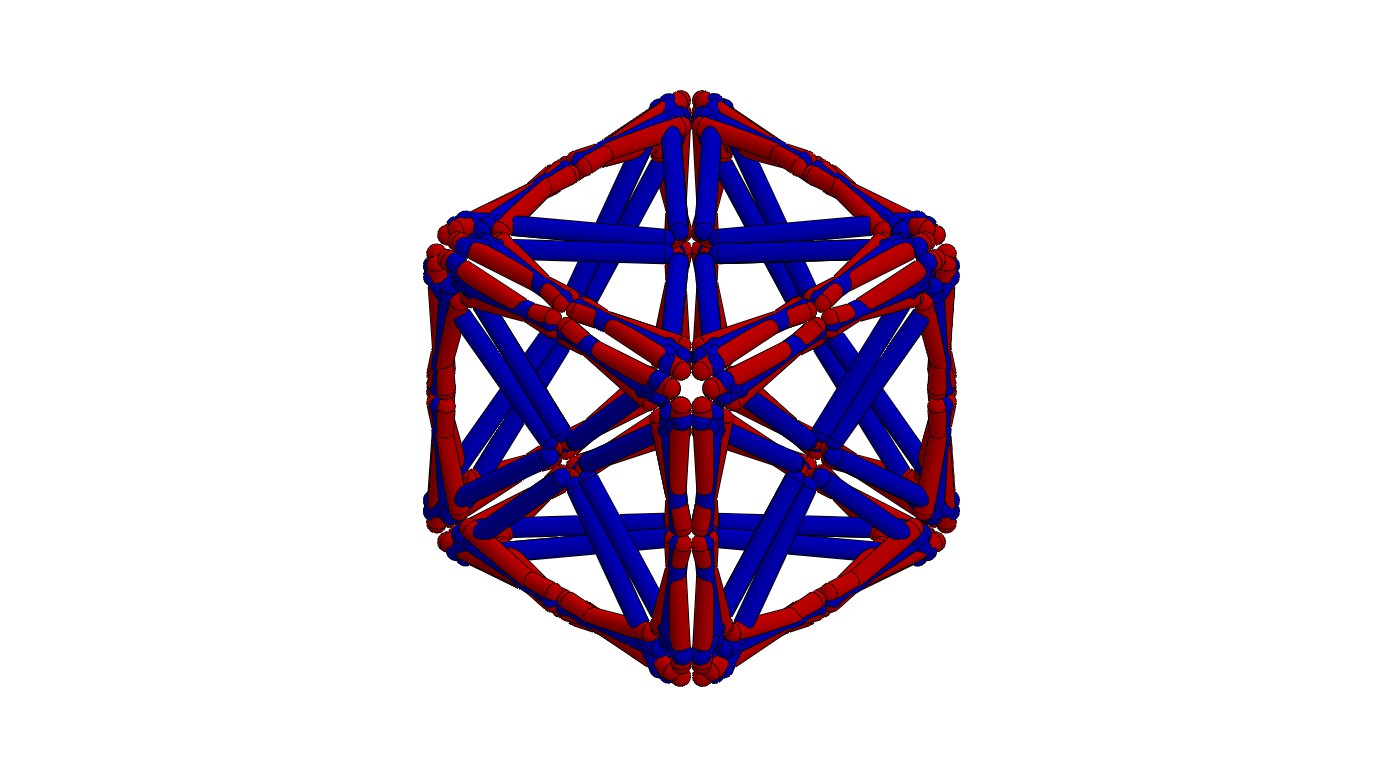} &  \includegraphics[width=1.8cm, bb=250 50 1000 700, clip=true]{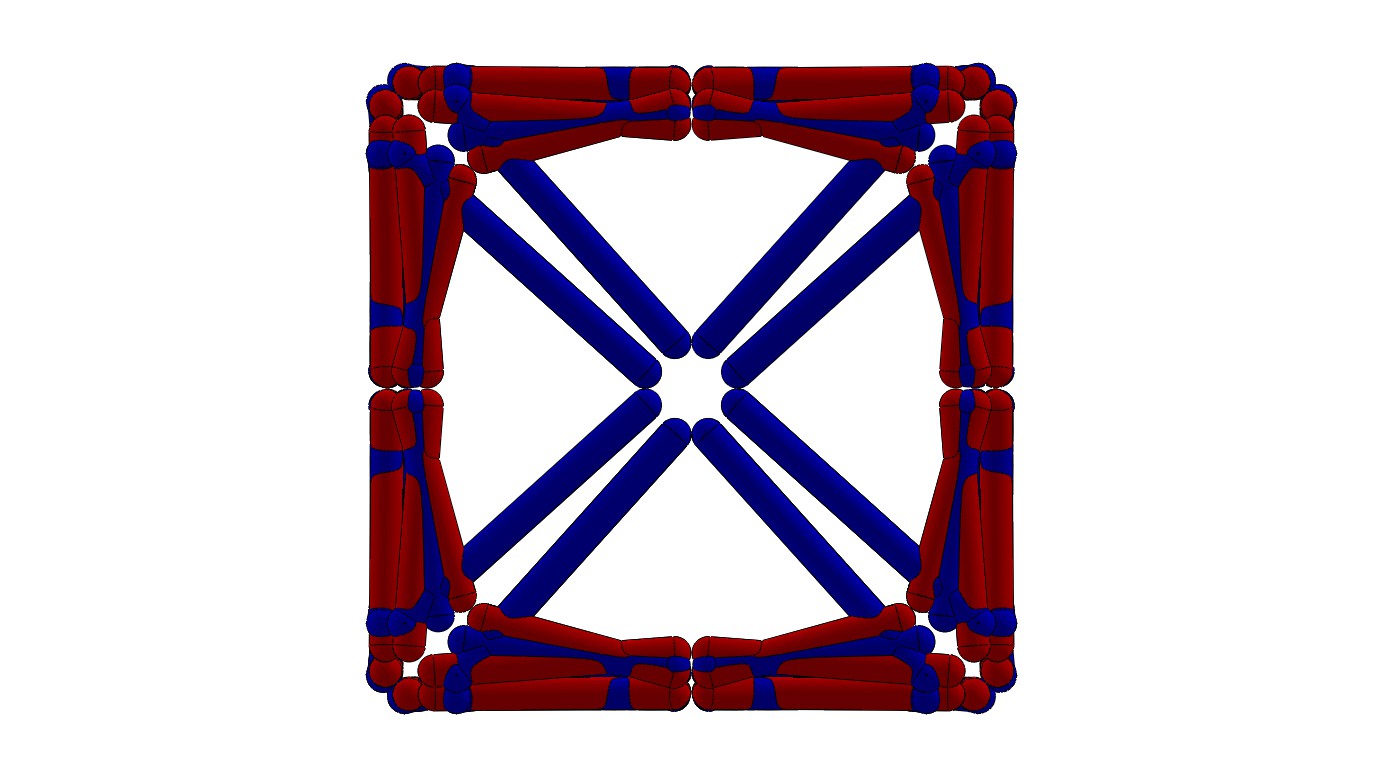}  &$ 0.12034  $ & $0.0833$ & \includegraphics[width=1.8cm, bb=250 50 1000 700, clip=true]{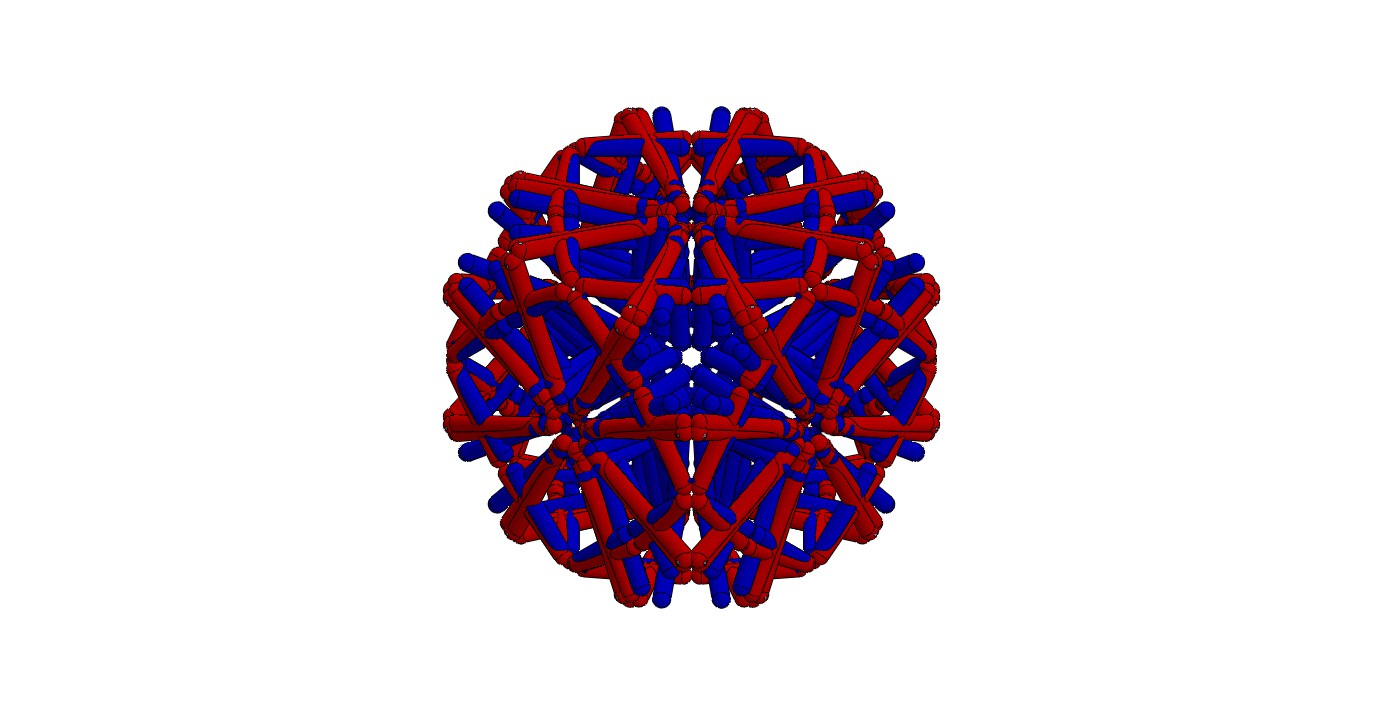} &  \includegraphics[width=1.8cm, bb=250 50 1000 700, clip=true]{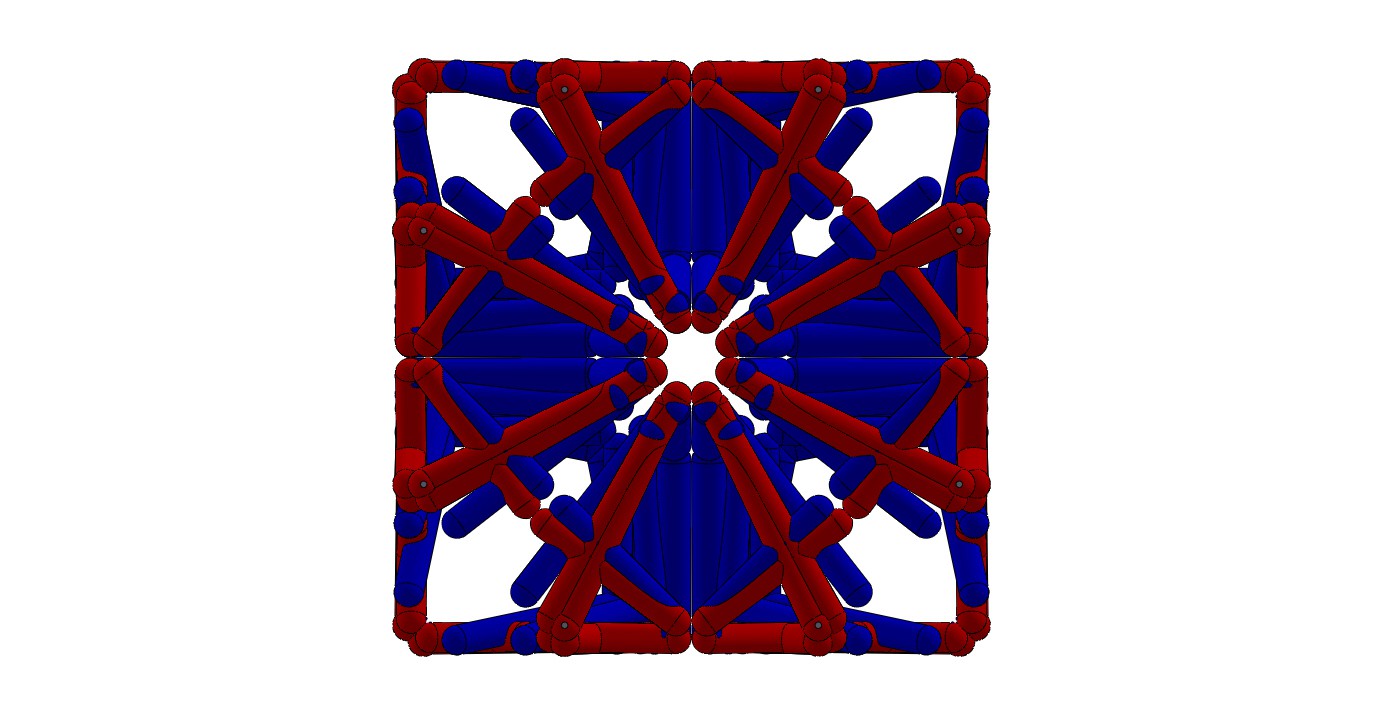} \tabularnewline
$ 0.09229 $ & $0.0556$ & \includegraphics[width=1.8cm, bb=250 50 1000 700, clip=true]{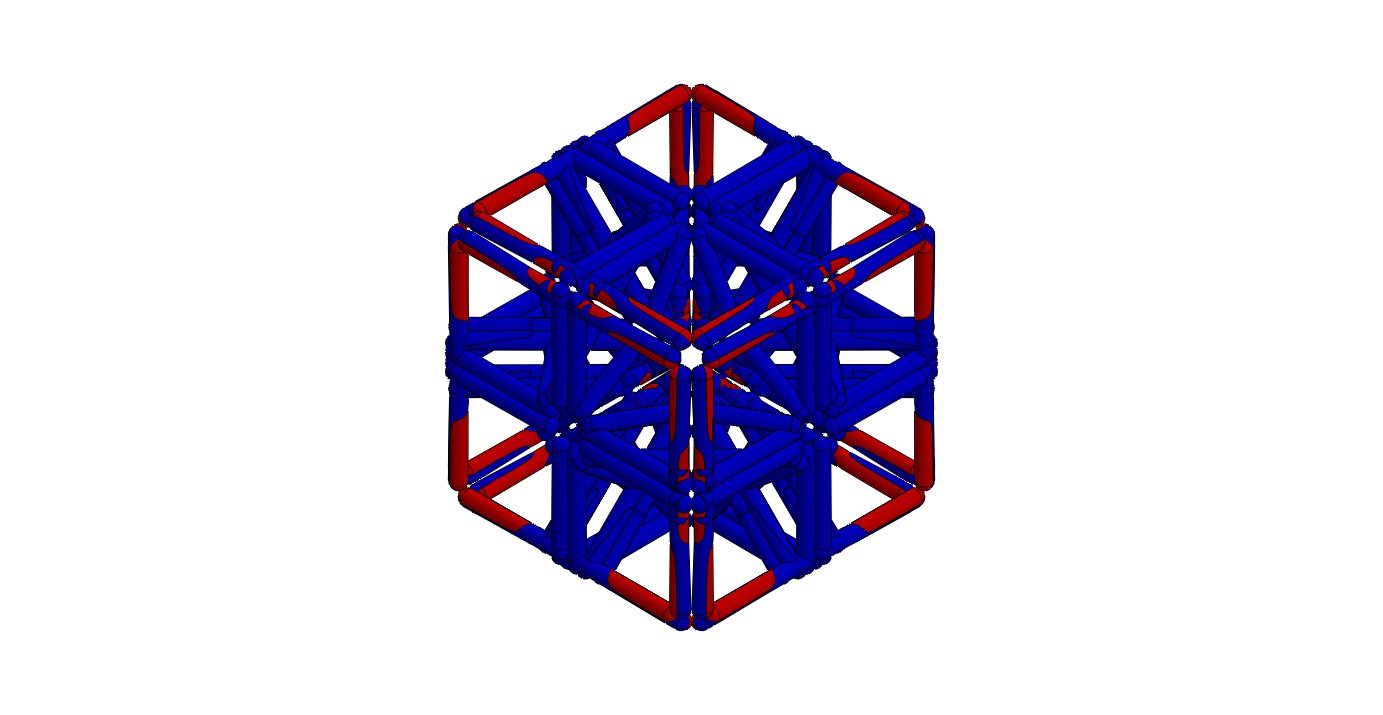} &  \includegraphics[width=1.8cm, bb=250 50 1000 700, clip=true]{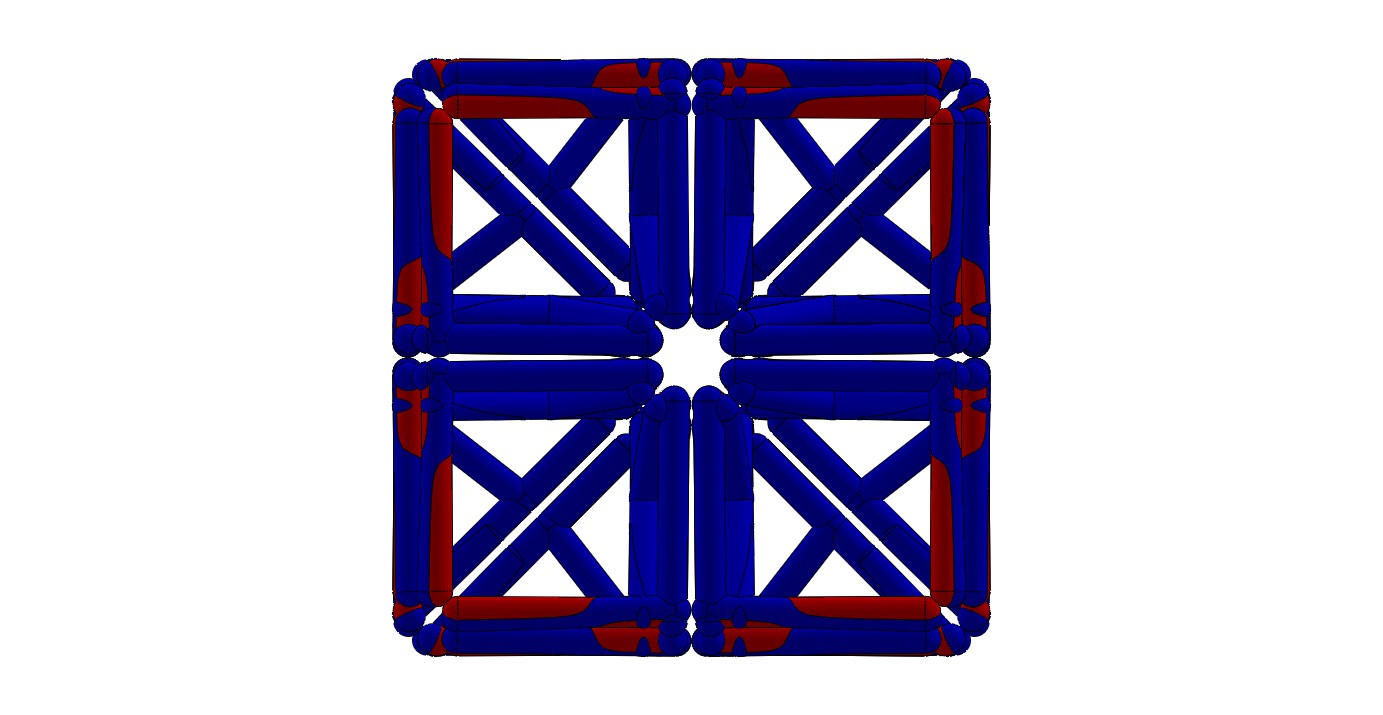}
&$ 0.13908 $ & $0.0889$ & \includegraphics[width=1.8cm, bb=250 50 1000 700, clip=true]{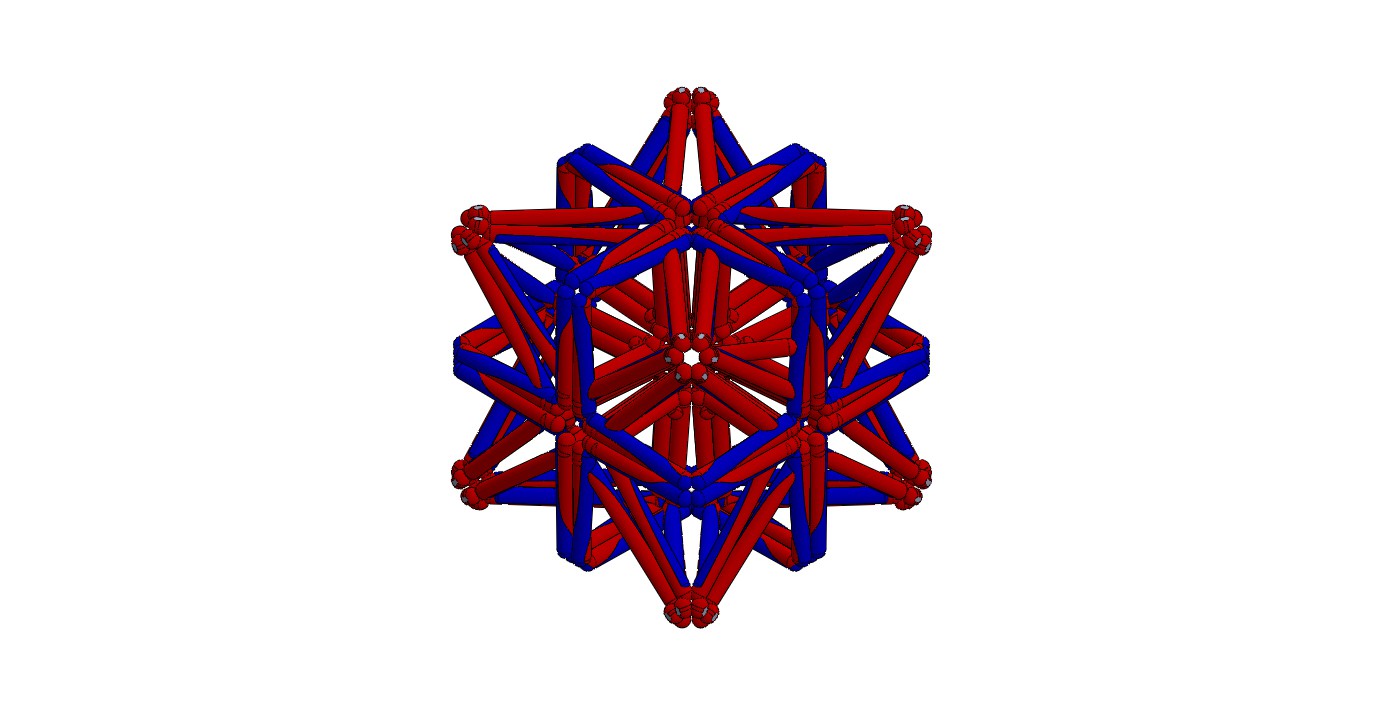} &  \includegraphics[width=1.8cm, bb=250 50 1000 700, clip=true]{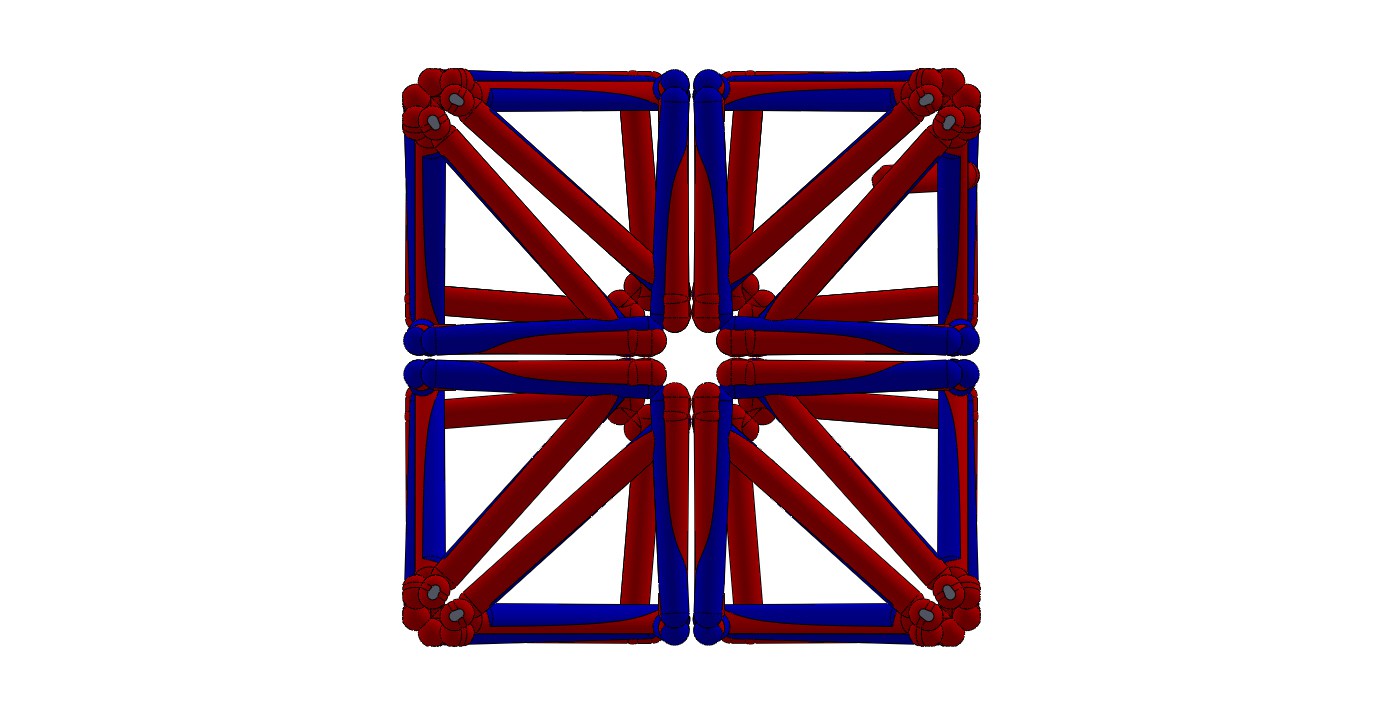} 
  \tabularnewline
$ 0.1171 $ & $0.0611$ & \includegraphics[width=1.8cm, bb=250 50 1000 700, clip=true]{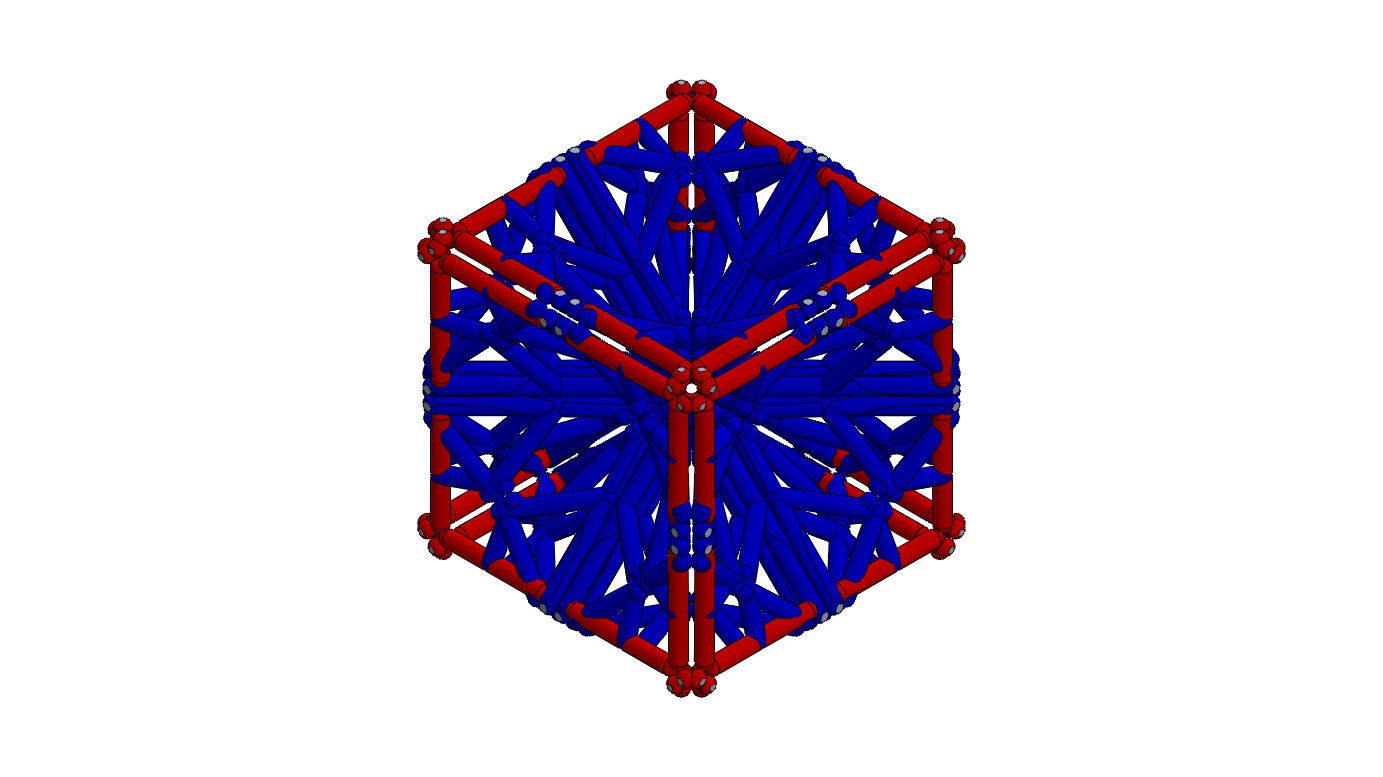} &  \includegraphics[width=1.8cm, bb=250 50 1000 700, clip=true]{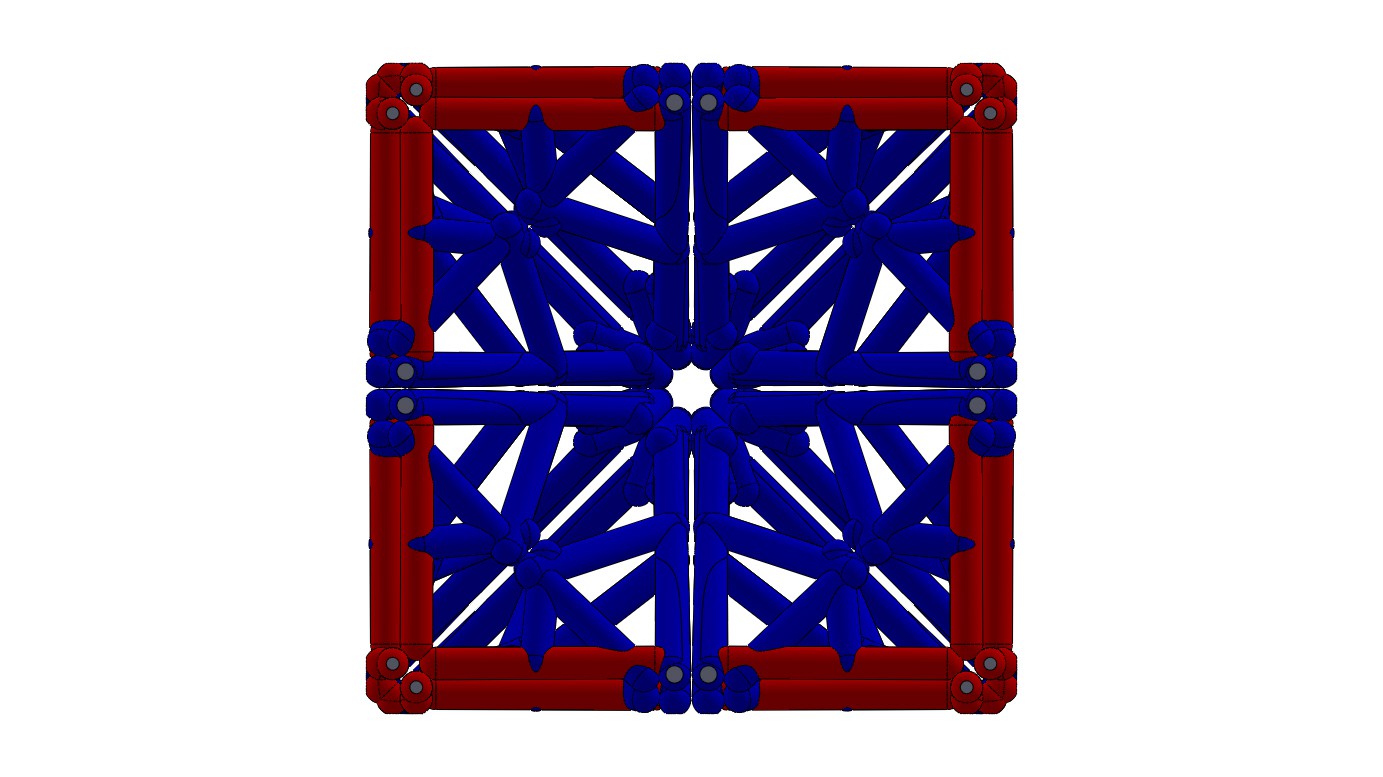}
& $ 0.15736 $ & $0.0944$ & \includegraphics[width=1.8cm, bb=250 50 1000 700, clip=true]{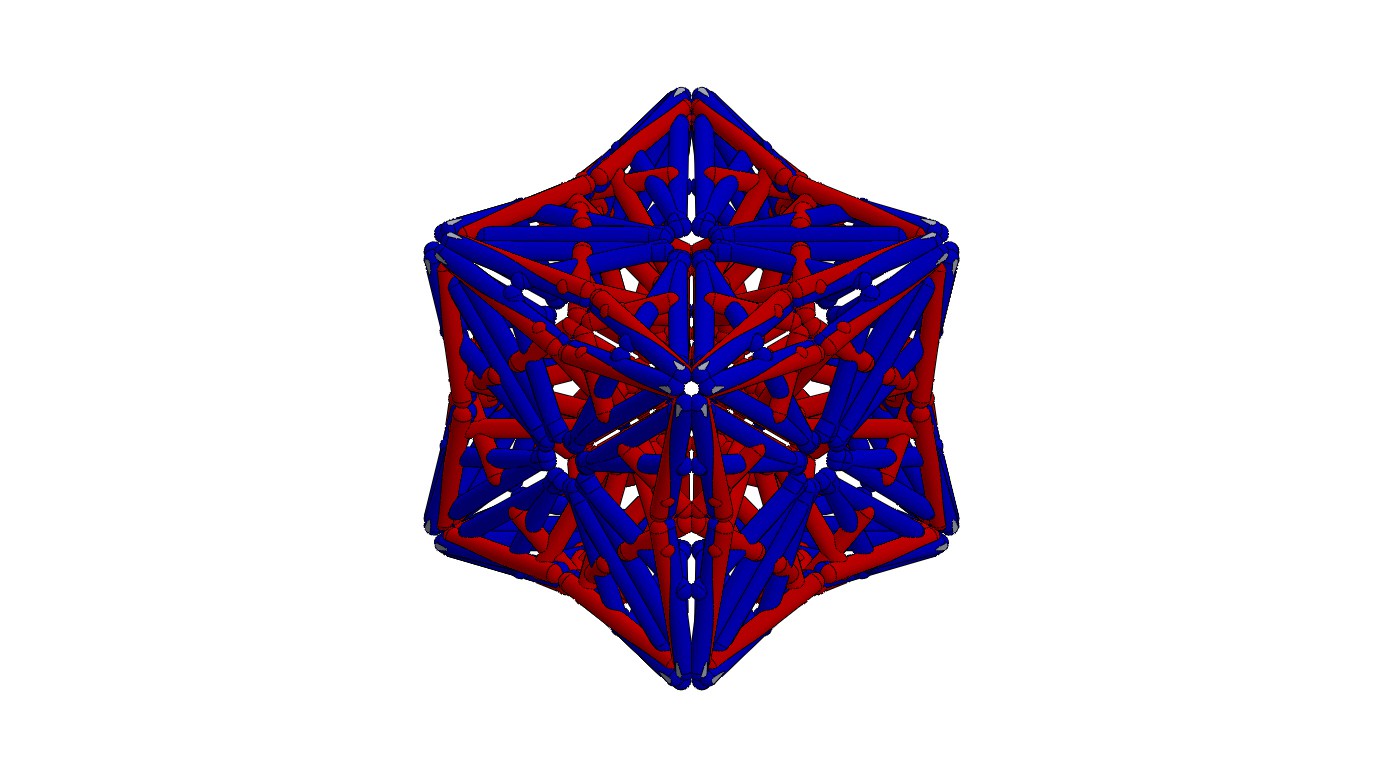} &  \includegraphics[width=1.8cm, bb=250 50 1000 700, clip=true]{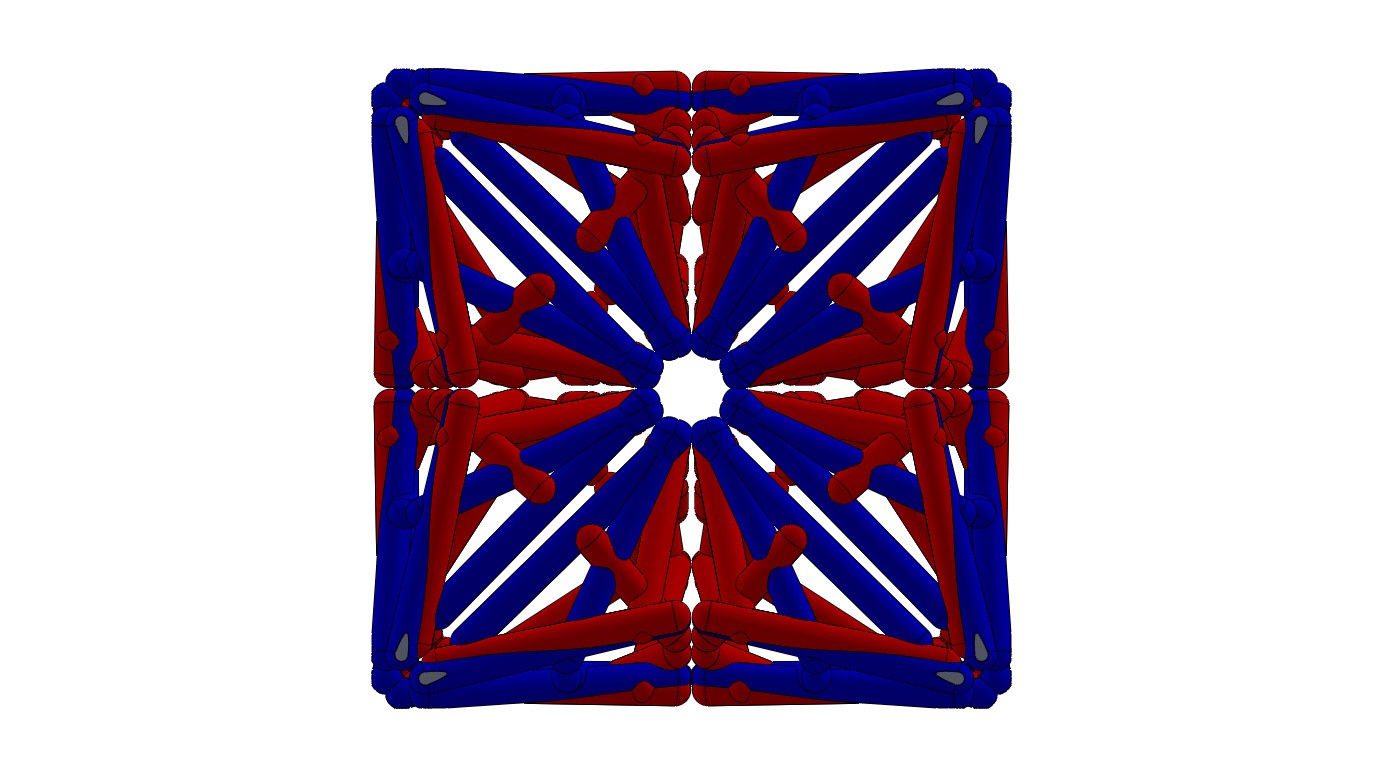} \tabularnewline
$ 0.09543 $ & $0.0667$ & \includegraphics[width=1.8cm, bb=250 50 1000 700, clip=true]{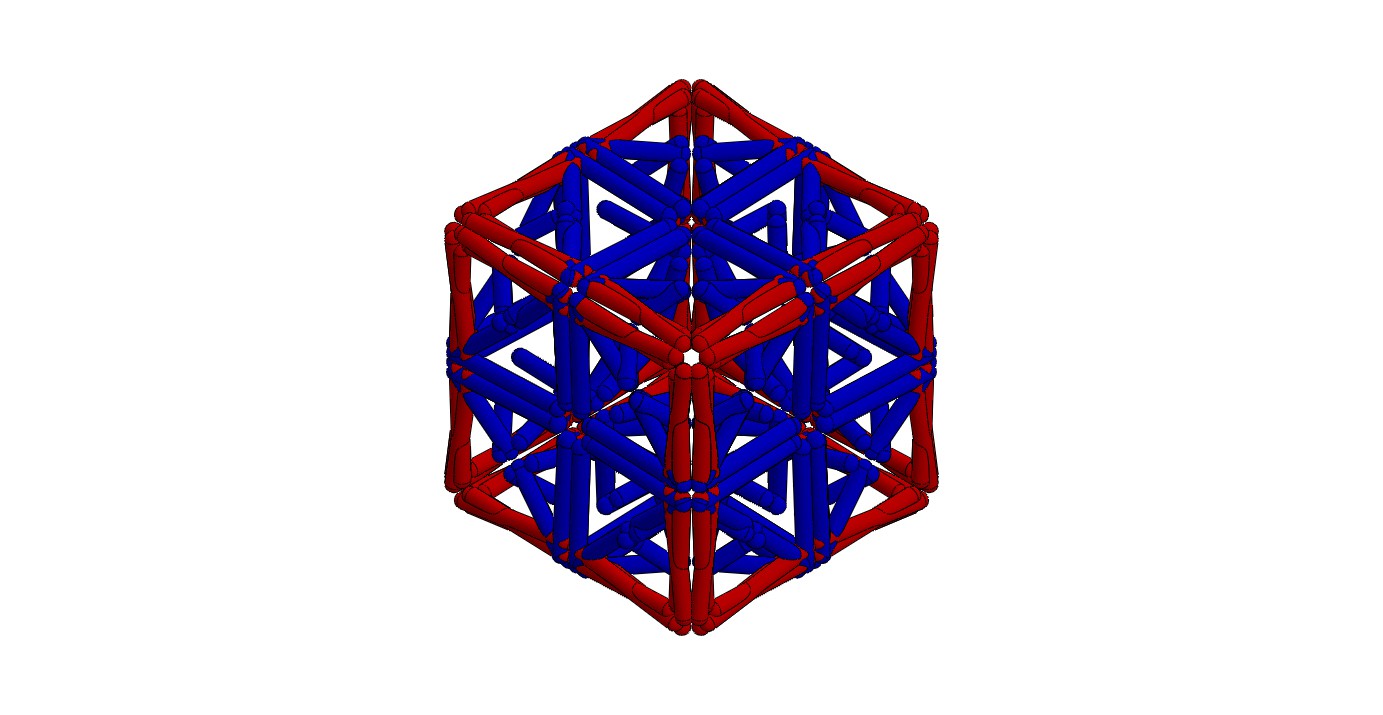} &  \includegraphics[width=1.8cm, bb=250 50 1000 700, clip=true]{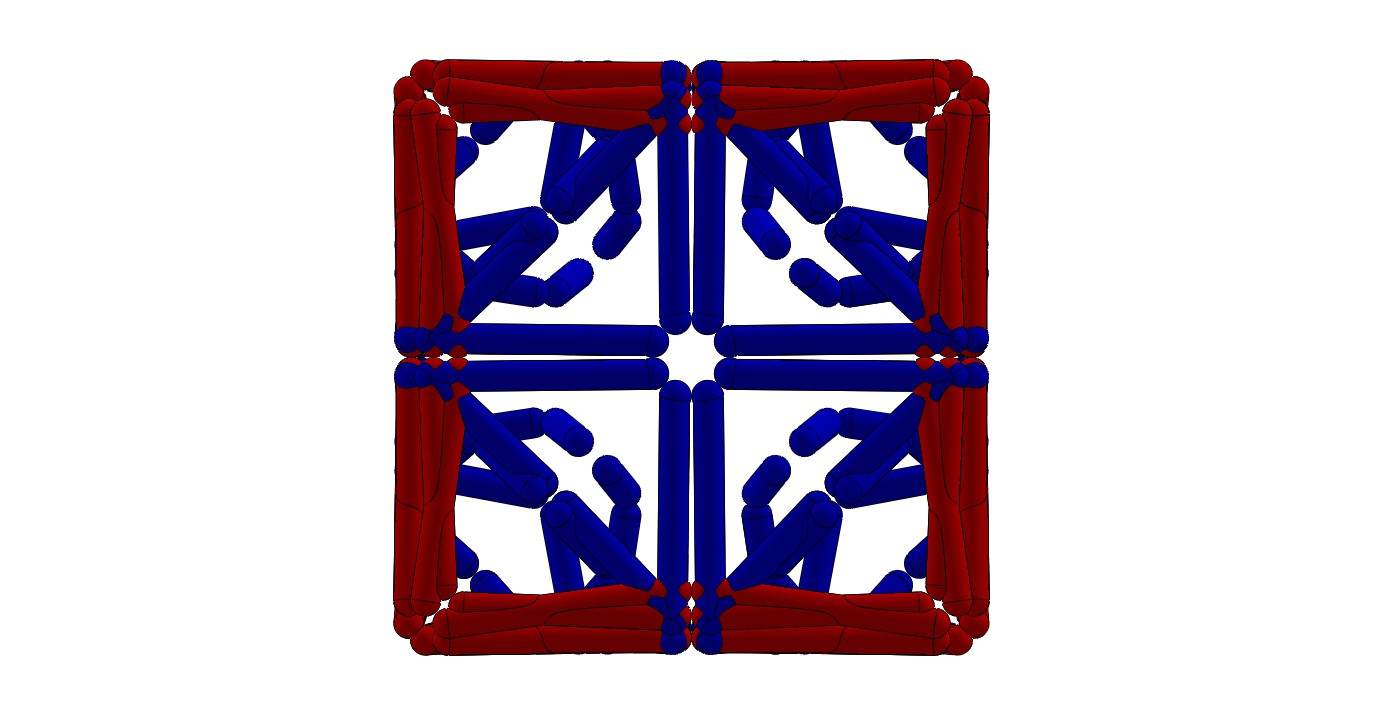}
&$ 0.18651 $ & $0.1$ & \includegraphics[width=1.8cm, bb=250 50 1000 700, clip=true]{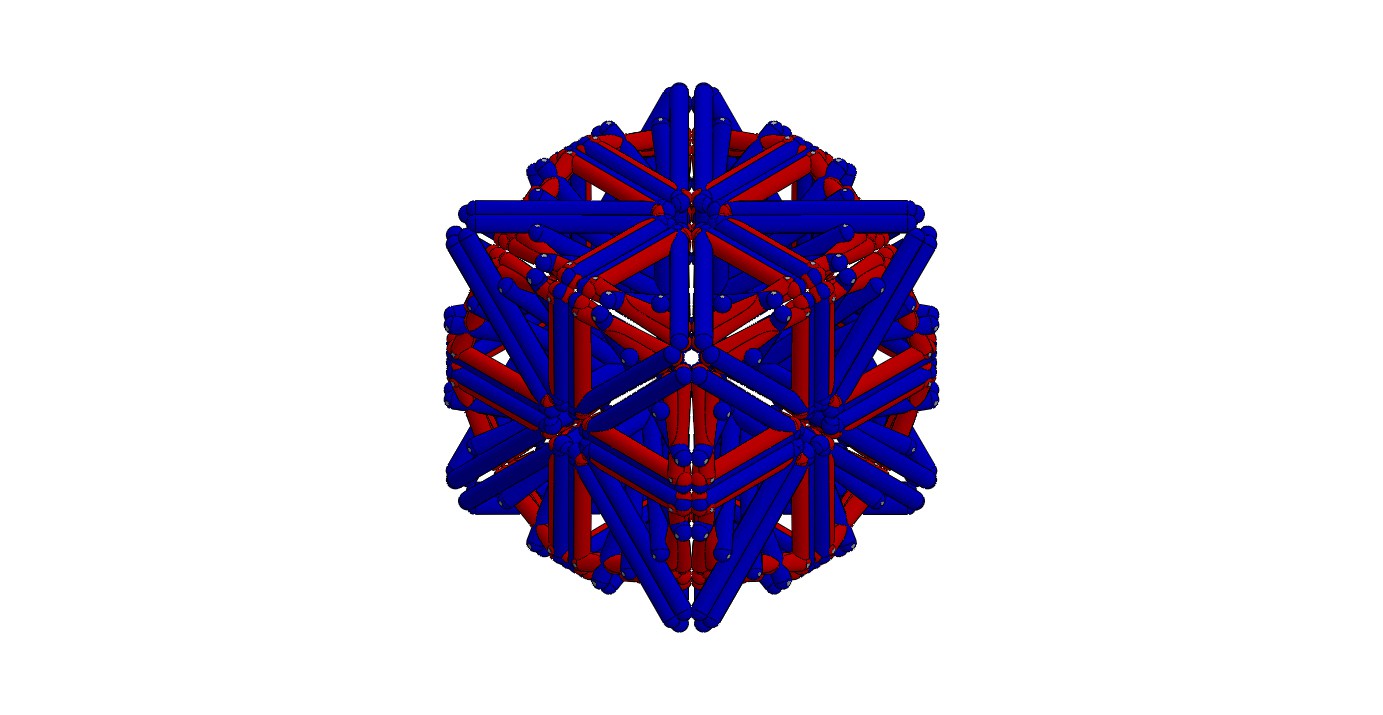} &  \includegraphics[width=1.8cm, bb=250 50 1000 700, clip=true]{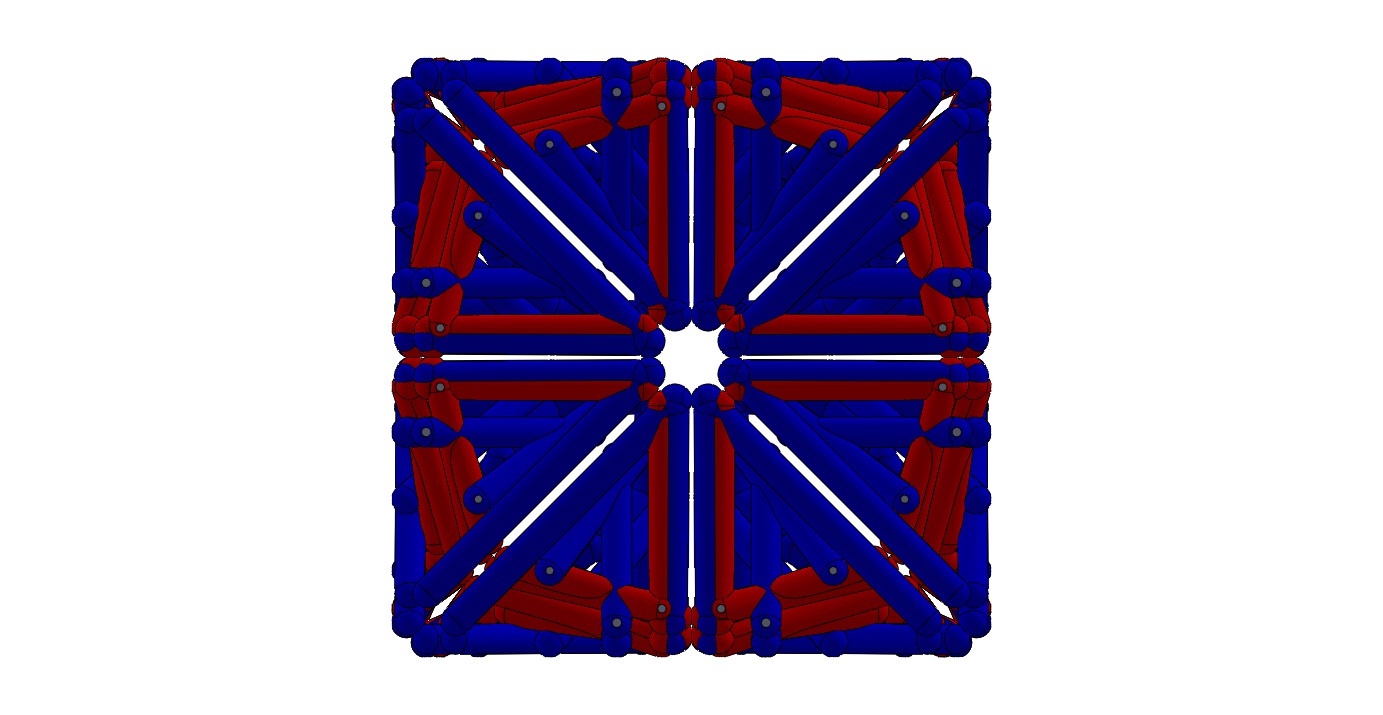} \tabularnewline
$ 0.12366 $ & $0.0722$ & \includegraphics[width=1.8cm, bb=250 50 1000 700, clip=true]{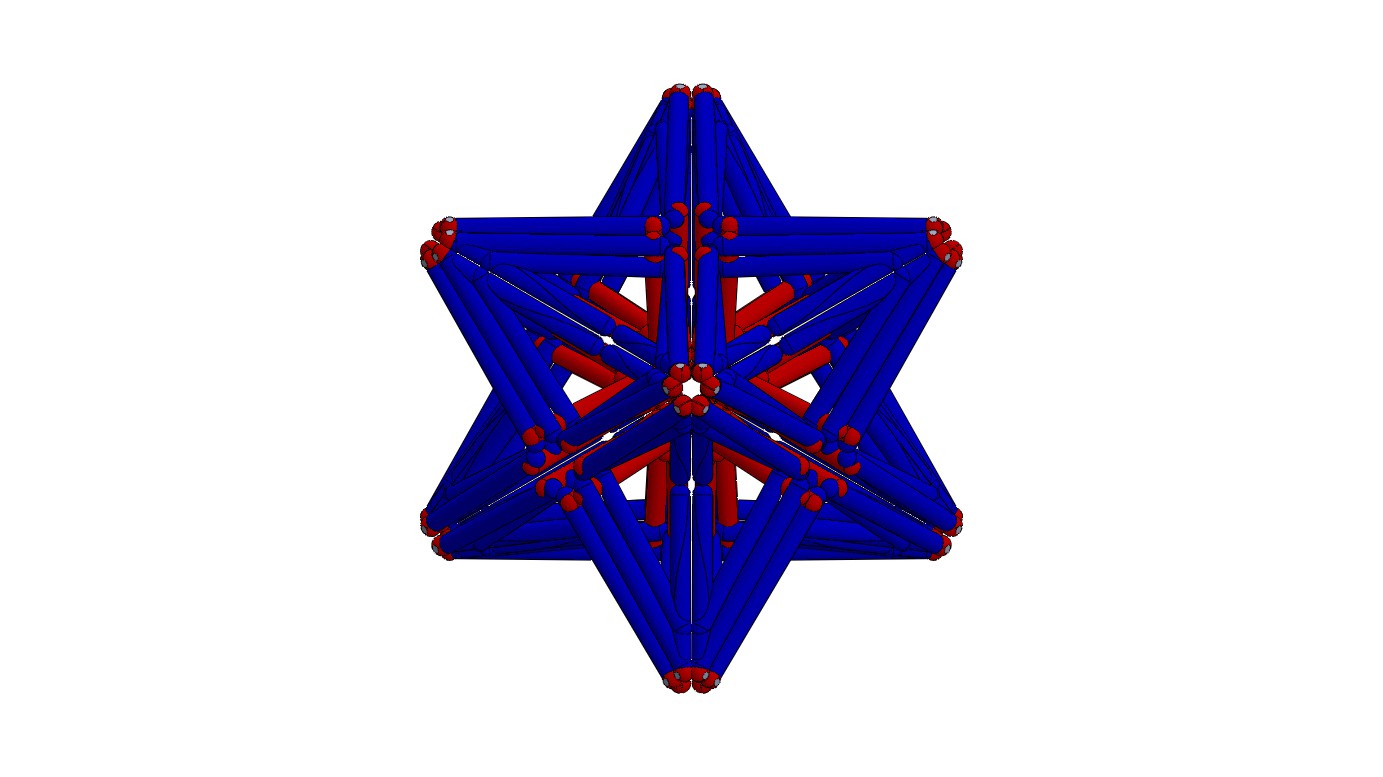} &  \includegraphics[width=1.8cm, bb=250 50 1000 700, clip=true]{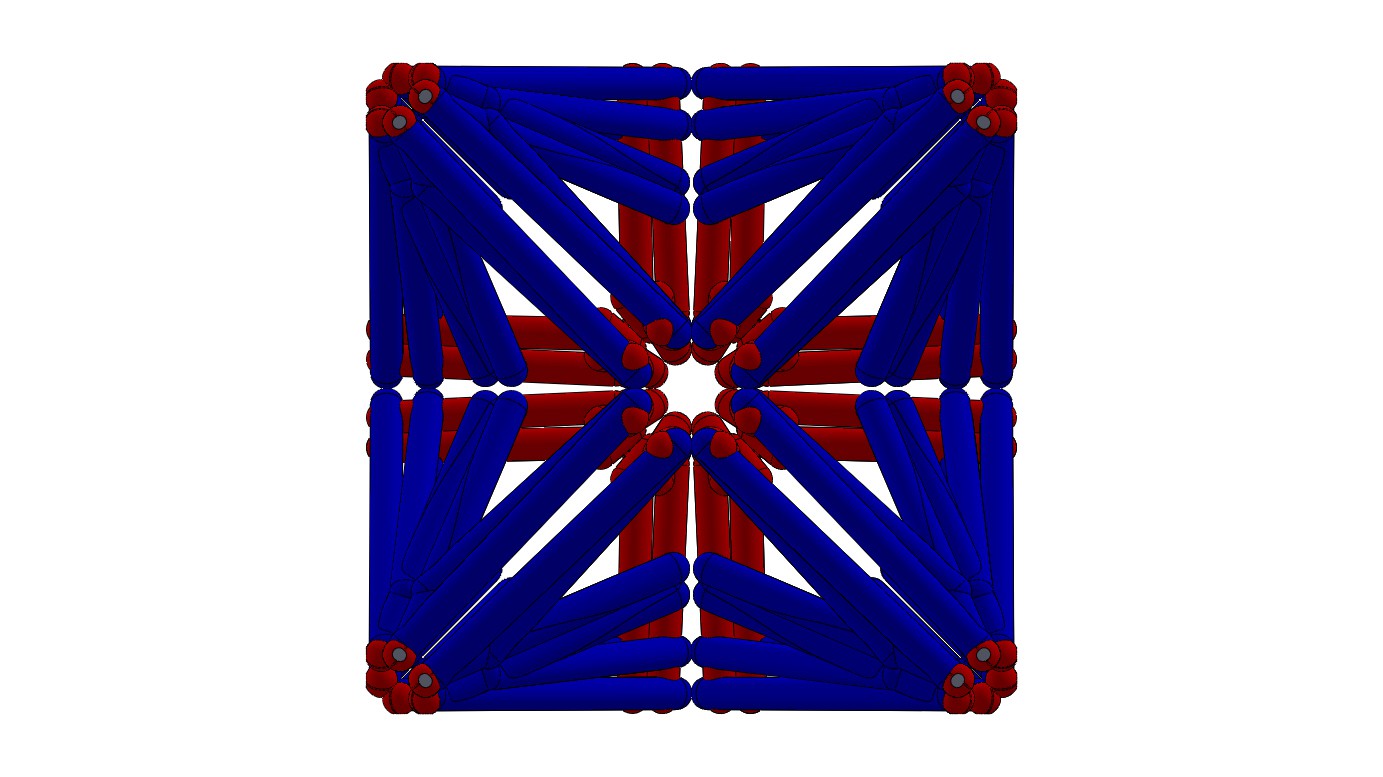}
 &$ 0.17525 $ & $0.1056$ & \includegraphics[width=1.8cm, bb=250 50 1000 700, clip=true]{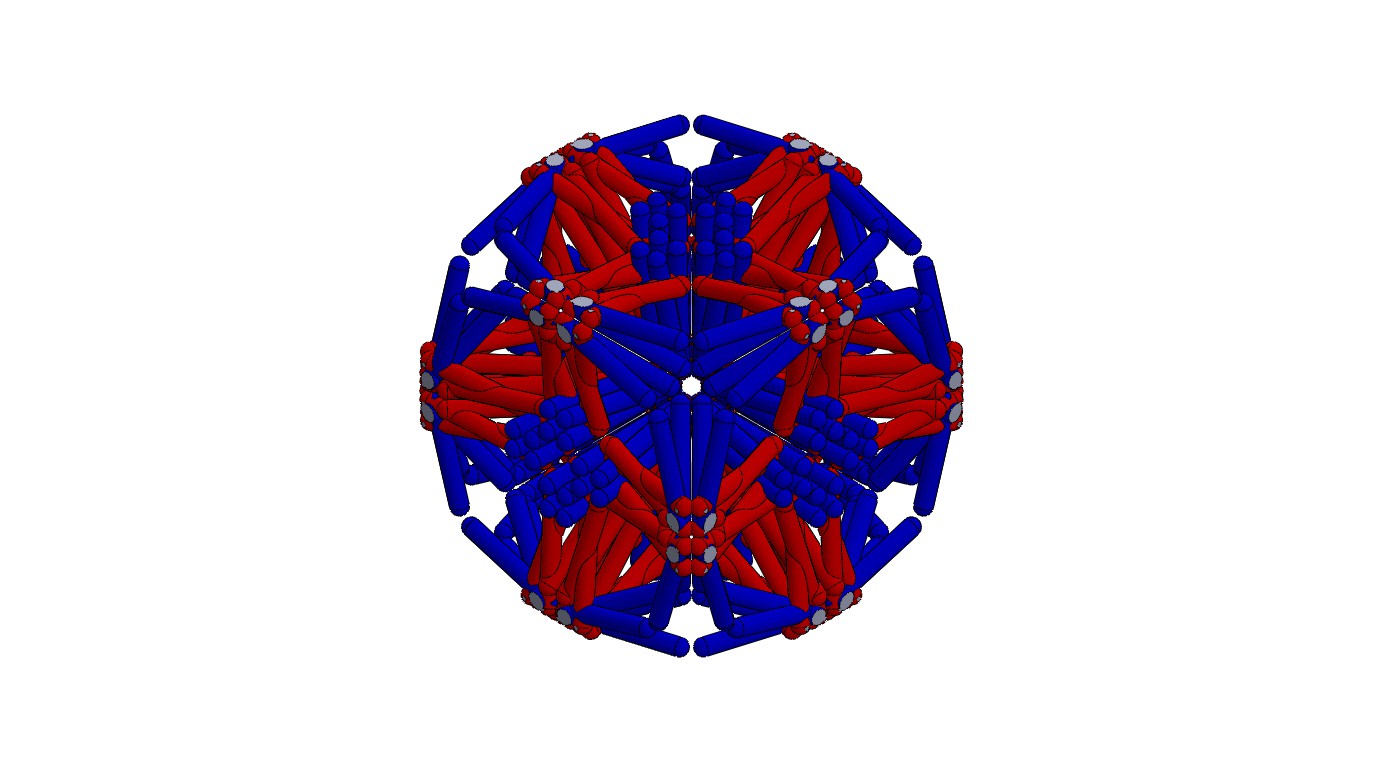} &  \includegraphics[width=1.8cm, bb=250 50 1000 700, clip=true]{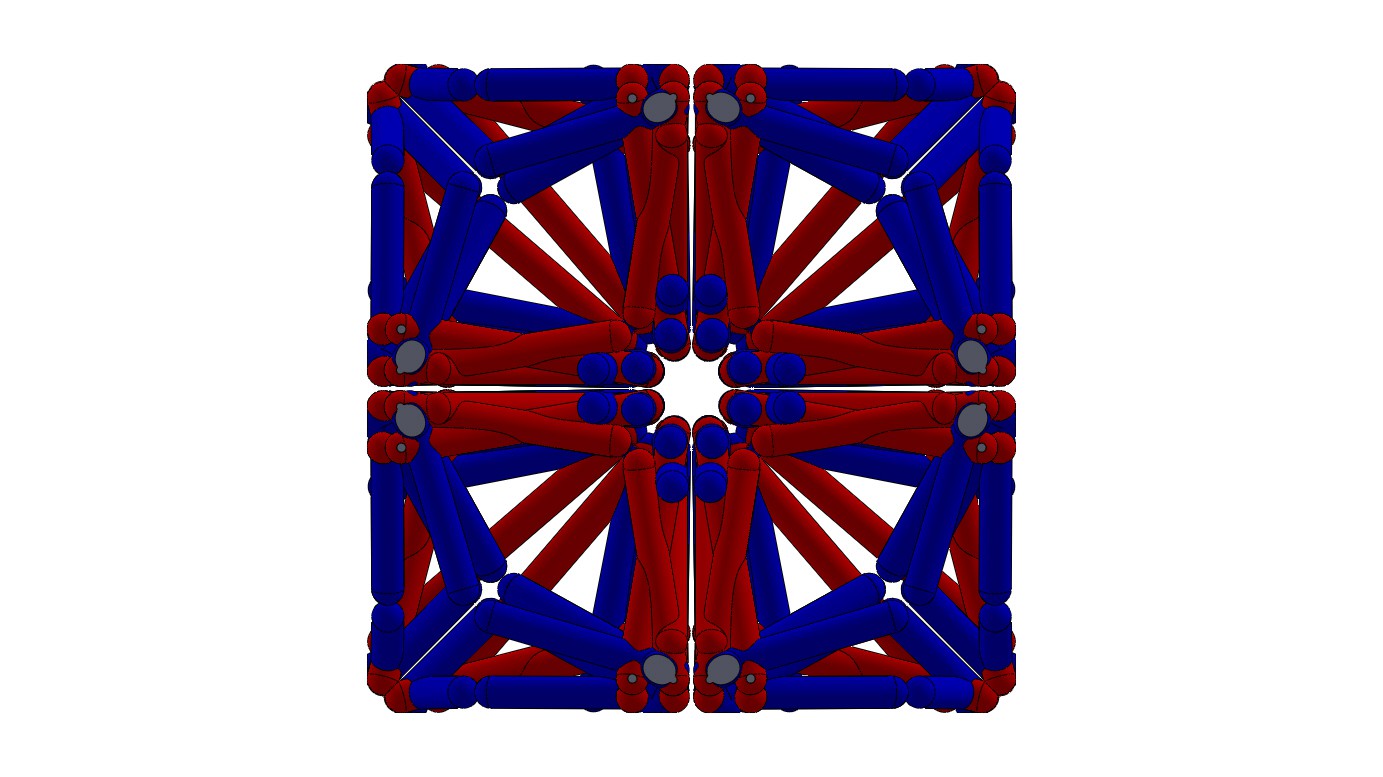} \tabularnewline
\hline 
\end{tabular}
\par\end{centering}
\caption{Maximal bulk modulus designs for cubic two-material lattices and different weight fraction limits $w_f^*$. Red bars are made of material 1, blue bars are made of material 2, and bars that have been removed from the design (i.e., with $\alpha_1^q, \alpha_2^q \approx 0$) are not shown.}
\label{table:cubic_bulk}
\end{table*}

\begin{figure}[h]
\centering
 \begin{subfigure}[c]{0.45\textwidth}
    \includegraphics[width=\columnwidth]{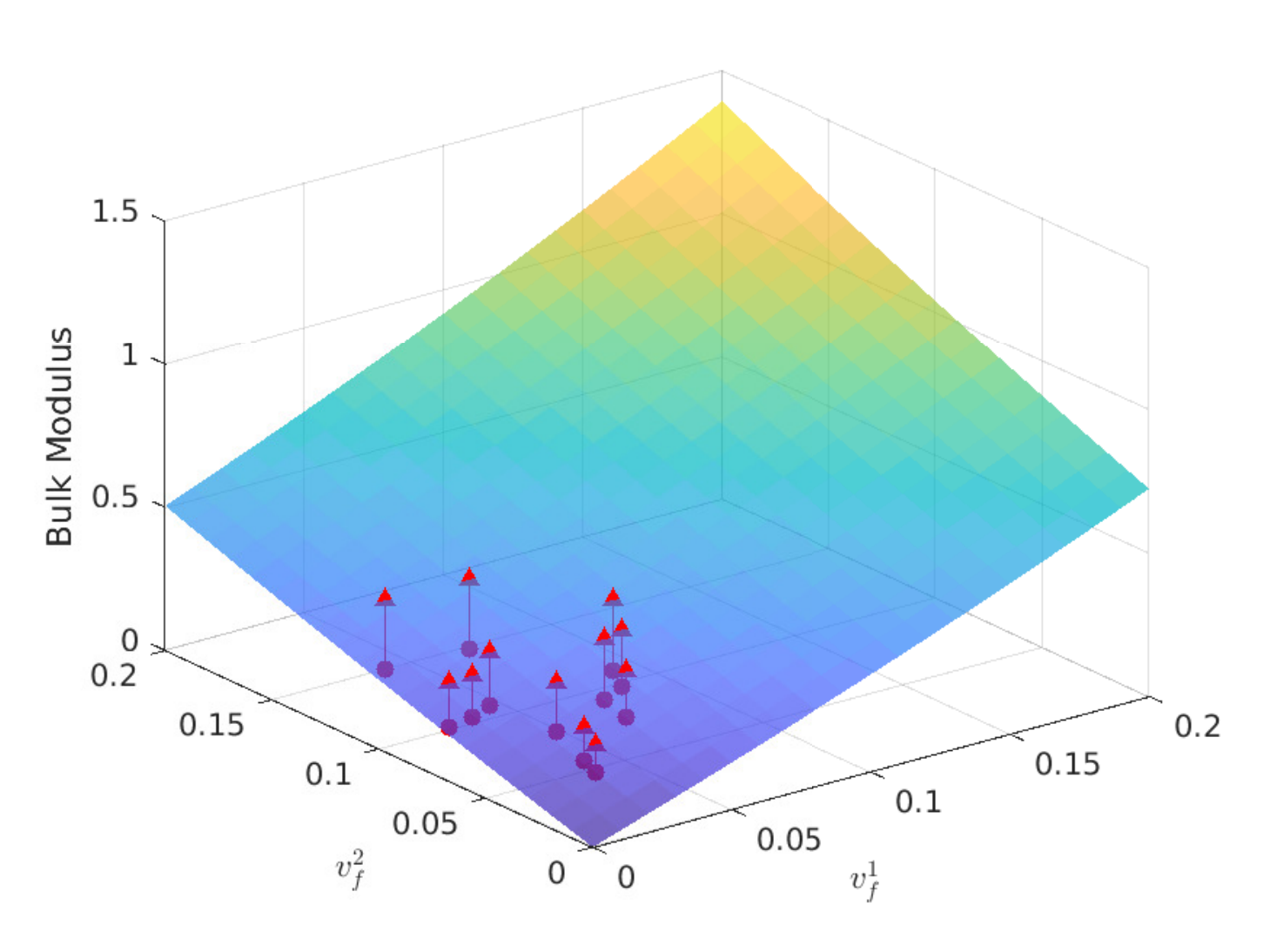} 
    \caption{Effective bulk modulus}
    \label{fig:HSW-K}
  \end{subfigure}
  \begin{subfigure}[c]{0.45\textwidth}
    \includegraphics[width=\columnwidth]{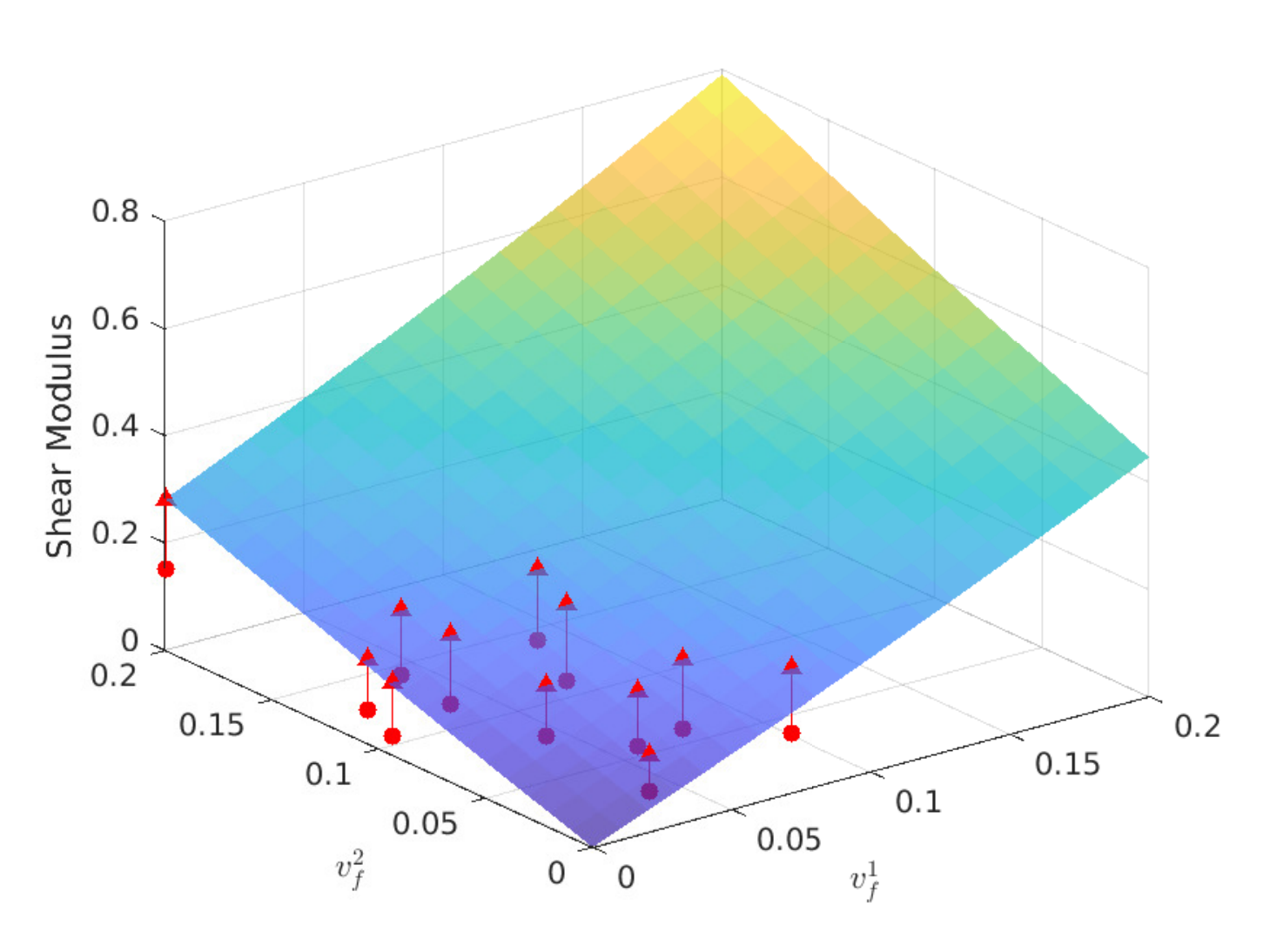} 
    \caption{Effective shear modulus}
    \label{fig:HSW-G}
  \end{subfigure} 
\caption{Comparison of effective moduli to HSW-bounds for the designs corresponding to Tables \ref{table:cubic_bulk} and \ref{table:cubic_shear}.  Arrows point from the optimal design to the HSW-bound surface.}
\label{fig:HSW-comparison}
\end{figure}

In Fig.~\ref{fig:HSW-K}, we compare the effective bulk moduli for the designs of Table \ref{table:cubic_bulk} to the Hashin-Shtrikman-Walpole bounds for three-phase materials (with one phase being void) \cite{gibiansky2000multiphase}.  As expected, all the bulk moduli for the optimal designs are below the bounds.  We also note that the moduli are not close to the bounds.  This is contrary to what has been shown for designs obtained using density-based topology optimization \cite{gibiansky2000multiphase}.  The reason for this is that the design representation is significantly more restrictive (i.e., a truss made of cylindrical struts of constant diameter) and we impose an additional geometric requirement (the no-cut constraint). Moreover, the lattice representation ensures an open-cell design (which we desire, as justified in Section \ref{sec:introduction}); however, closed-cell designs are known to render higher bulk moduli \cite{bendsoe2013topology}.

We now perform the optimization for this same problem by adding a third material with elastic modulus $E_3 = 7.5$ and physical density $\gamma_3 = 0.675$. The results for different weight fraction limits are shown in Table \ref{table:cubic_bulk_3mat}. An important note about these results is that the range of weight fraction limits that produces three-material designs is narrower than for two-material designs. This is expected, since above and below that range two-material designs (and eventually, single-material designs) are more weight-efficient.  

%
\begin{table*}
\begin{centering}
\begin{tabular}{>{\centering}m{1.4cm}>{\centering}m{0.6cm}>{\centering}m{1.8cm}>{\centering}m{1.8cm}}
$K$ & $w_f^*$ & iso & side
\tabularnewline
\hline 
$ 0.09399$ & $0.0667$ & \includegraphics[width=1.8cm, bb=250 50 1000 700, clip=true]{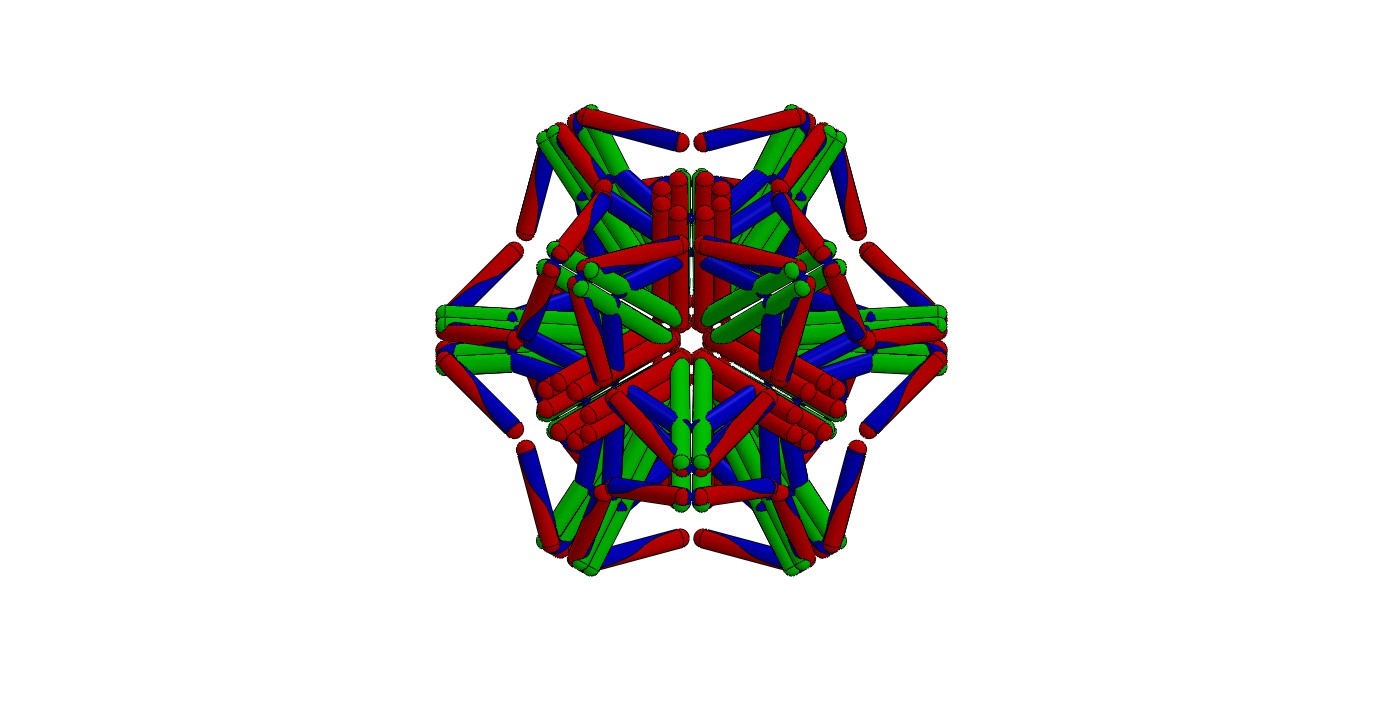} &  \includegraphics[width=1.8cm, bb=250 50 1000 700, clip=true]{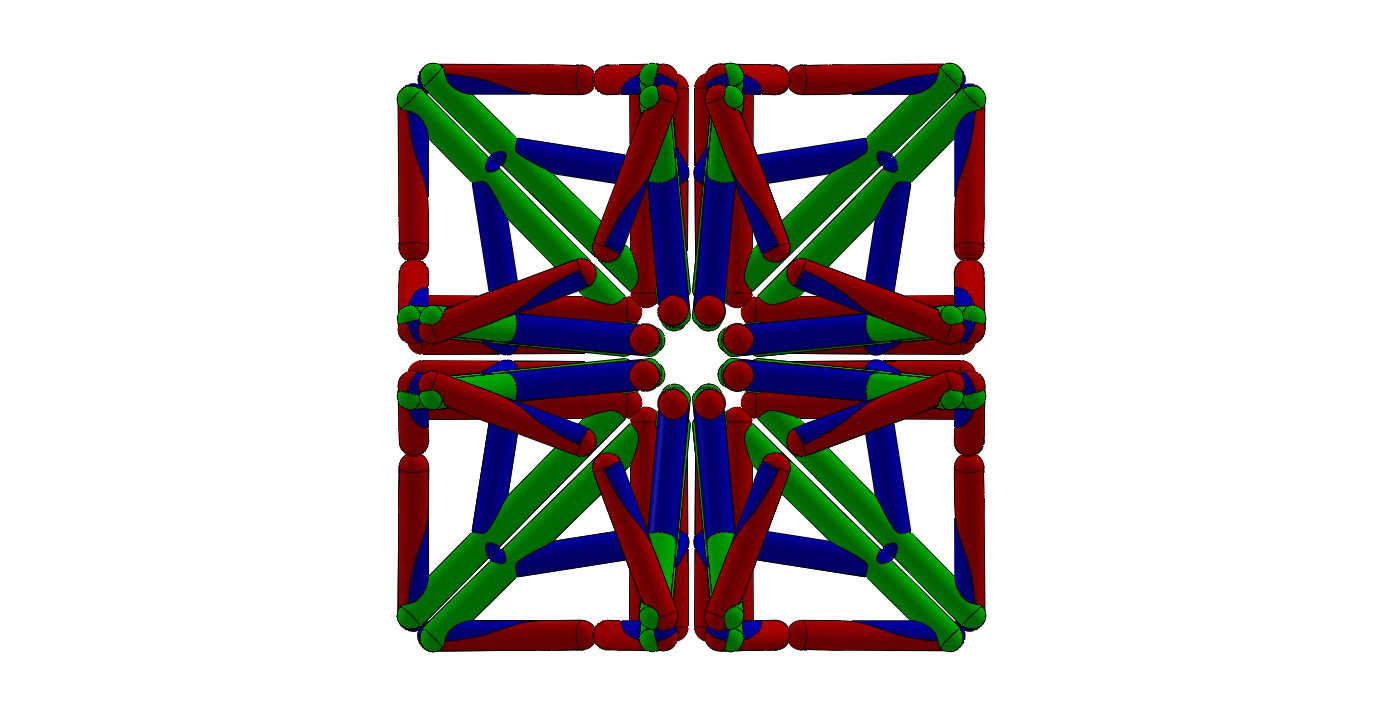}   \tabularnewline
$ 0.12024 $ & $0.0889$ & \includegraphics[width=1.8cm, bb=250 50 1000 700, clip=true]{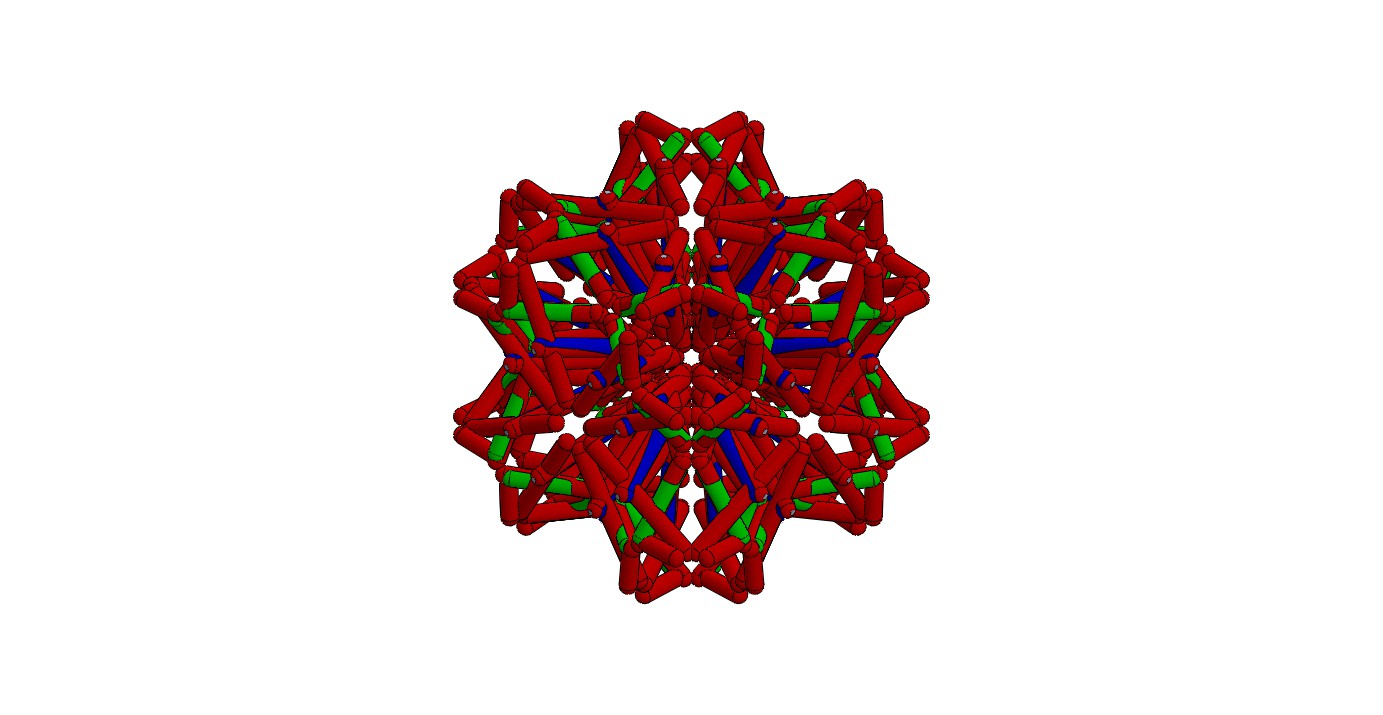} &  \includegraphics[width=1.8cm, bb=250 50 1000 700, clip=true]{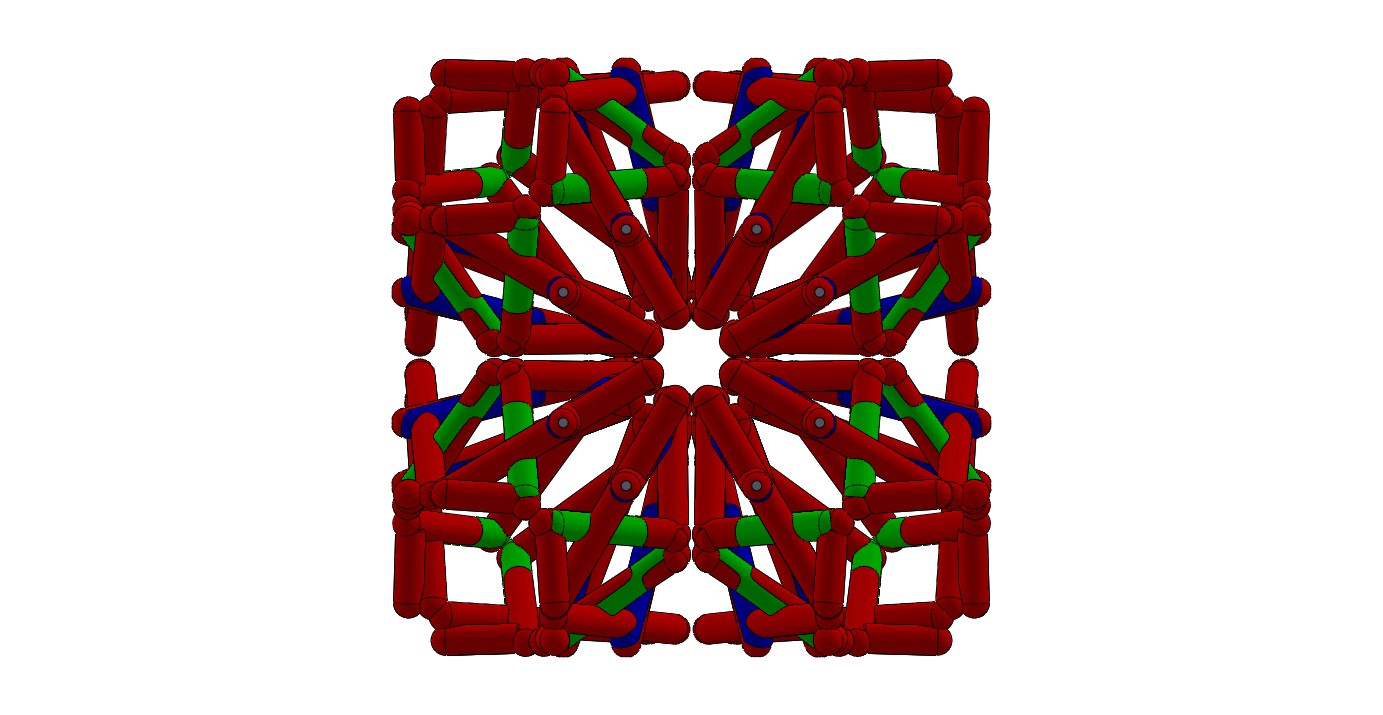}   \tabularnewline
$ 0.16937 $ & $0.0944$ & \includegraphics[width=1.8cm, bb=250 50 1000 700, clip=true]{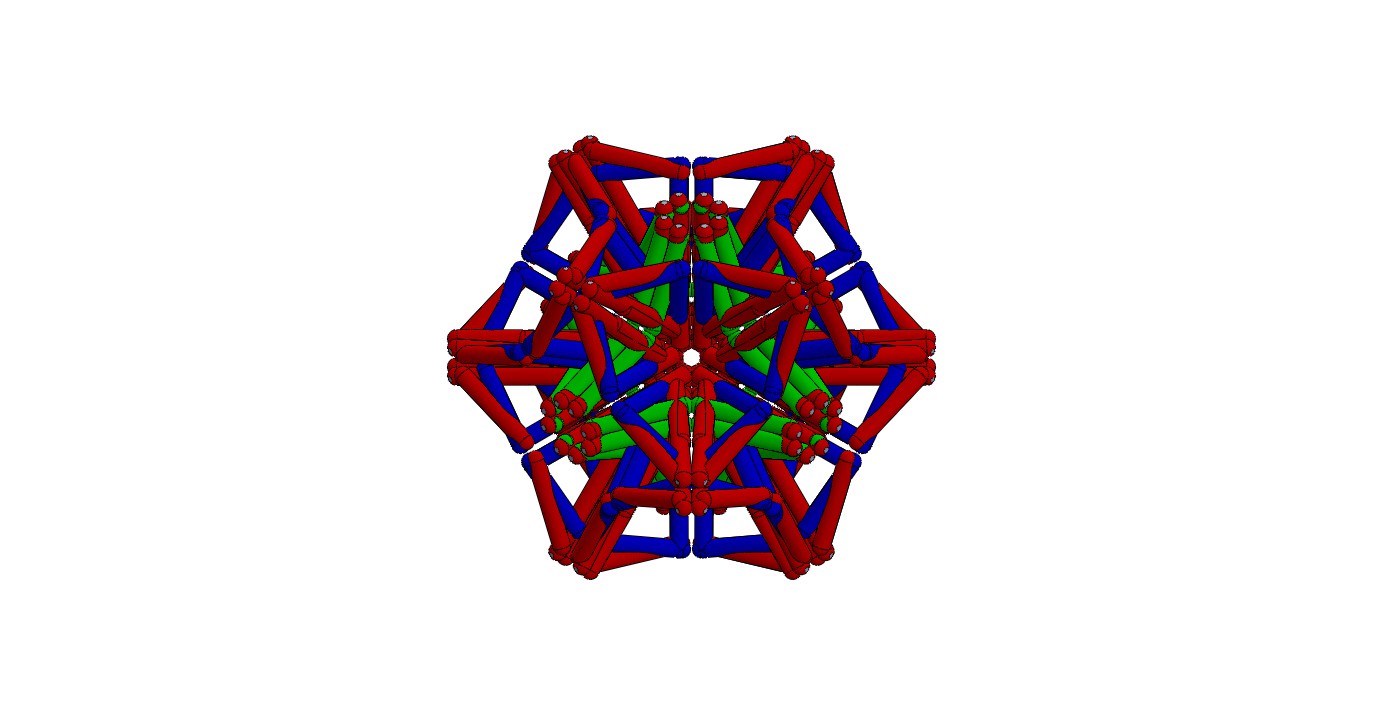} &  \includegraphics[width=1.8cm, bb=250 50 1000 700, clip=true]{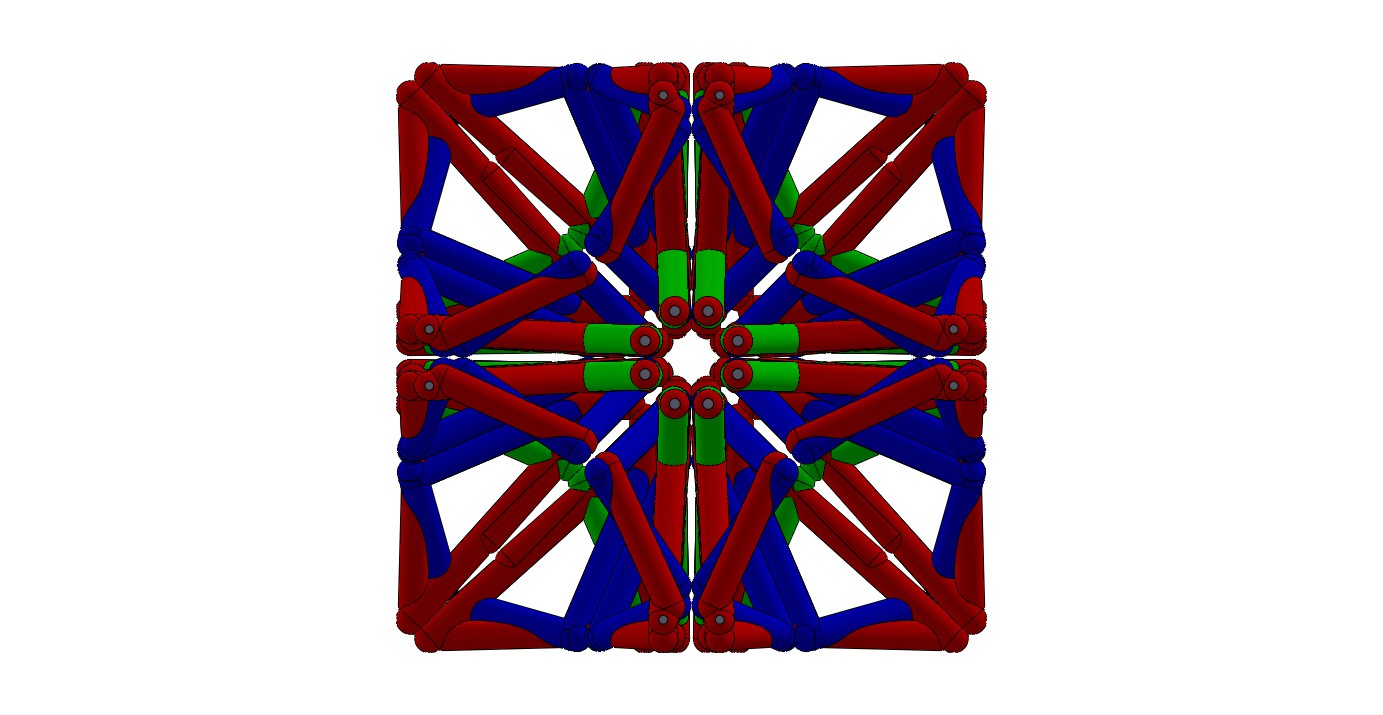}   \tabularnewline
$ 0.15466 $ & $0.1$ & \includegraphics[width=1.8cm, bb=250 50 1000 700, clip=true]{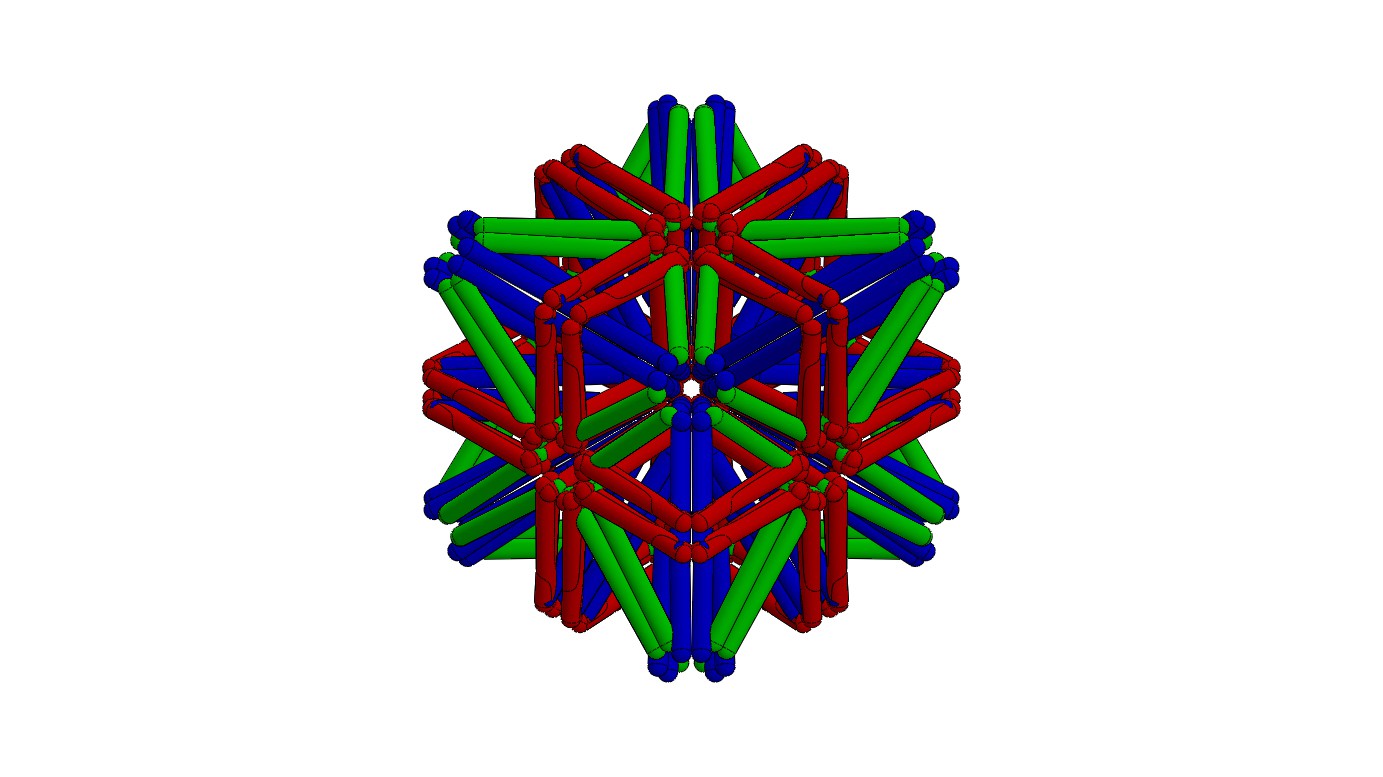} &  \includegraphics[width=1.8cm, bb=250 50 1000 700, clip=true]{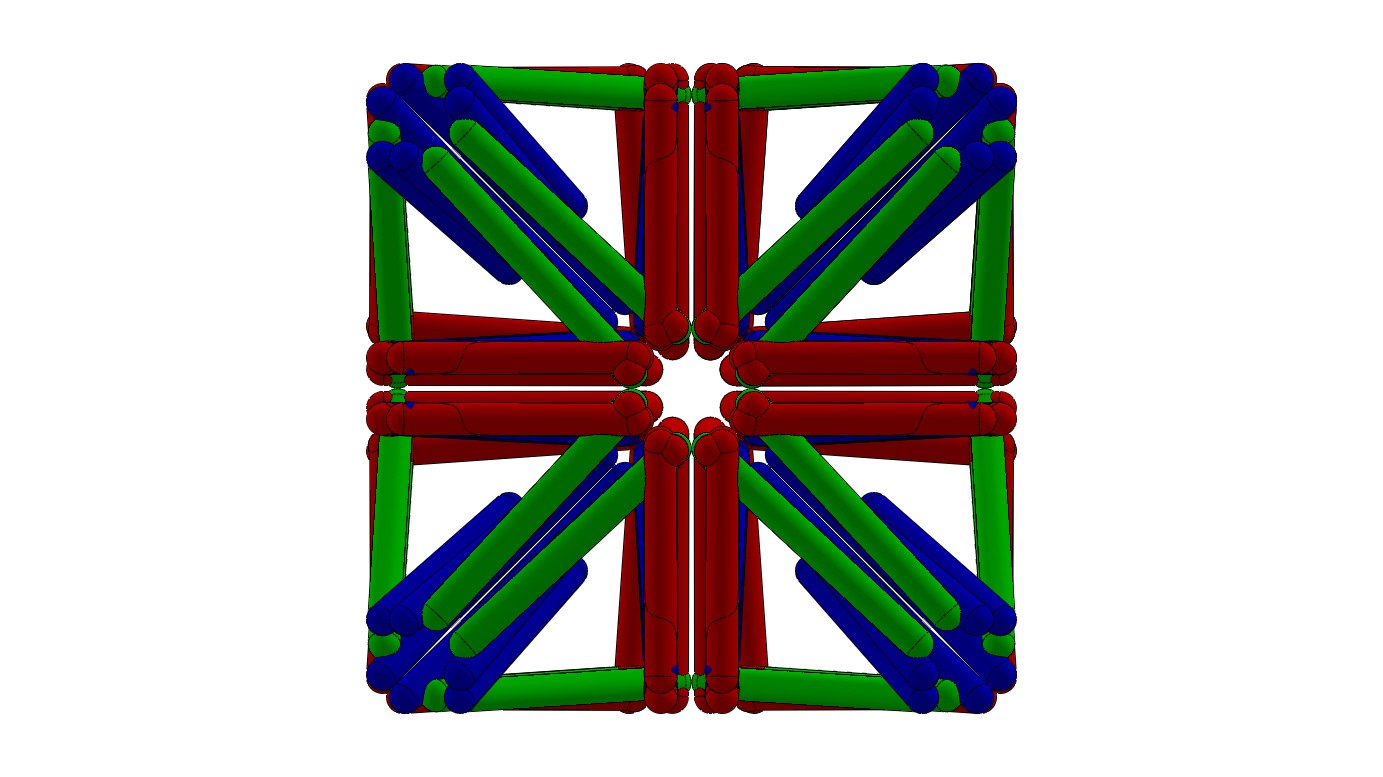}   \tabularnewline
\hline 
\end{tabular}
\par\end{centering}
\caption{Maximal bulk modulus designs for cubic three-material lattices and different weight fraction limits $w_f^*$. Red, blue and green bars are made of materials 1, 2, and 3, respectively;  and bars that have been removed from the design (i.e., with $\alpha_1^q, \alpha_2^q, \alpha_3^q \approx 0$) are not shown.}
\label{table:cubic_bulk_3mat}
\end{table*}
%

\subsection{Two- and Three-material Lattices with Maximal Shear Modulus and Cubic Symmetry}
\label{sec:cubic} 

In this section we present results for maximization of the effective shear modulus, both for two-material and three-material lattices.  The material properties are the same as in the previous section. 
The results of the optimization of two-material lattices for different weight fraction limits are presented in Table ~\ref{table:cubic_shear}. As before, we note that small changes in the weight fraction limit produce different designs. The overall trend is again that the maximal shear modulus increases as we increase the weight fraction limit as expected, however convergence to local minima likely prevents strict monotonicity. In Fig.~\ref{fig:HSW-G}, we compare the effective shear moduli for the designs of Table \ref{table:cubic_shear} to the Hashin-Shtrikman-Walpole bounds for three-phase materials.  As with the bulk modulus, all the effective shear moduli for the optimal designs are below the bounds.  

The results of the optimization for this same problem with three materials (with the same properties as before) are shown in Table \ref{table:cubic_shear_3mat}. Once again, the range of weight fraction limits that produces three-material designs is narrower than for two-material designs.   

\begin{table*}
\begin{centering}
\begin{tabular}{>{\centering}m{1.4cm}>{\centering}m{0.6cm}>{\centering}m{1.8cm}>{\centering}m{1.8cm}|>{\centering}m{1.4cm}>{\centering}m{.6cm}>{\centering}m{1.8cm}>{\centering}m{1.8cm}}
$G$ & $w_f^*$ & iso & side & $G$ & $w_f^*$ & iso & side 
\tabularnewline
\hline \\[-1ex]
$ 0.0242$ & $0.0444$ & \includegraphics[width=1.8cm, bb=250 50 1000 700, clip=true]{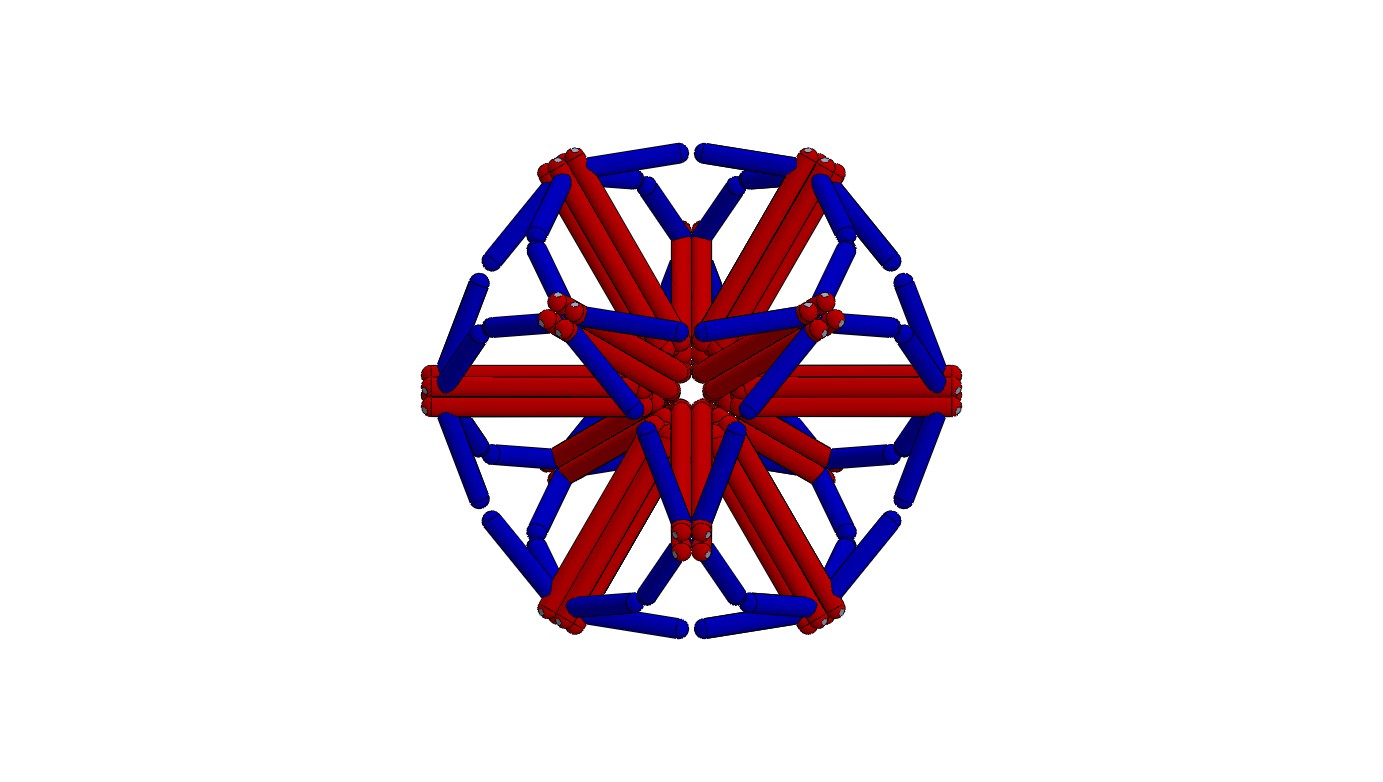} &  \includegraphics[width=1.8cm, bb=250 50 1000 700, clip=true]{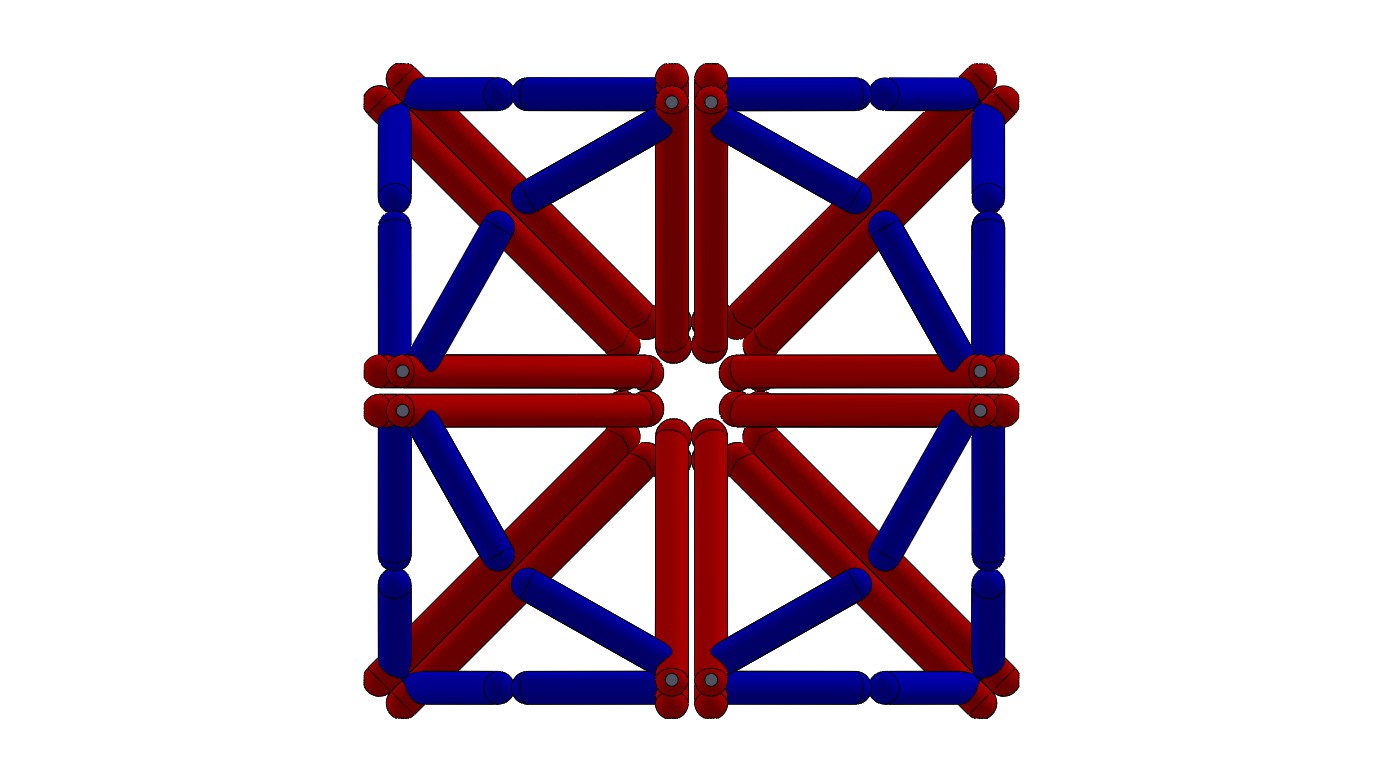}  &$ 0.08252 $ & $0.0778$ & \includegraphics[width=1.8cm, bb=250 50 1000 700, clip=true]{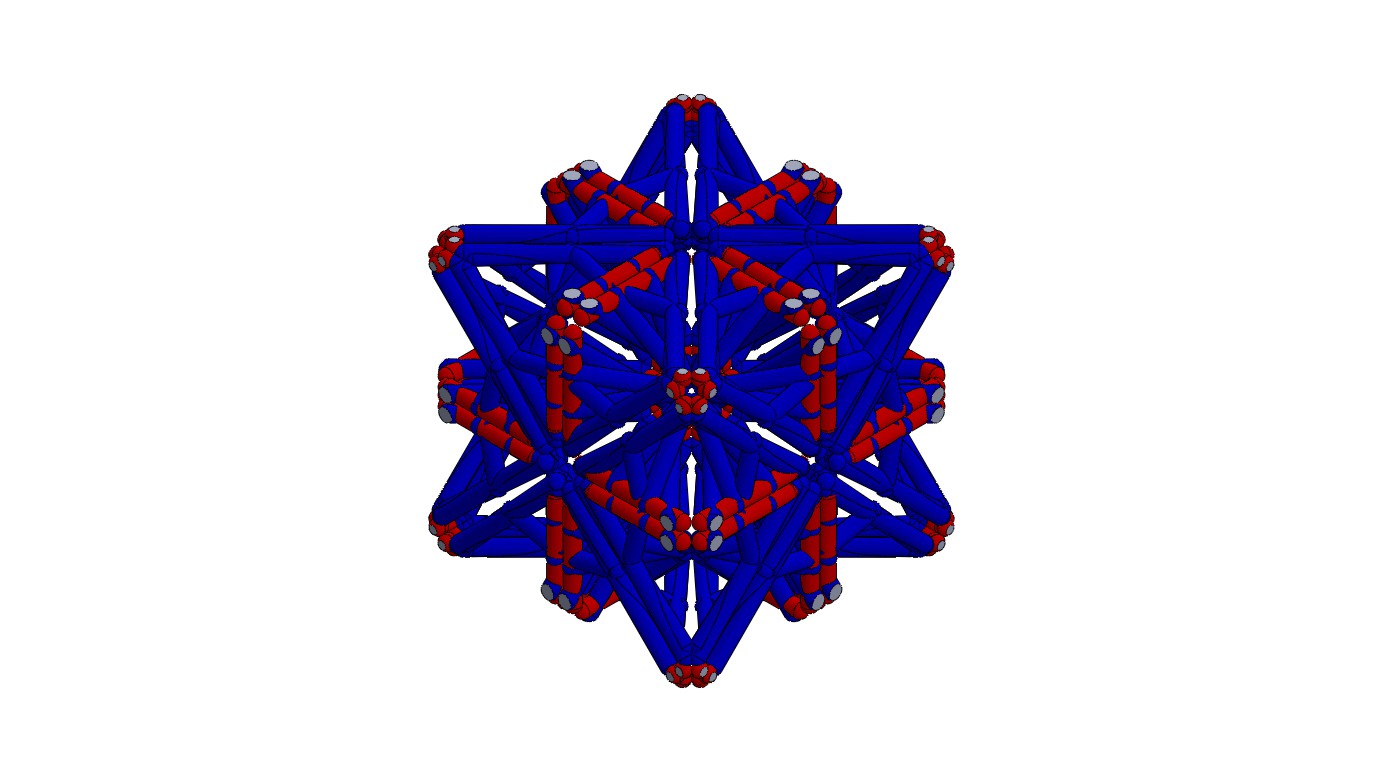} &  \includegraphics[width=1.8cm, bb=250 50 1000 700, clip=true]{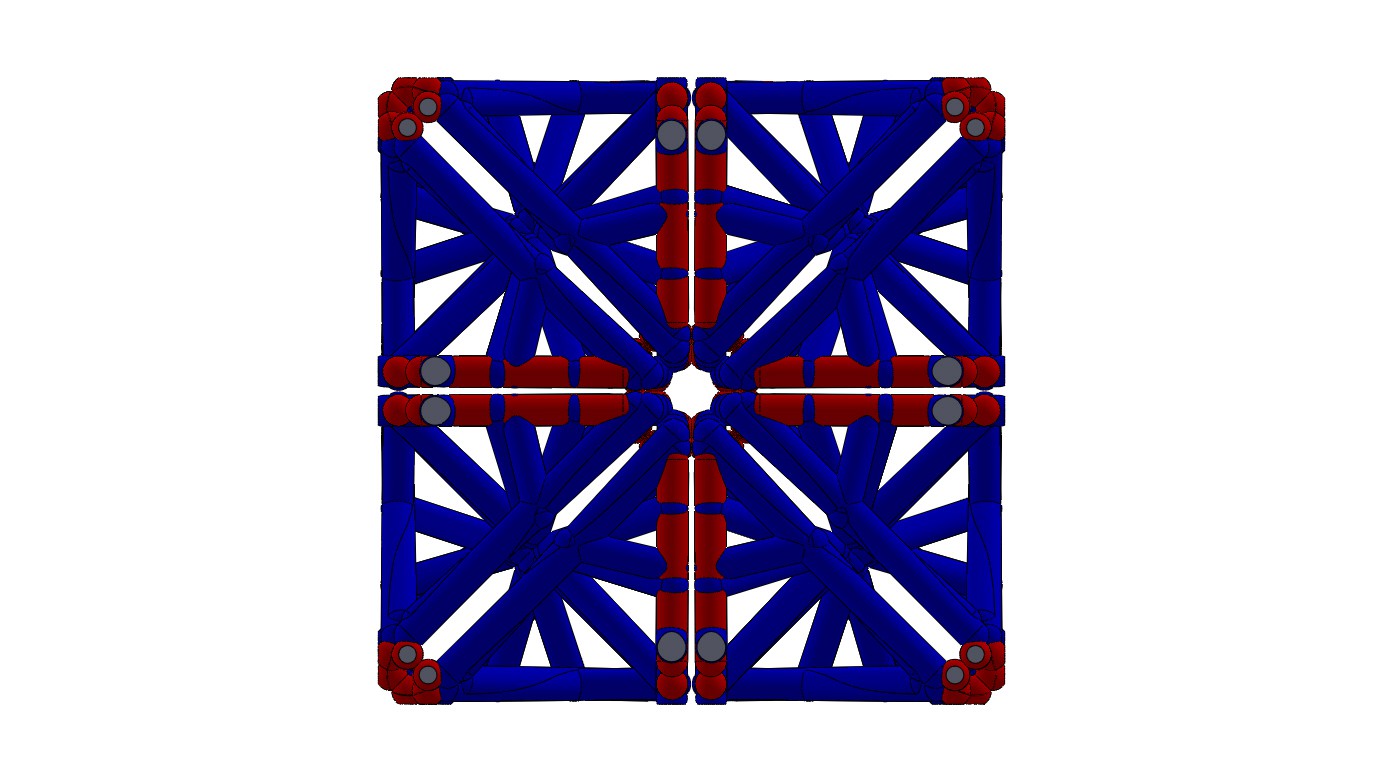}  \tabularnewline
$ 0.03494$ & $0.05$ & \includegraphics[width=1.8cm, bb=250 50 1000 700, clip=true]{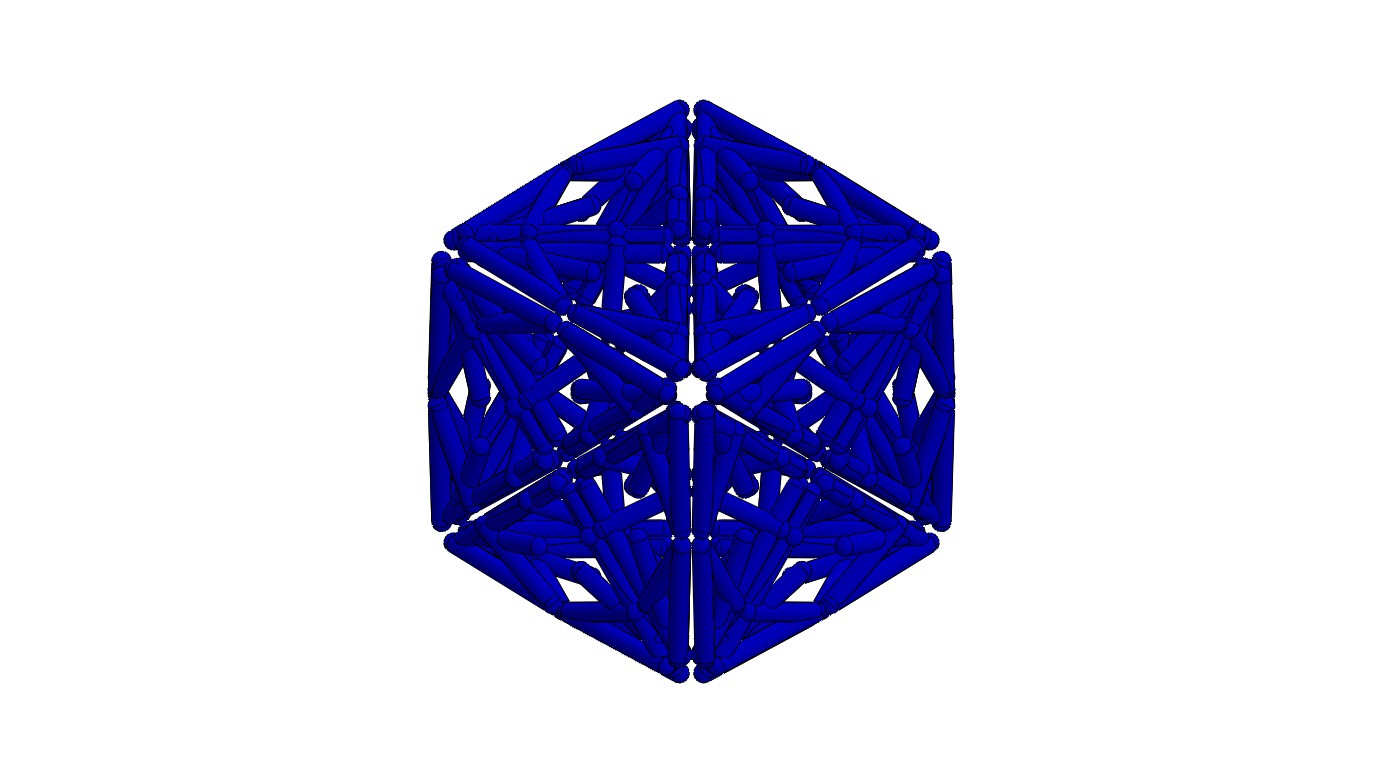} &  \includegraphics[width=1.8cm, bb=250 50 1000 700, clip=true]{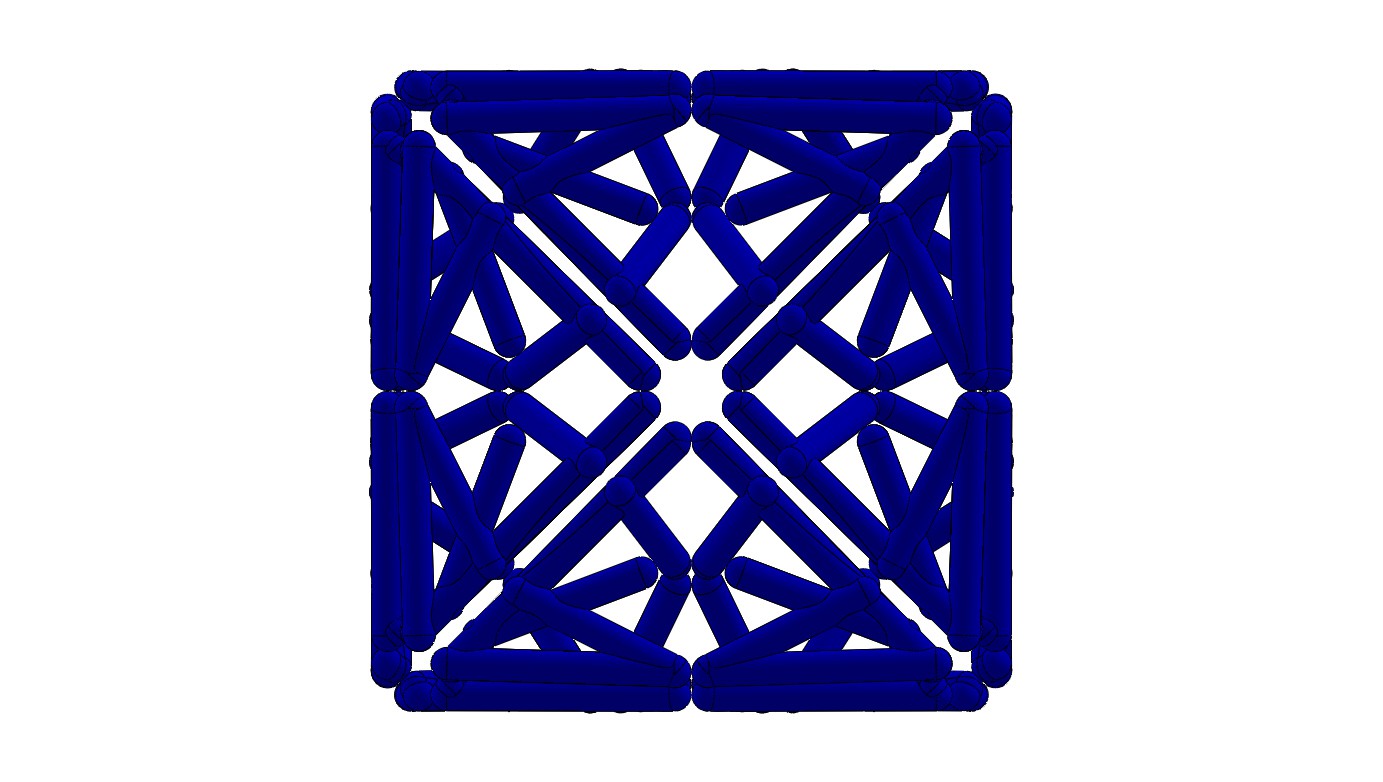}
 &$ 0.05861$ & $0.0833$ & \includegraphics[width=1.8cm, bb=250 50 1000 700, clip=true]{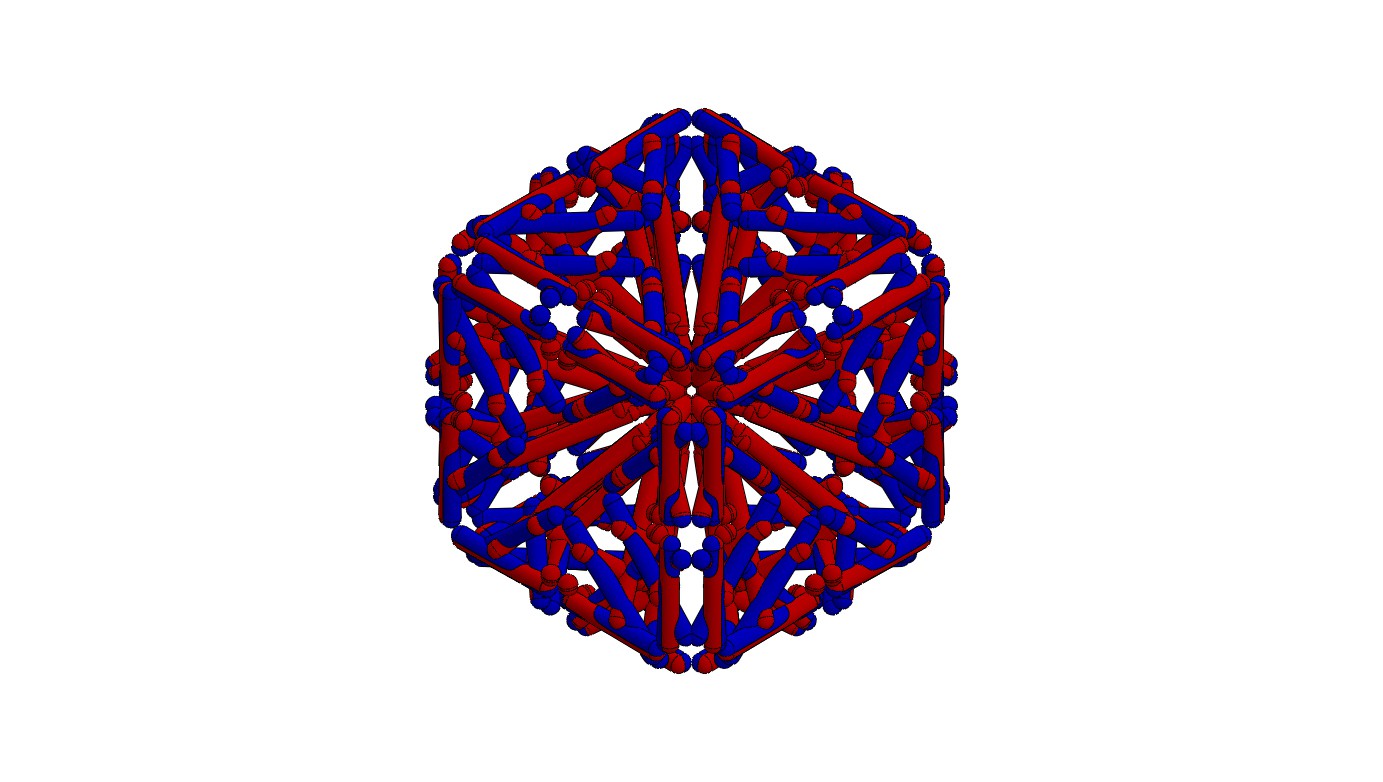} &  \includegraphics[width=1.8cm, bb=250 50 1000 700, clip=true]{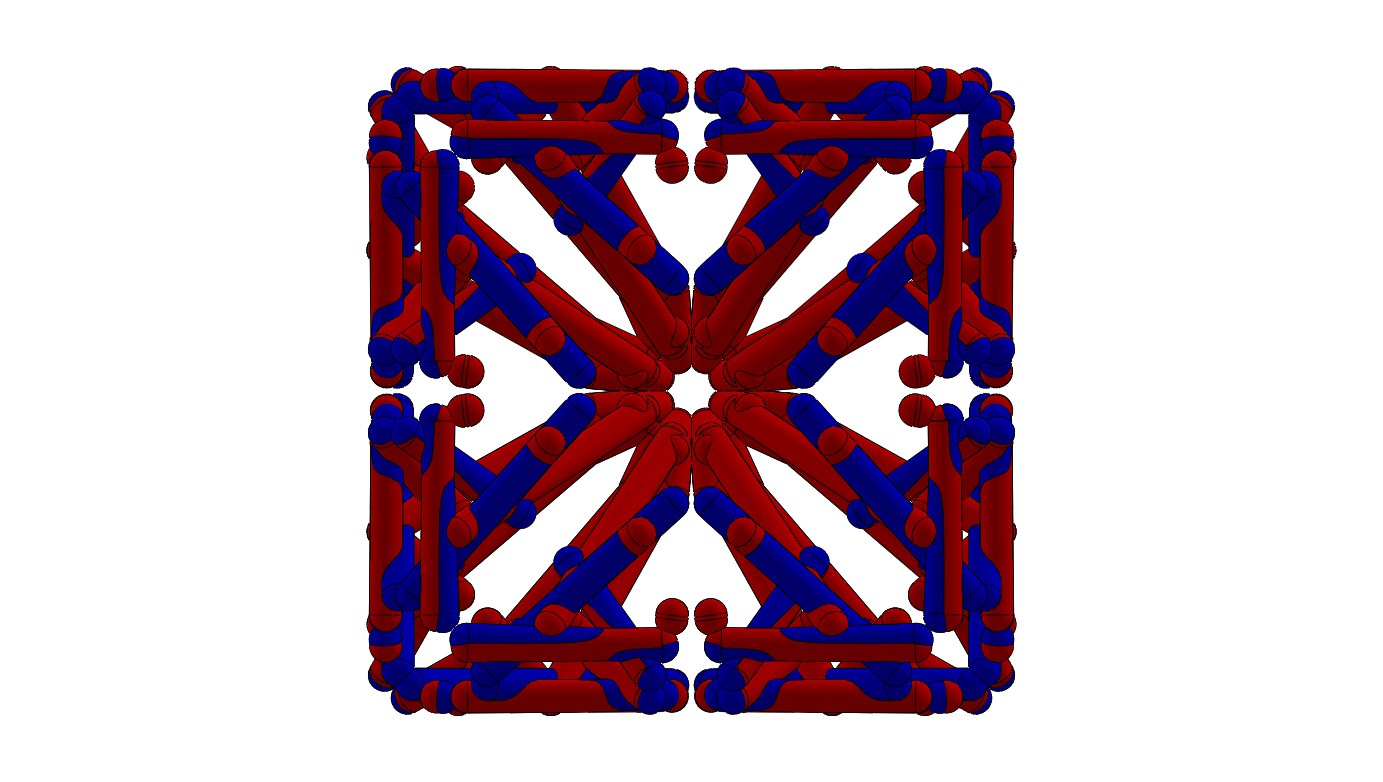}  \tabularnewline
$ 0.05653 $ & $0.0556$ & \includegraphics[width=1.8cm, bb=250 50 1000 700, clip=true]{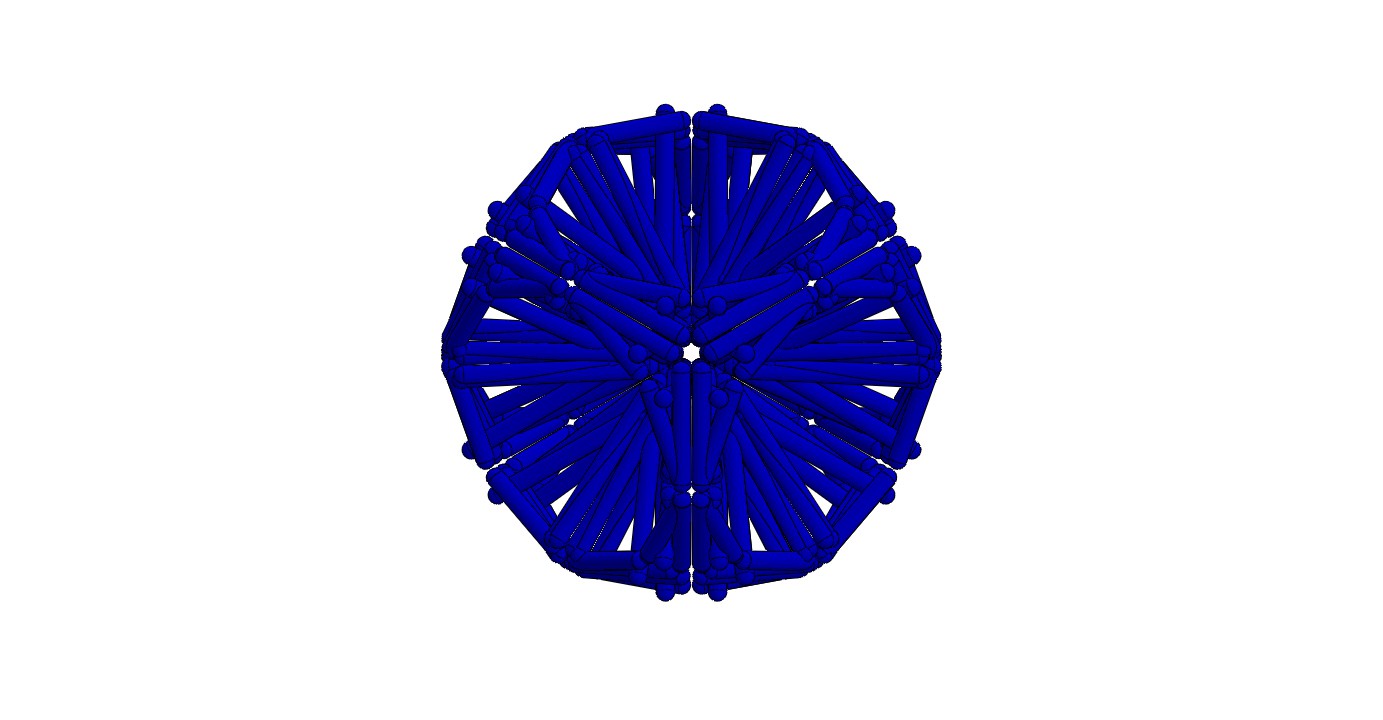} &  \includegraphics[width=1.8cm, bb=250 50 1000 700, clip=true]{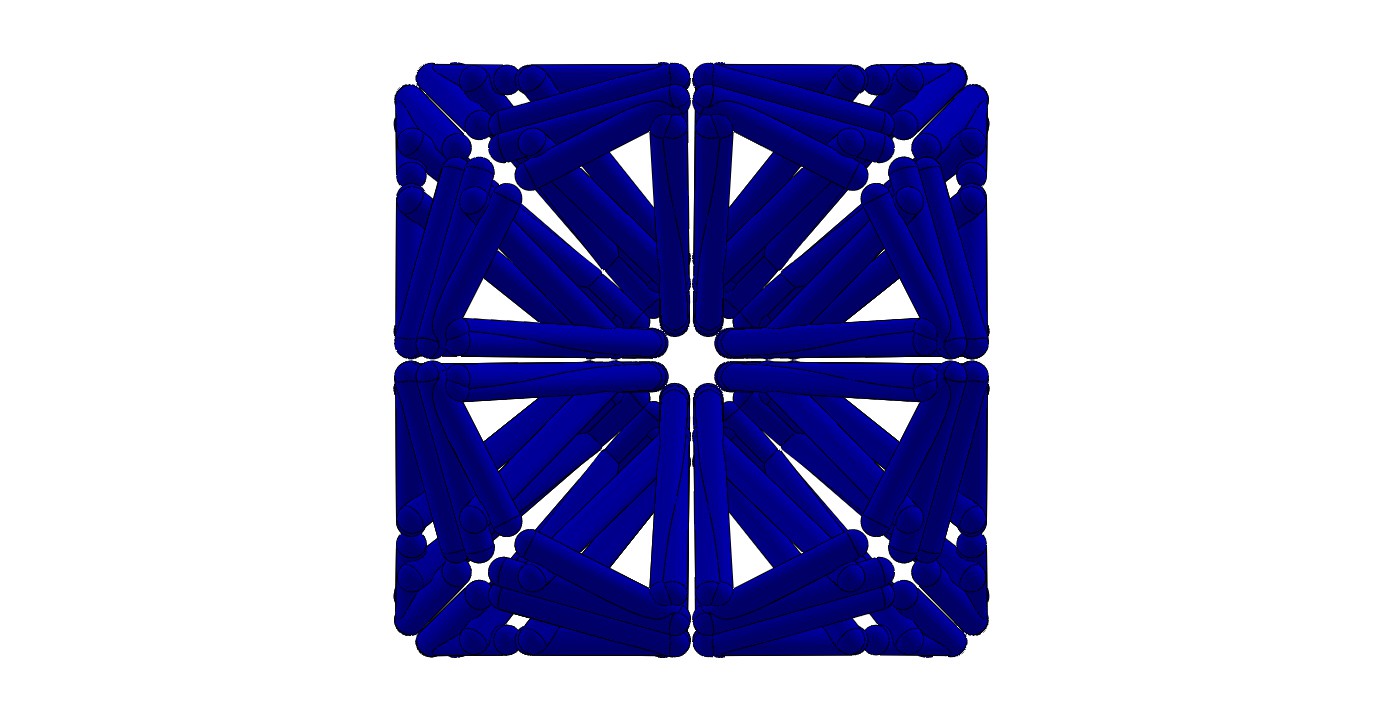}
 & $ 0.0709 $ & $0.0889$ & \includegraphics[width=1.8cm, bb=250 50 1000 700, clip=true]{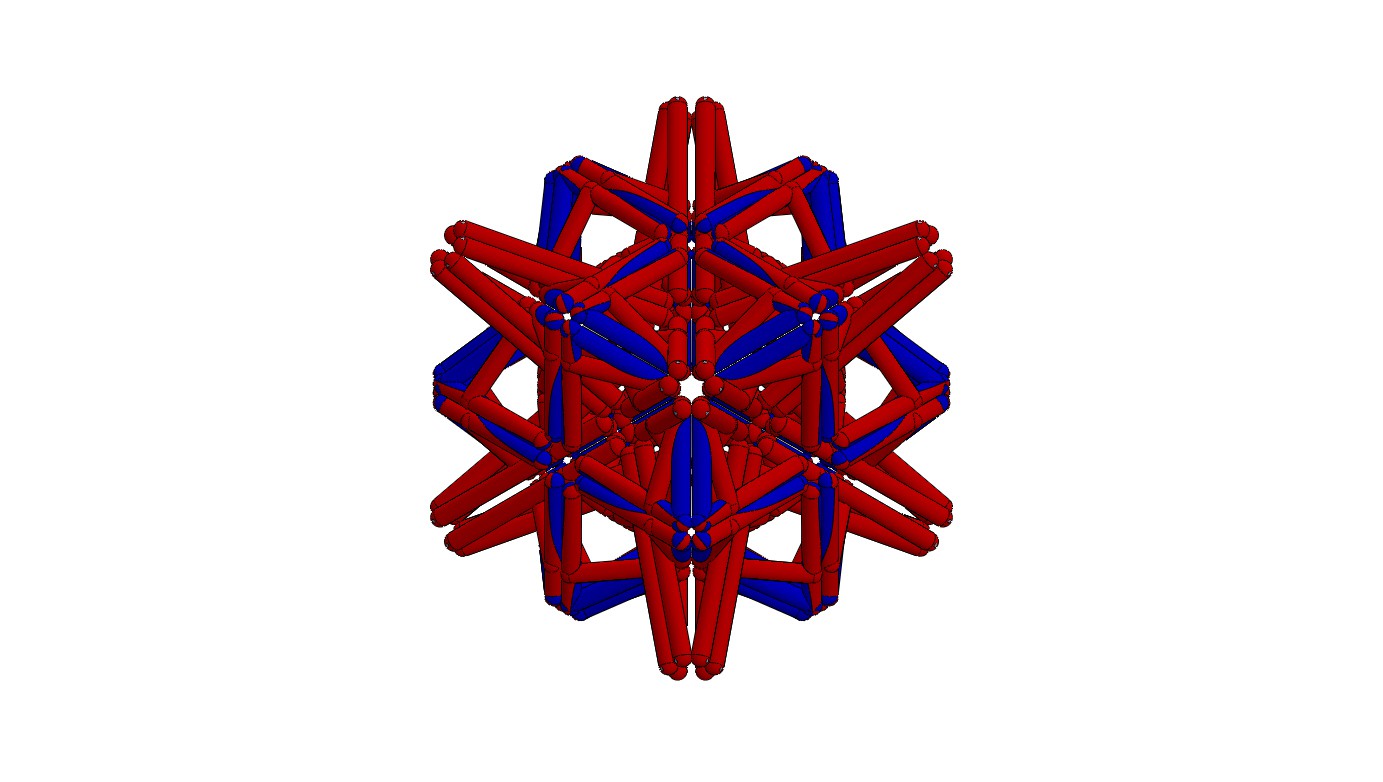} &  \includegraphics[width=1.8cm, bb=250 50 1000 700, clip=true]{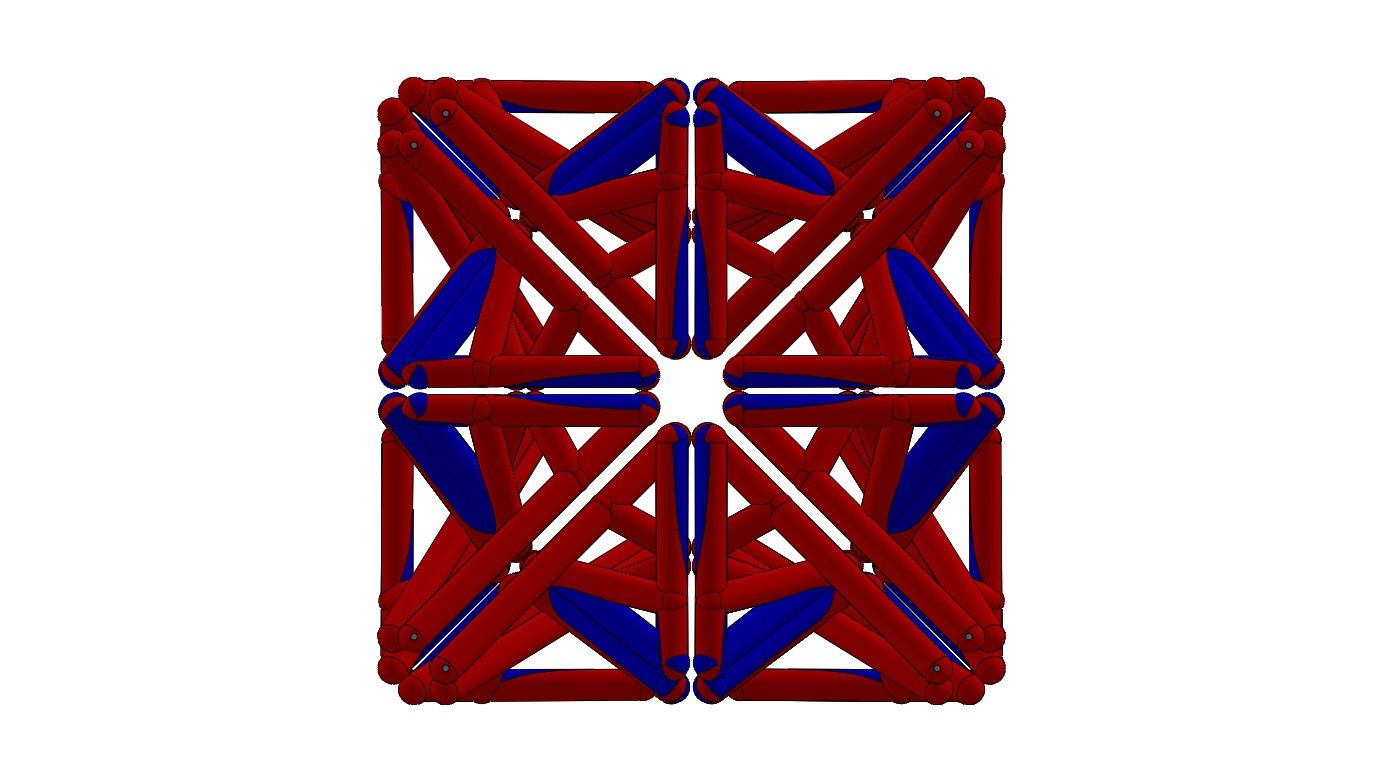}   \tabularnewline
$ 0.0526 $ & $0.0611$ & \includegraphics[width=1.8cm, bb=250 50 1000 700, clip=true]{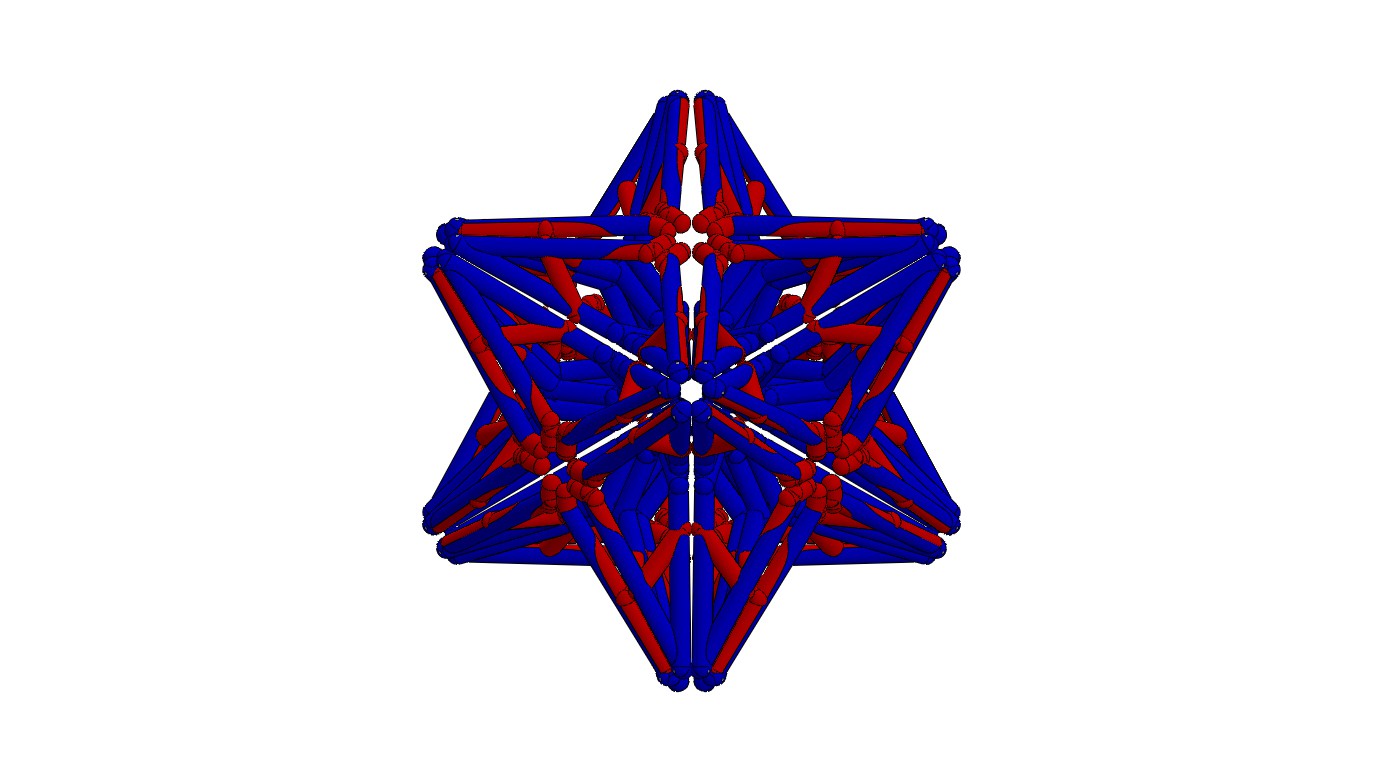} &  \includegraphics[width=1.8cm, bb=250 50 1000 700, clip=true]{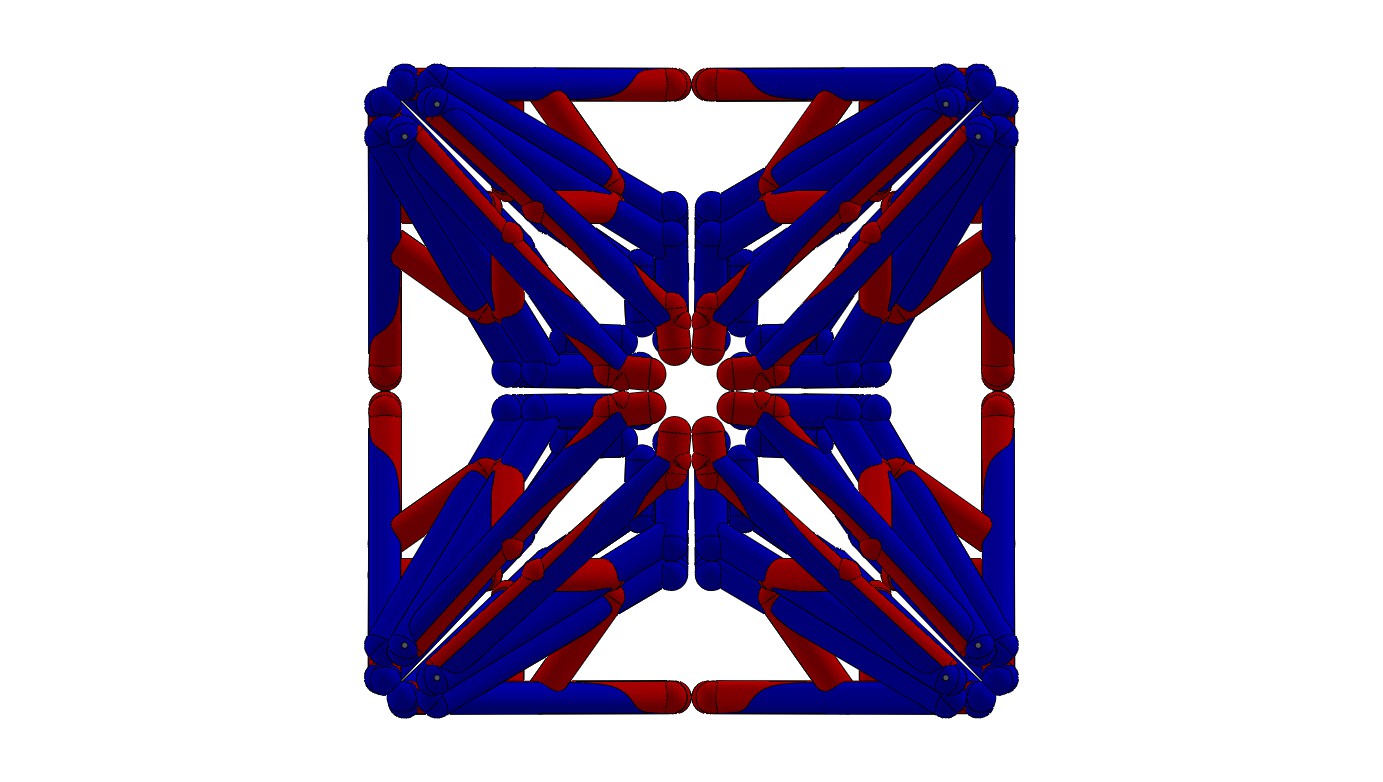}
&$ 0.08605$ & $0.0944$ & \includegraphics[width=1.8cm, bb=250 50 1000 700, clip=true]{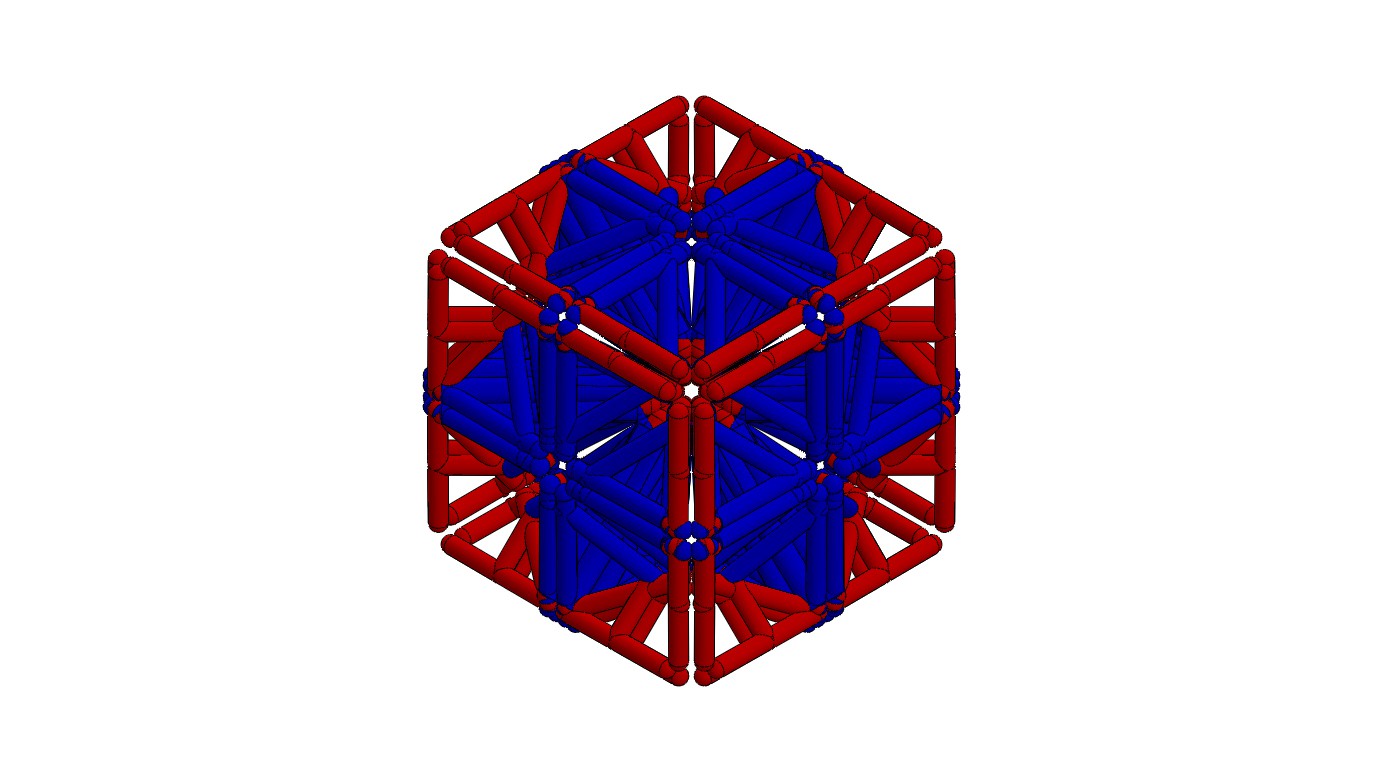} &  \includegraphics[width=1.8cm, bb=250 50 1000 700, clip=true]{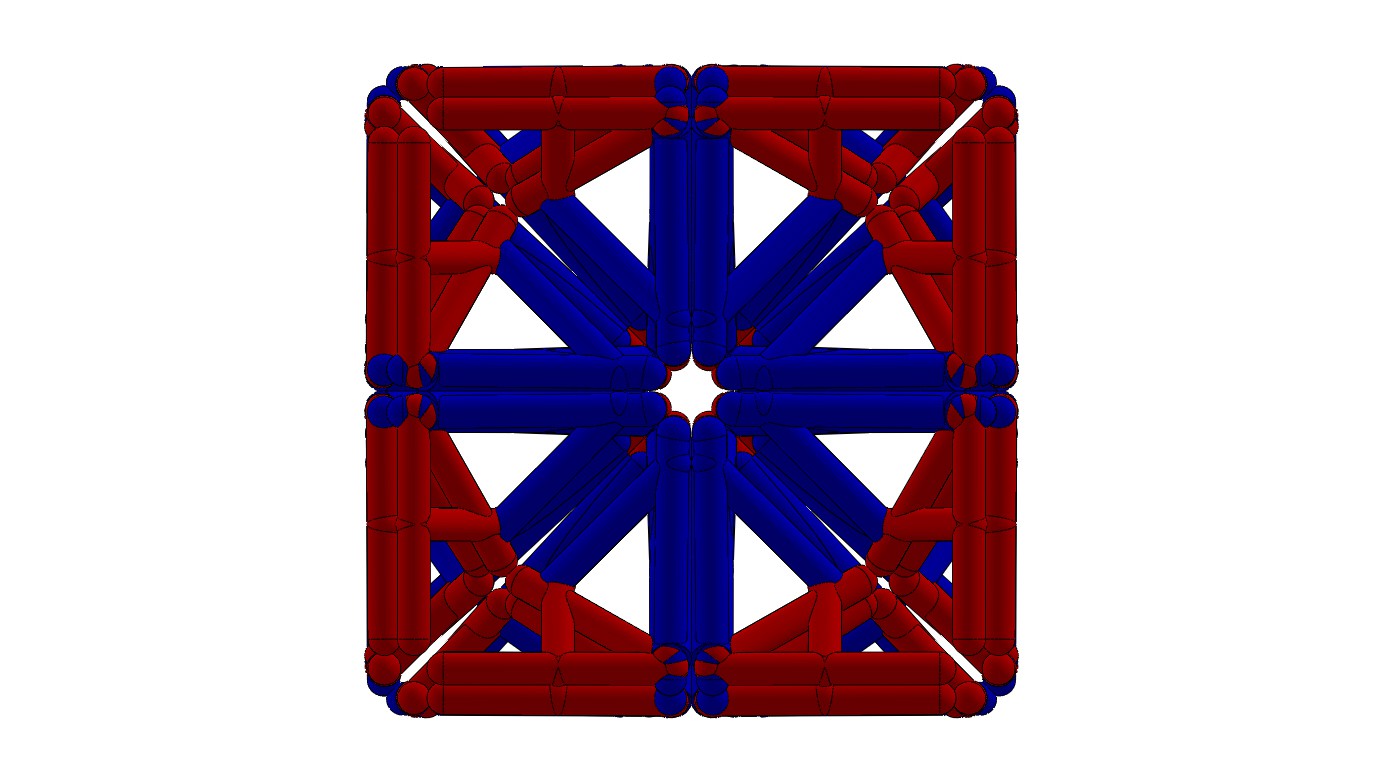}\tabularnewline
$ 0.05019 $ & $0.0667$ & \includegraphics[width=1.8cm, bb=250 50 1000 700, clip=true]{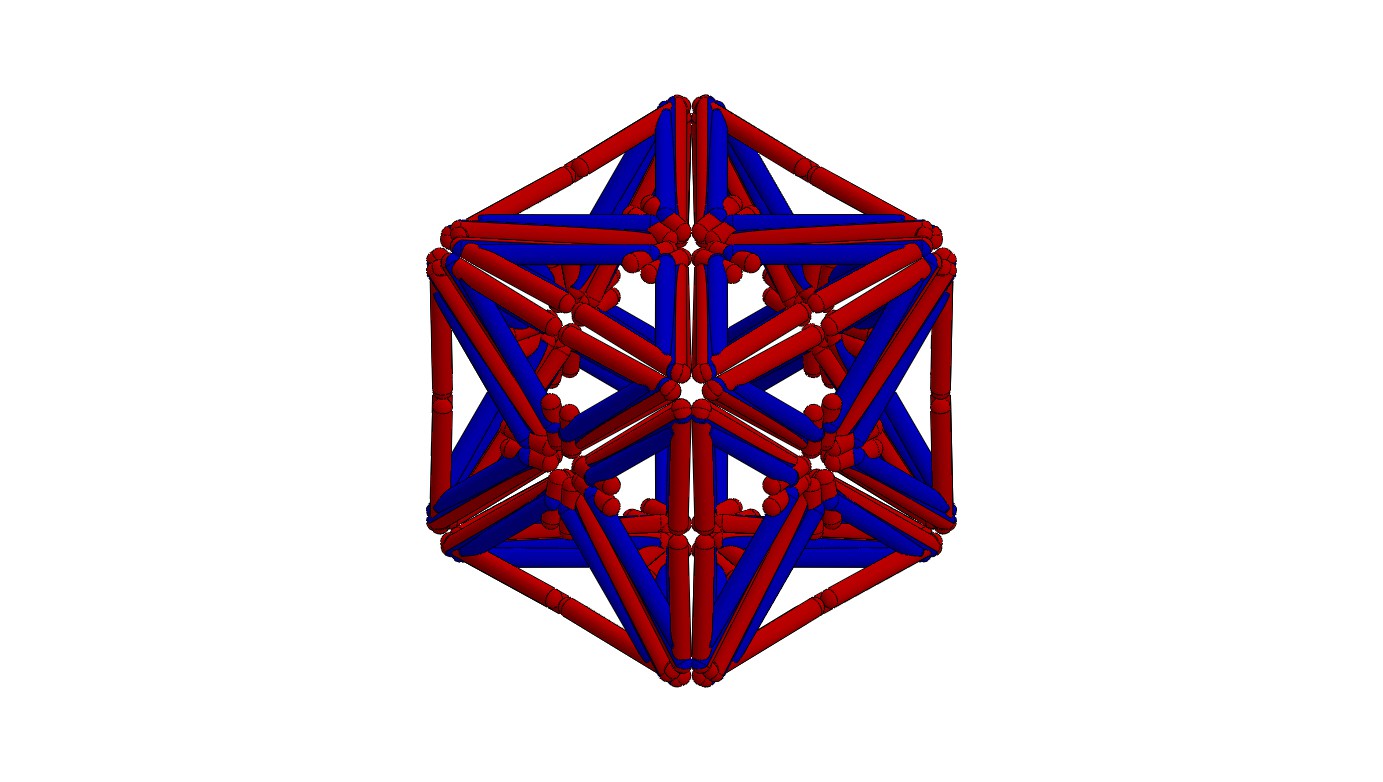} &  \includegraphics[width=1.8cm, bb=250 50 1000 700, clip=true]{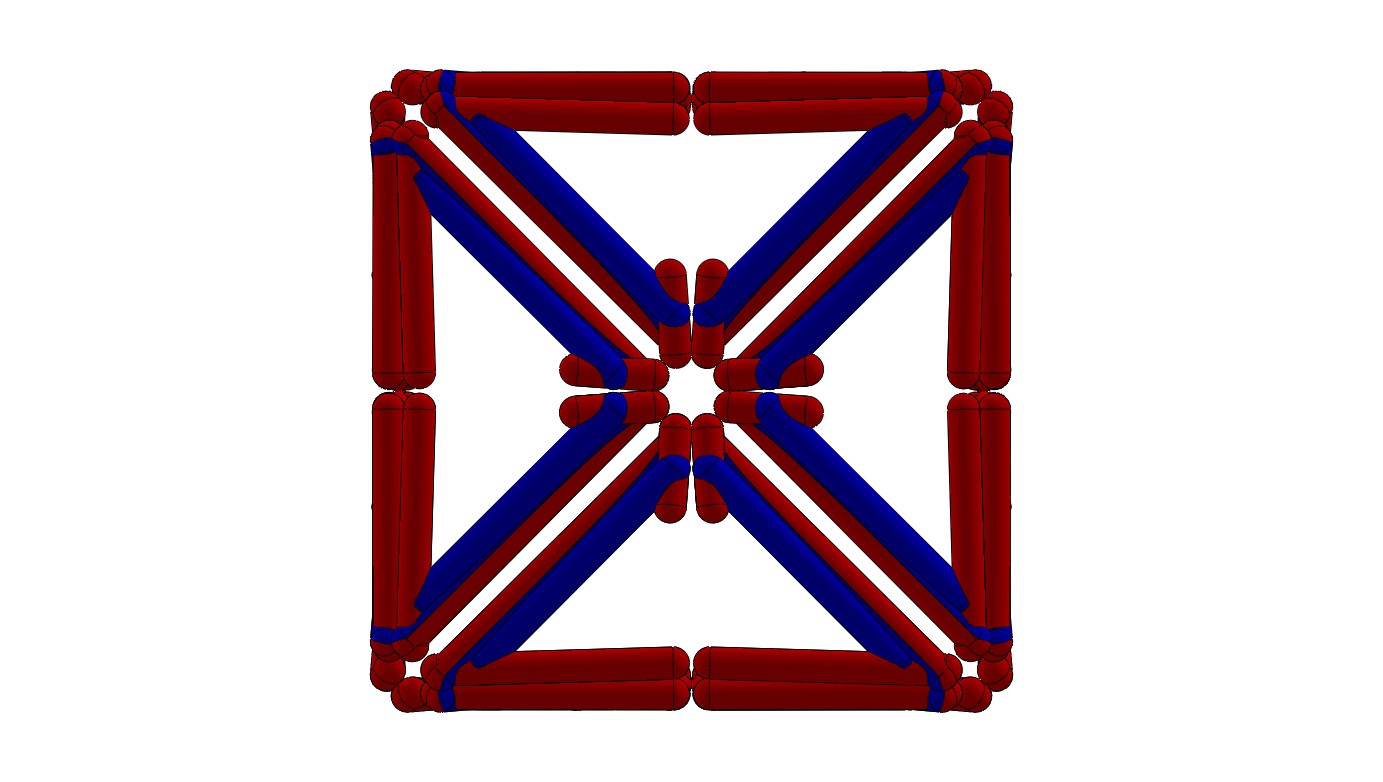}
&$ 0.15178$ & $0.1$ & \includegraphics[width=1.8cm, bb=250 50 1000 700, clip=true]{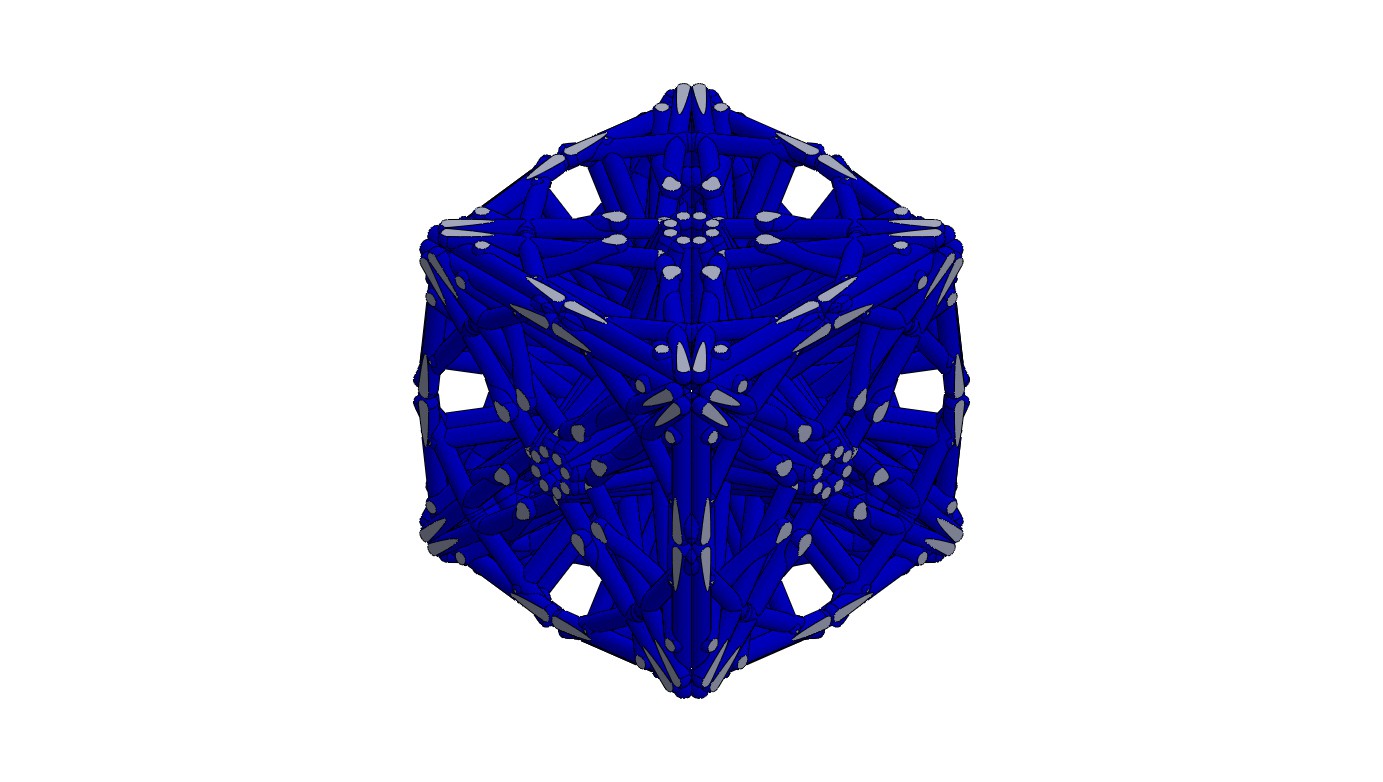} &  \includegraphics[width=1.8cm, bb=250 50 1000 700, clip=true]{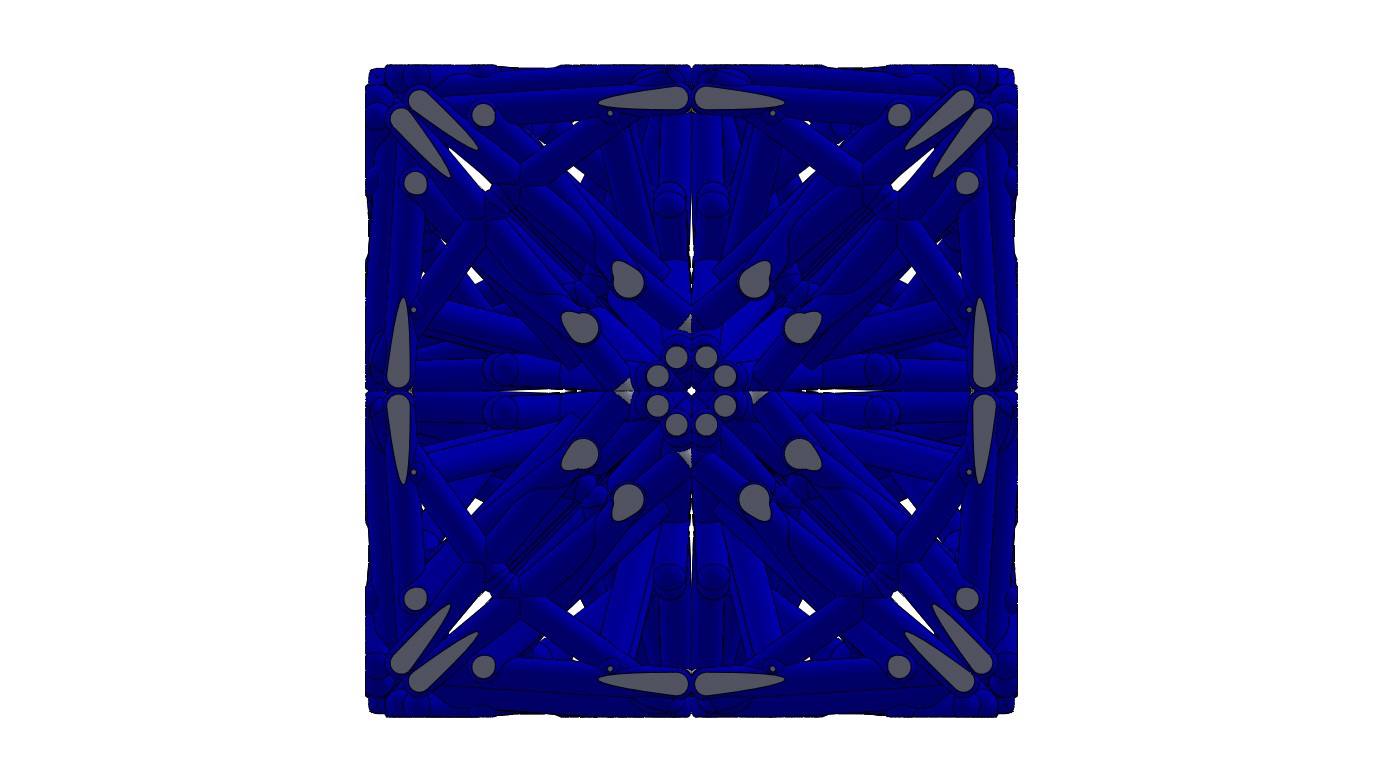} \tabularnewline
$ 0.0563 $ & $0.0722$ & \includegraphics[width=1.8cm, bb=250 50 1000 700, clip=true]{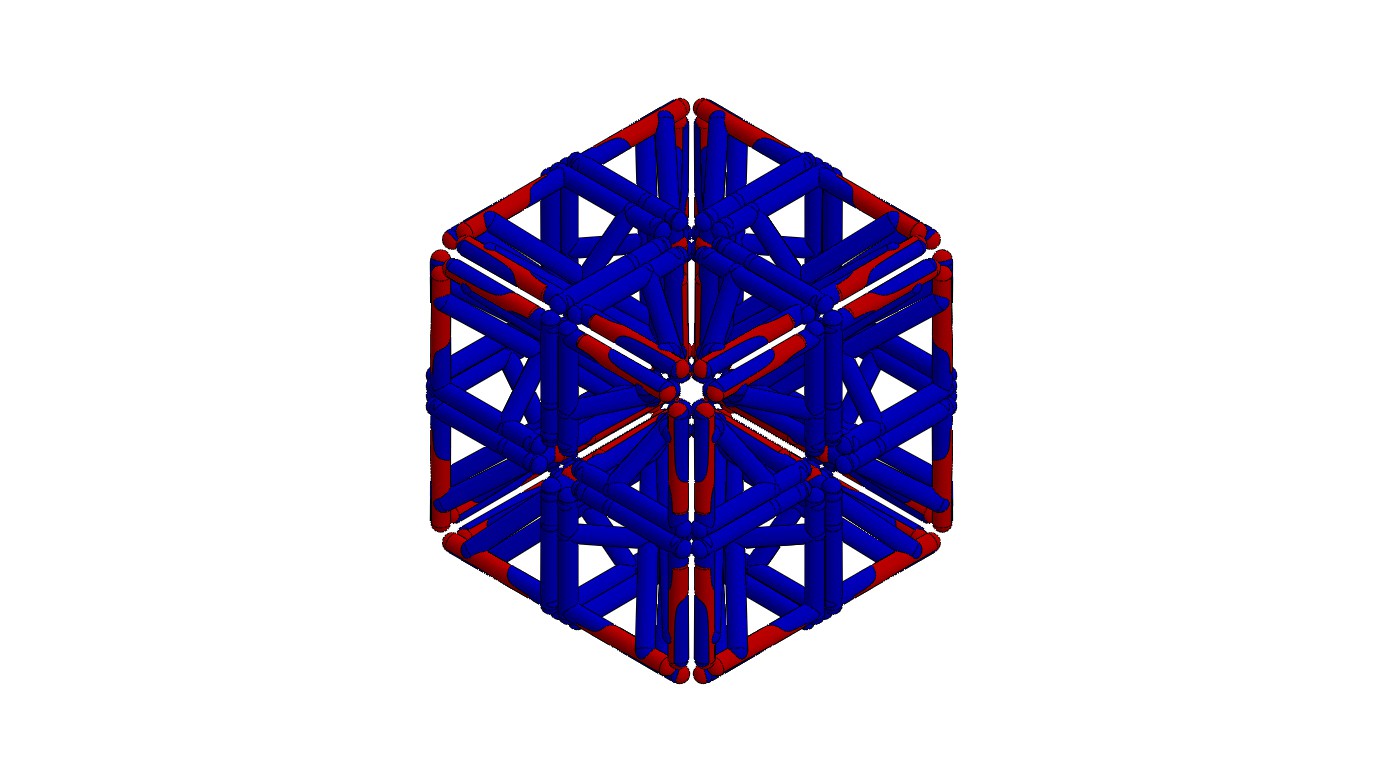} &  \includegraphics[width=1.8cm, bb=250 50 1000 700, clip=true]{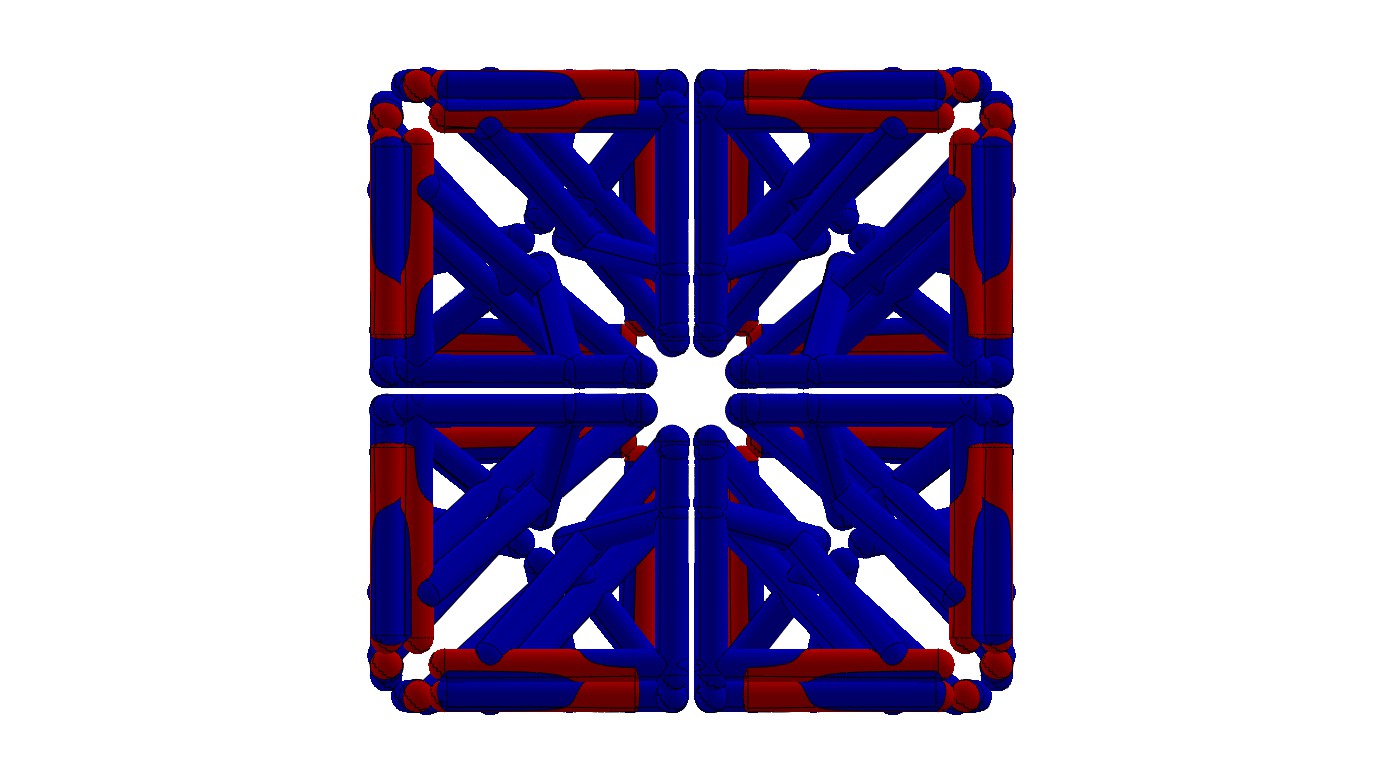}
  &$ 0.12434 $ & $0.1056$ & \includegraphics[width=1.8cm, bb=250 50 1000 700, clip=true]{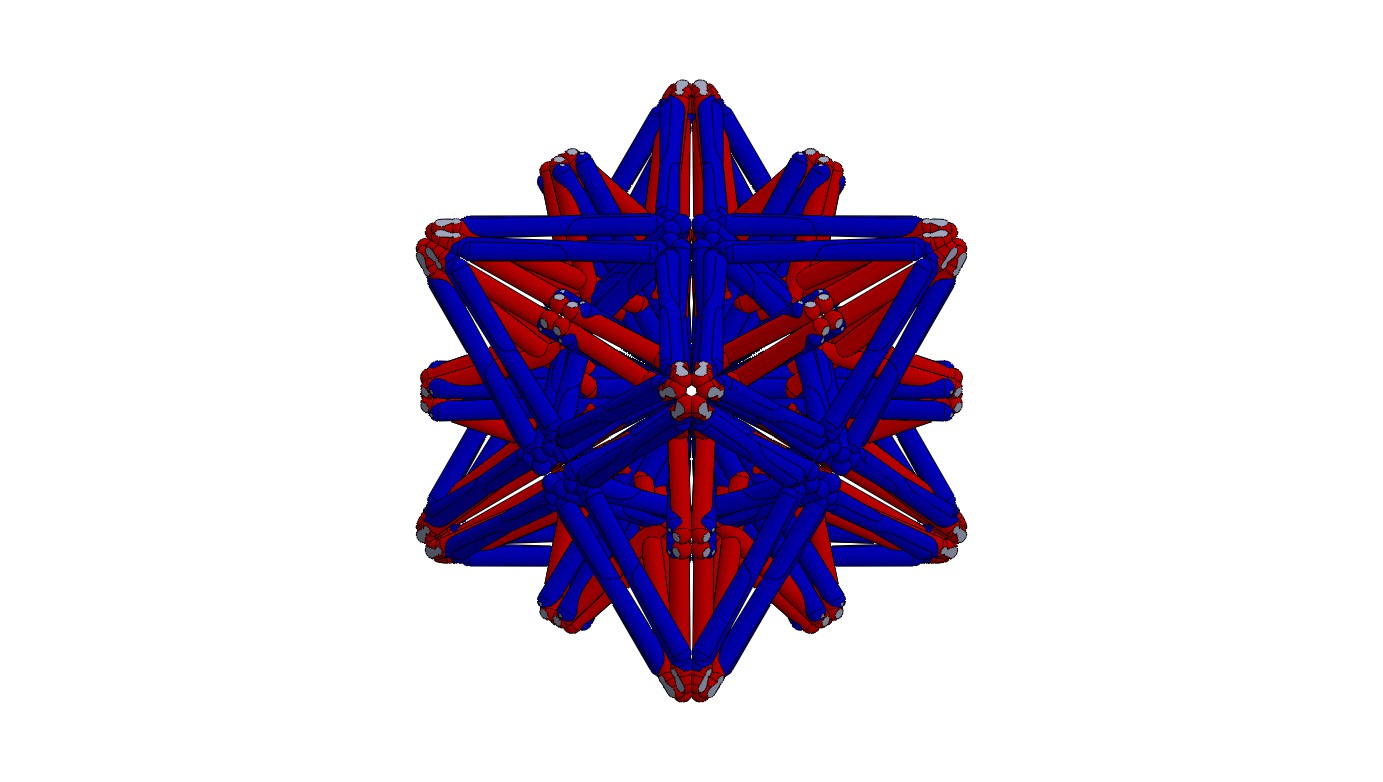} &  \includegraphics[width=1.8cm, bb=250 50 1000 700, clip=true]{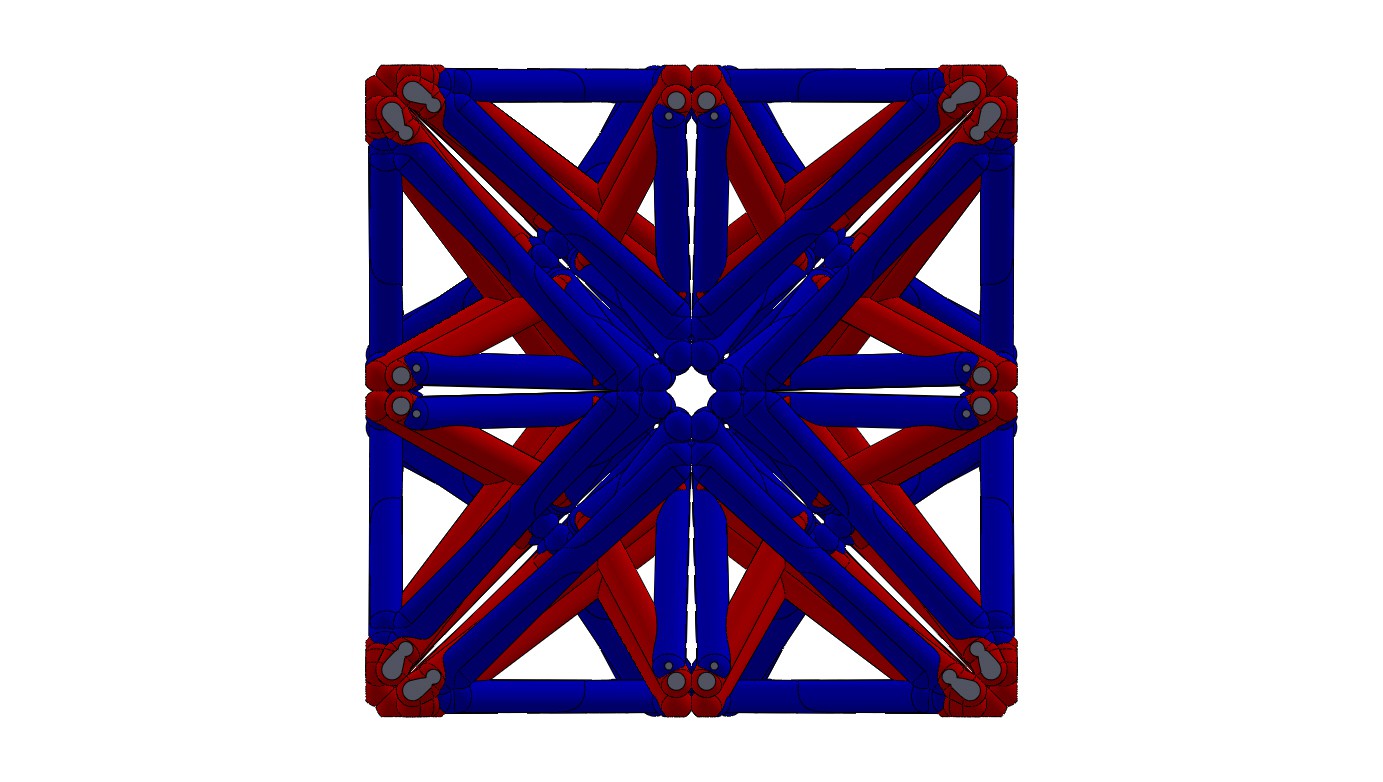} \tabularnewline

\hline 
\end{tabular}
\par\end{centering}
\caption{Maximal shear modulus designs for cubic two-material lattices and different weight fraction limits $w_f^*$. Red bars are made of material 1, blue bars are made of material 2, and bars that have been removed from the design (i.e., with $\alpha_1^q, \alpha_2^q \approx 0$) are not shown.}
\label{table:cubic_shear}
\end{table*}
\begin{table*}
\begin{centering}
\begin{tabular}{>{\centering}m{1.4cm}>{\centering}m{0.6cm}>{\centering}m{1.8cm}>{\centering}m{1.8cm}}
$G$ & $w_f^*$ & iso & side
\tabularnewline
\hline  \\[-1ex]
$ 0.04591 $ & $0.0556$ & \includegraphics[width=1.8cm, bb=250 50 1000 700, clip=true]{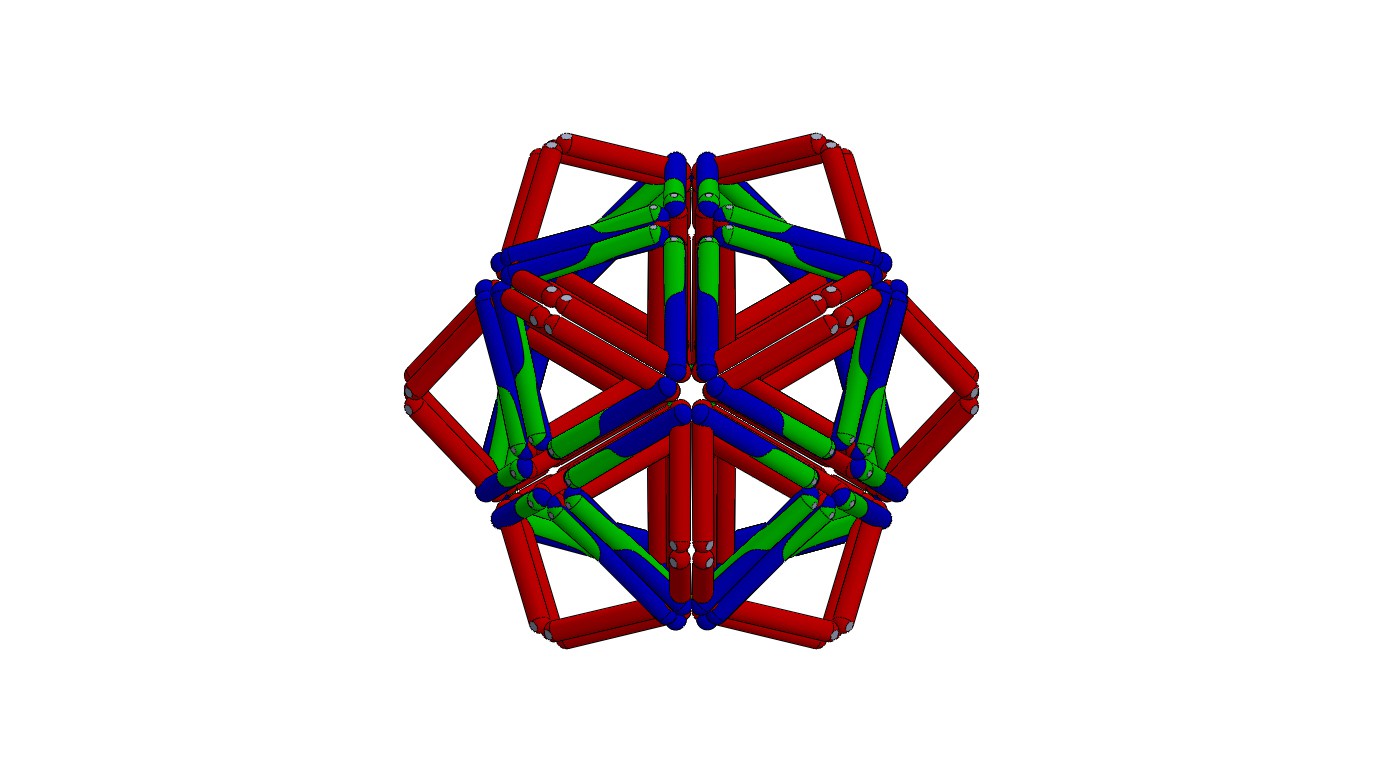} &  \includegraphics[width=1.8cm, bb=250 50 1000 700, clip=true]{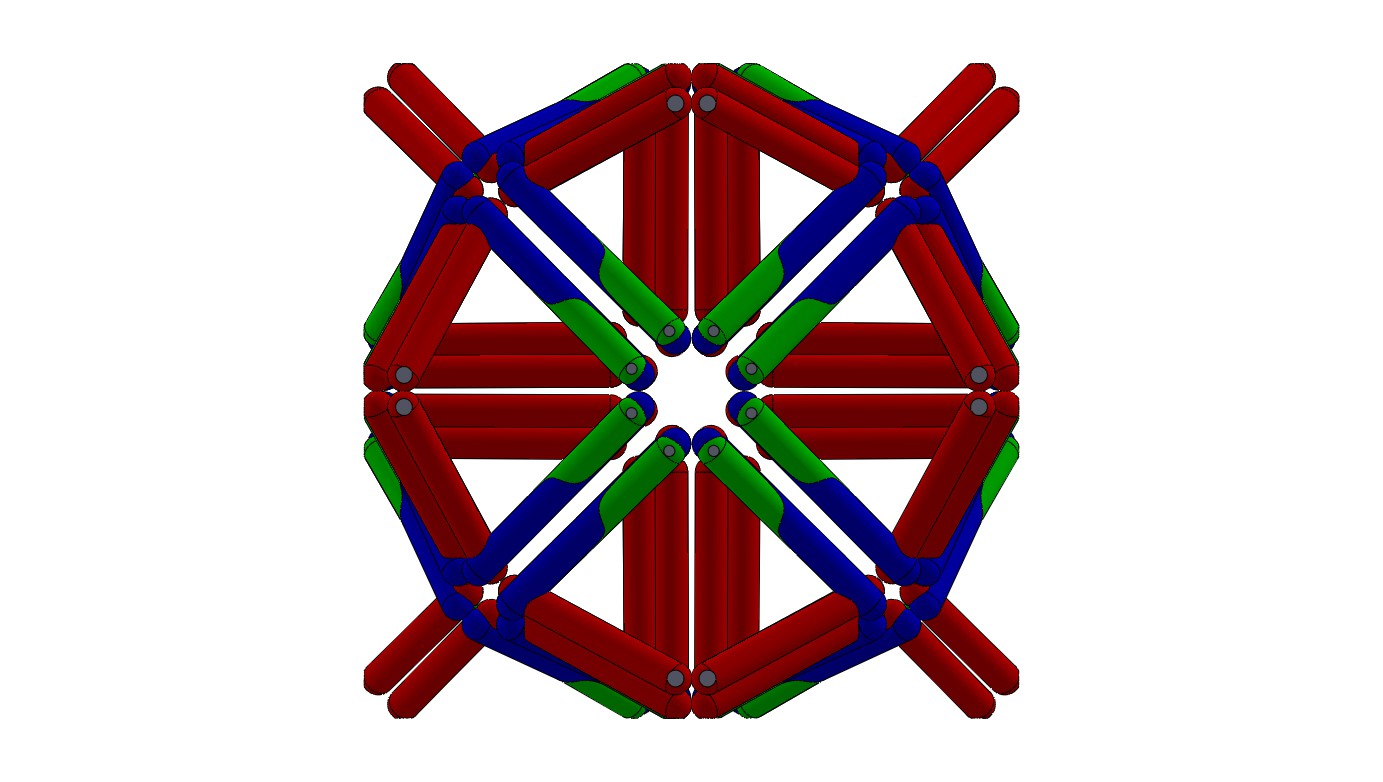}   \tabularnewline
$  0.03146$ & $0.0722$ & \includegraphics[width=1.8cm, bb=250 50 1000 700, clip=true]{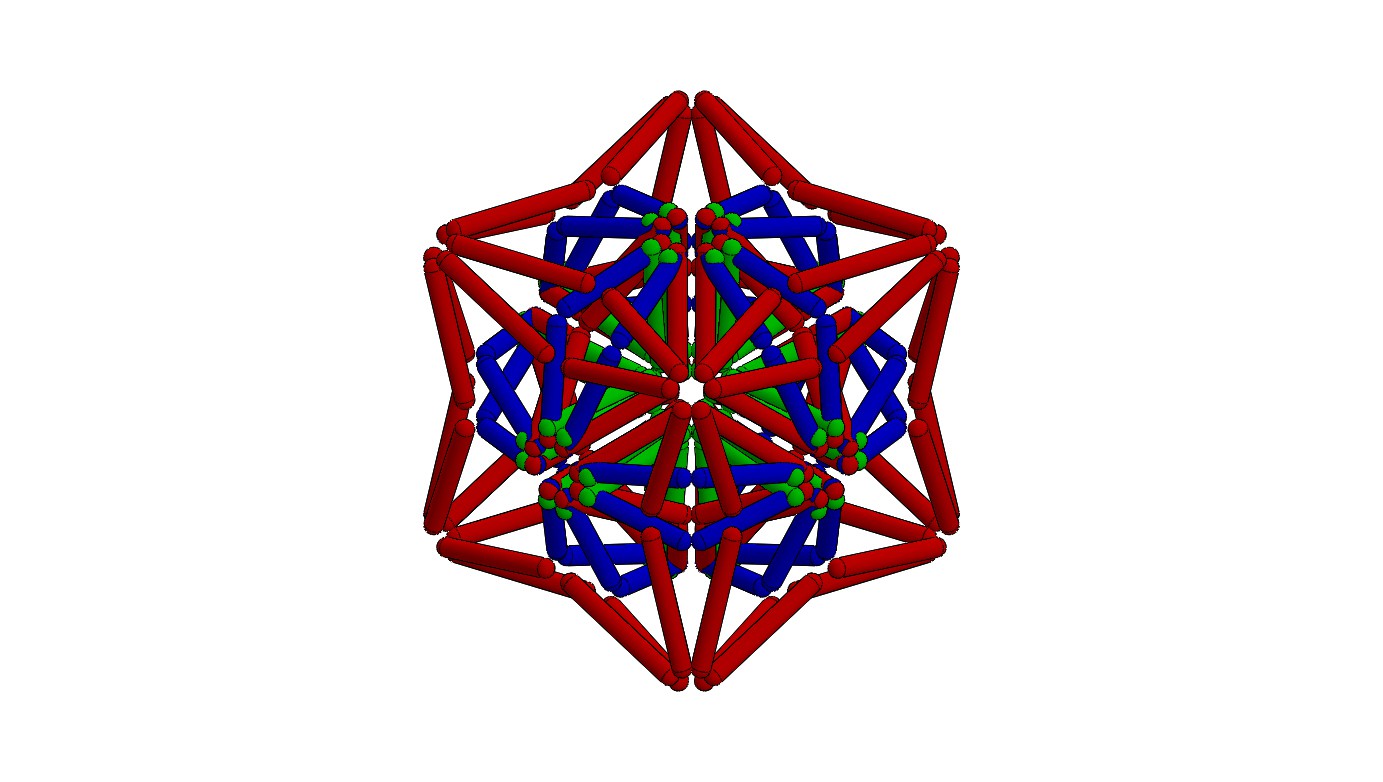} &  \includegraphics[width=1.8cm, bb=250 50 1000 700, clip=true]{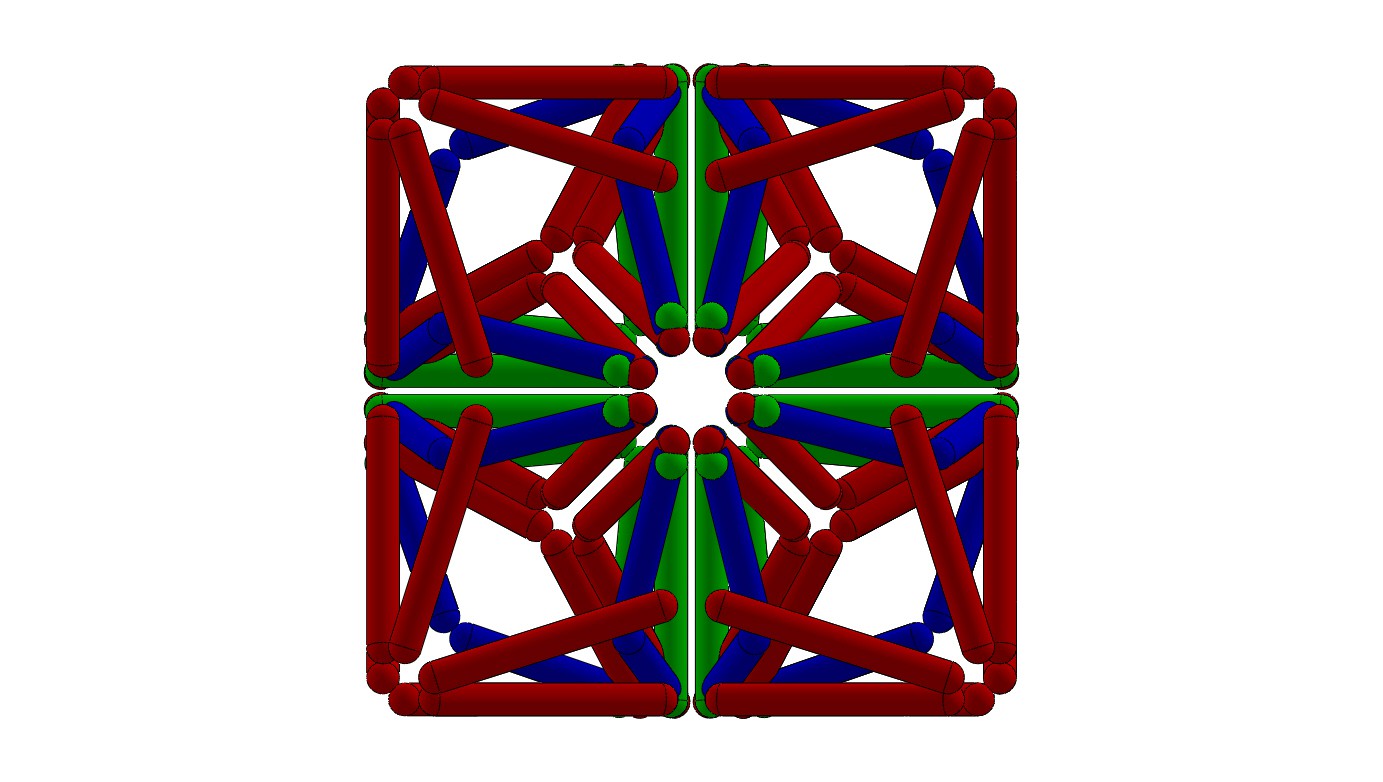}   \tabularnewline
$ 0.08593 $ & $0.0944$ & \includegraphics[width=1.8cm, bb=250 50 1000 700, clip=true]{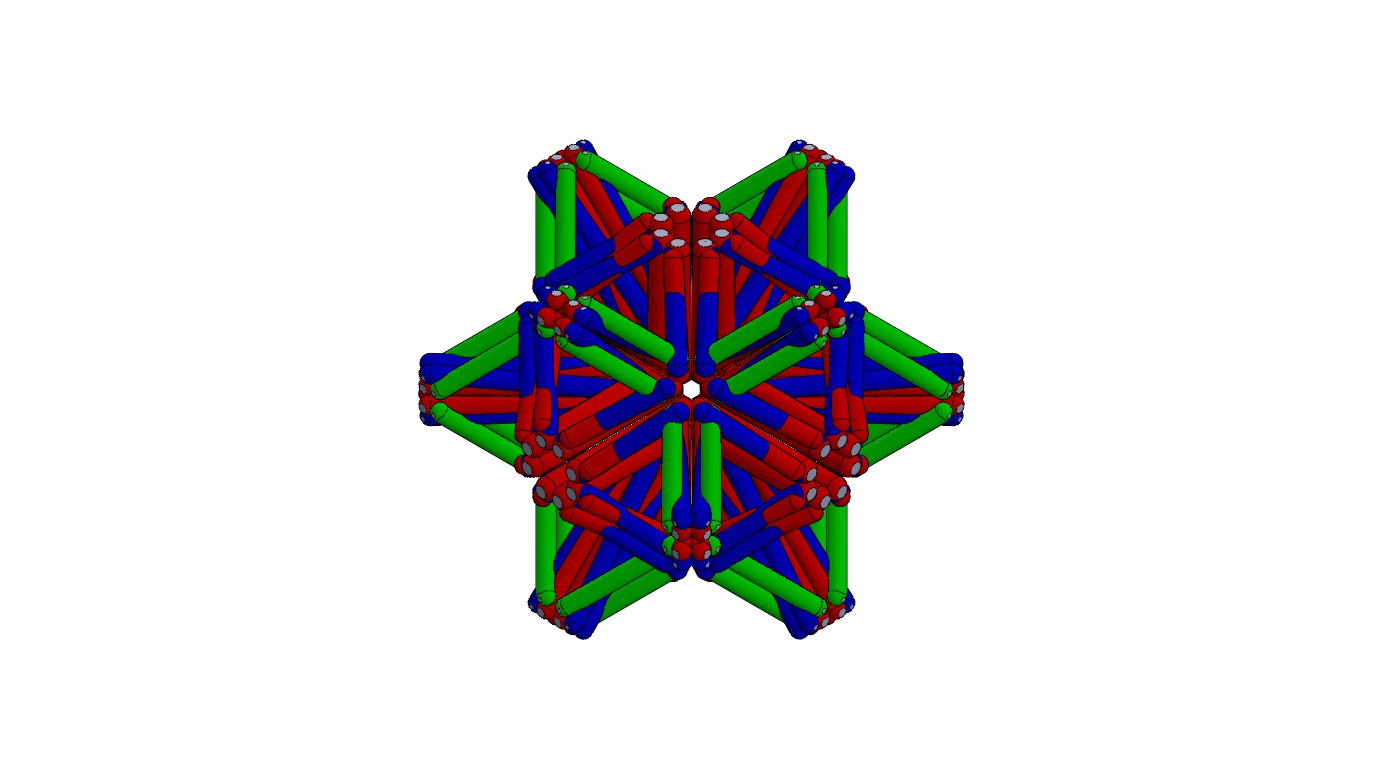} &  \includegraphics[width=1.8cm, bb=250 50 1000 700, clip=true]{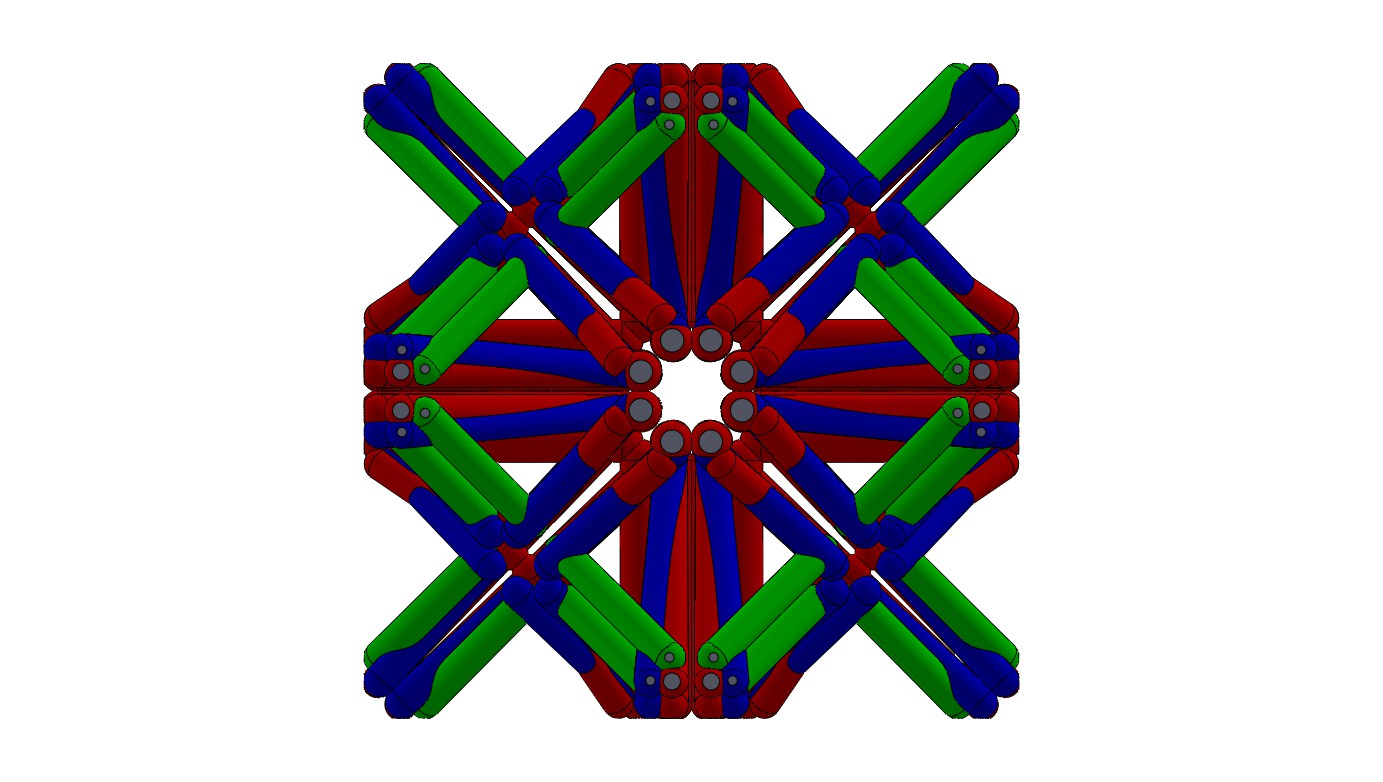}   \tabularnewline
$ 0.1288 $ & $0.1056$ & \includegraphics[width=1.8cm, bb=250 50 1000 700, clip=true]{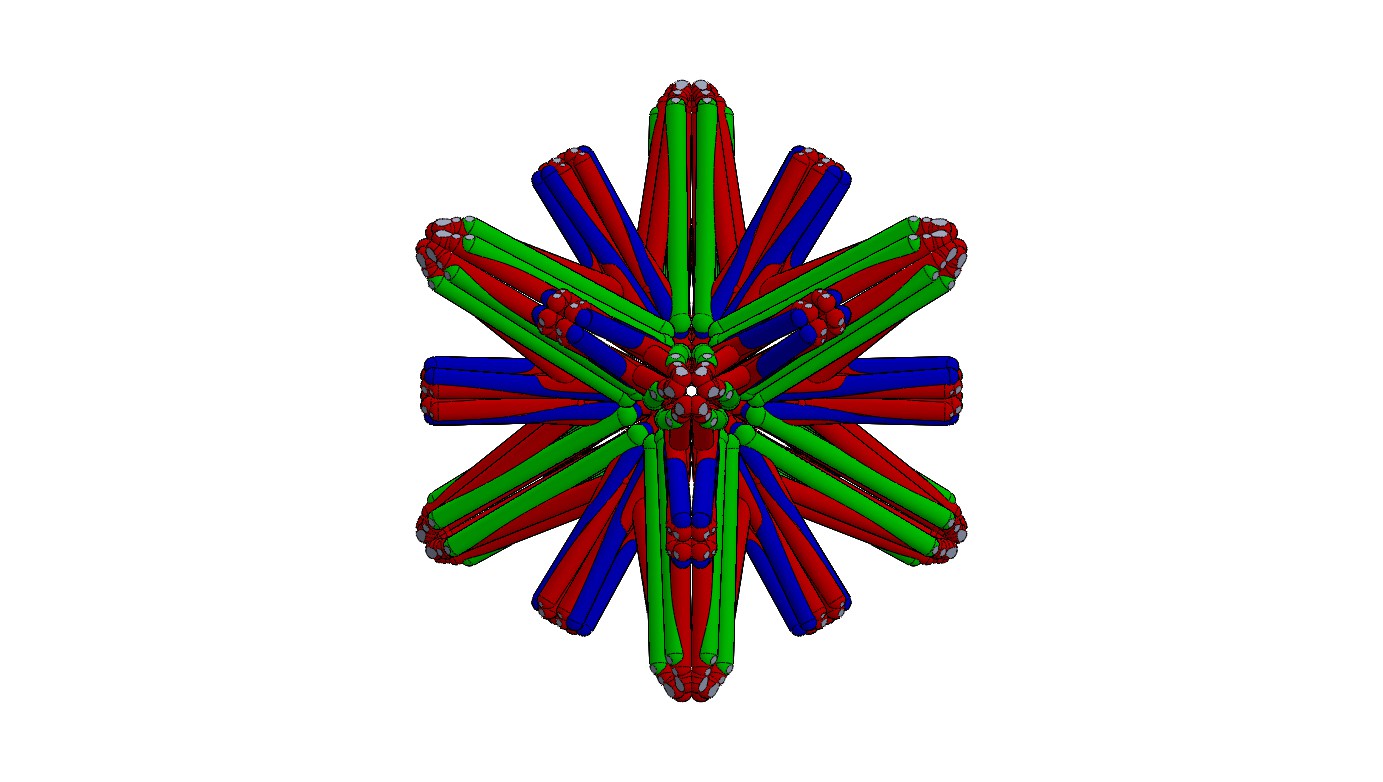} &  \includegraphics[width=1.8cm, bb=250 50 1000 700, clip=true]{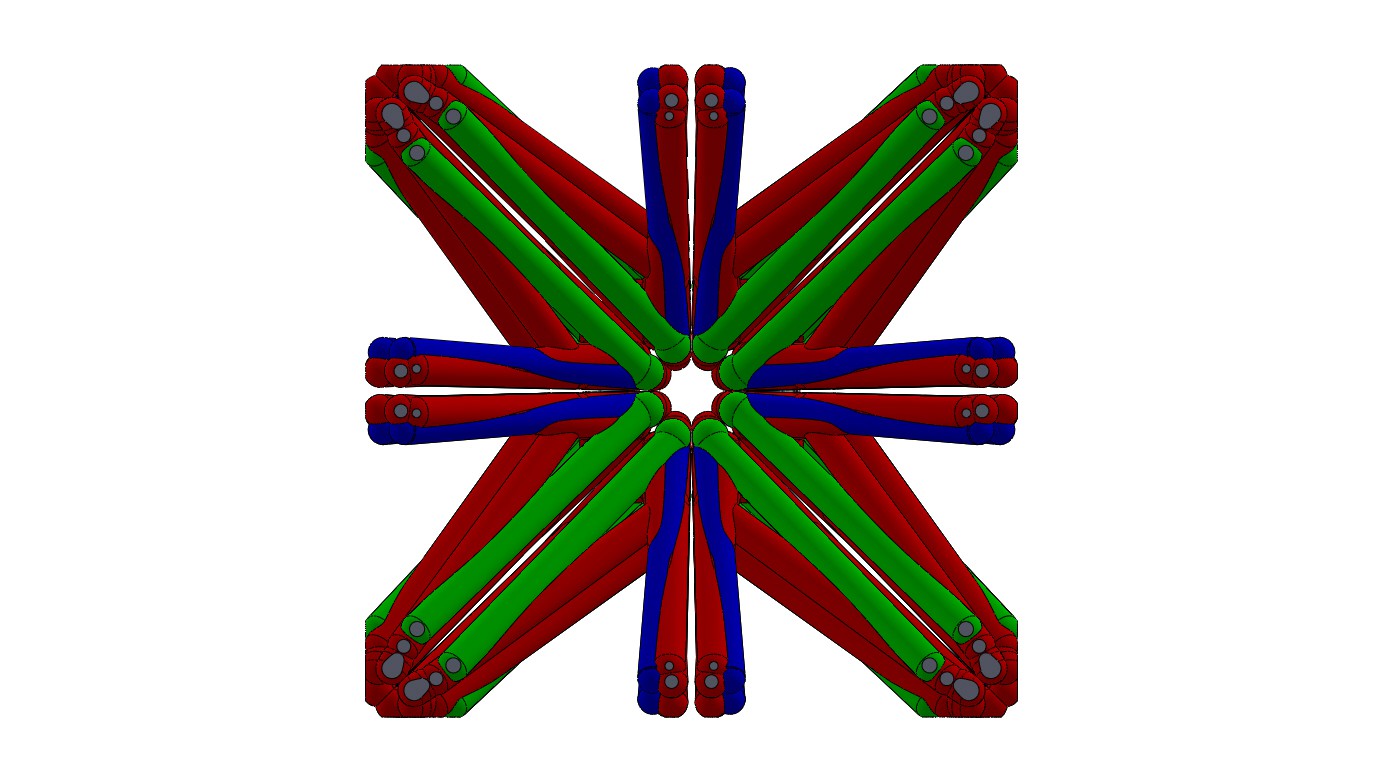}   \tabularnewline

\hline 
\end{tabular}
\par\end{centering}
\caption{Maximal shear modulus designs for cubic three-material lattices for different weight fraction limits $w_f^*$. Red, blue and green  bars are made of materials 1, 2, and 3, respectively; and bars that have been removed from the design (i.e., with $\alpha_1^q, \alpha_2^q, \alpha_3^q \approx 0$) are not shown.}
\label{table:cubic_shear_3mat}
\end{table*}
%
\subsection{Two-material Lattices with Negative Poisson Ratio and Cubic Symmetry}
\label{sec:cubic} 

Finally, we present results for the minimization of the effective Poisson's ratio for two- and three-material cubic lattices.  The material properties are the same as before. As detailed in Section \ref{sec:opt-problem}, for this problem we add a constraint that ensures a minimum bulk modulus of $K_{min}$, cf.\ Table \ref{table:parameters}. The results for two-material lattices are shown in Table \ref{table:poisson}. For the weight fraction limits we employed, the resulting effective Poisson's ratios are all negative, therefore the lattice is auxetic. In this case, the effective Poisson's ratio monotonically decreases as we increase the weight fraction limit.  We also present a three-material lattice design in Table \ref{table:poisson_3mat}. In this case, we could only find a narrow weight fraction limit range that would render a three-material design with negative Poisson's ratio.  We note, however, that the three-material design performs better than the two-material design for the same weight fraction limit shown in Table \ref{table:poisson}.

\begin{table*}
\begin{centering}
\begin{tabular}{>{\centering}m{1.4cm}>{\centering}m{0.6cm}>{\centering}m{1.8cm}>{\centering}m{1.8cm}|>{\centering}m{1.4cm}>{\centering}m{.6cm}>{\centering}m{1.8cm}>{\centering}m{1.8cm}}
$\nu$ & $w_f^*$ & iso & side & $\nu$ & $w_f^*$ & iso & side
\tabularnewline
\hline 
$ -0.05721 $ & $ 0.0333$ & \includegraphics[width=1.8cm, bb=250 50 1000 700, clip=true]{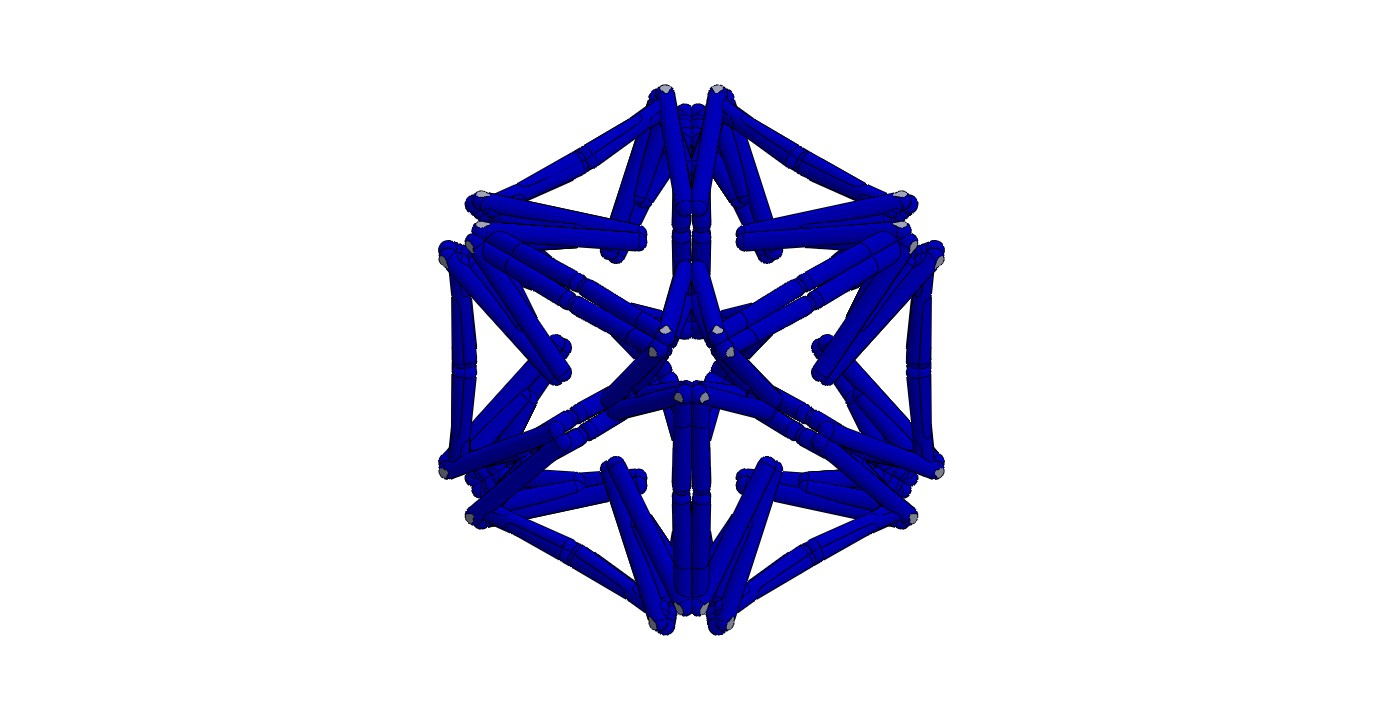} &  \includegraphics[width=1.8cm, bb=250 50 1000 700, clip=true]{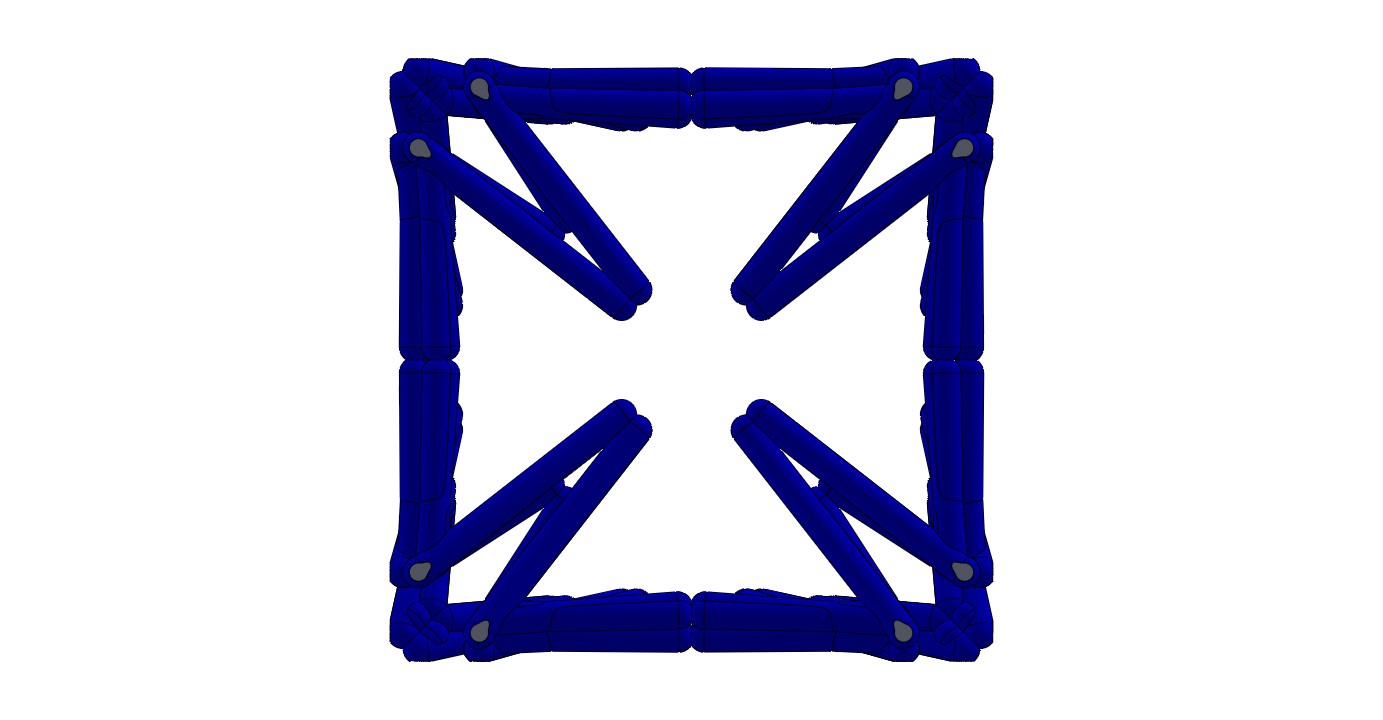} &$-0.63902 $ & $0.05$ & \includegraphics[width=1.8cm, bb=250 50 1000 700, clip=true]{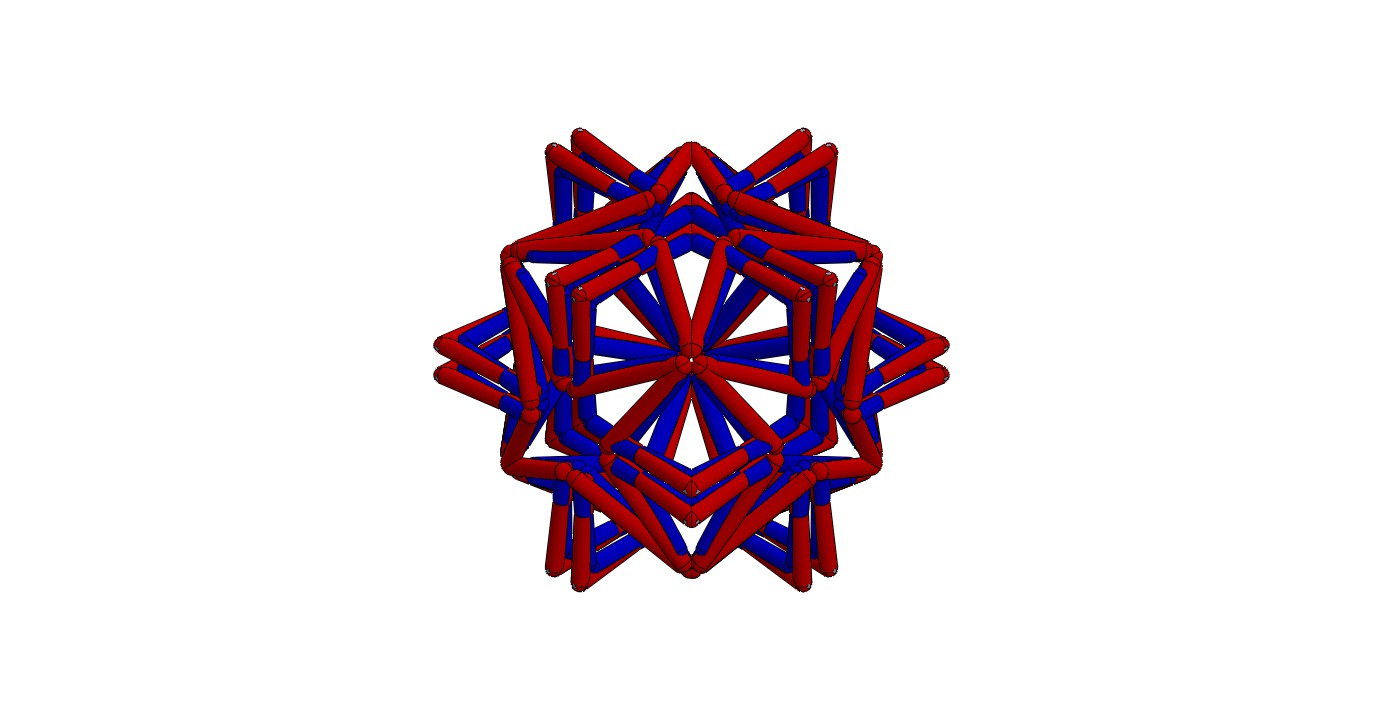} &  \includegraphics[width=1.8cm, bb=250 50 1000 700, clip=true]{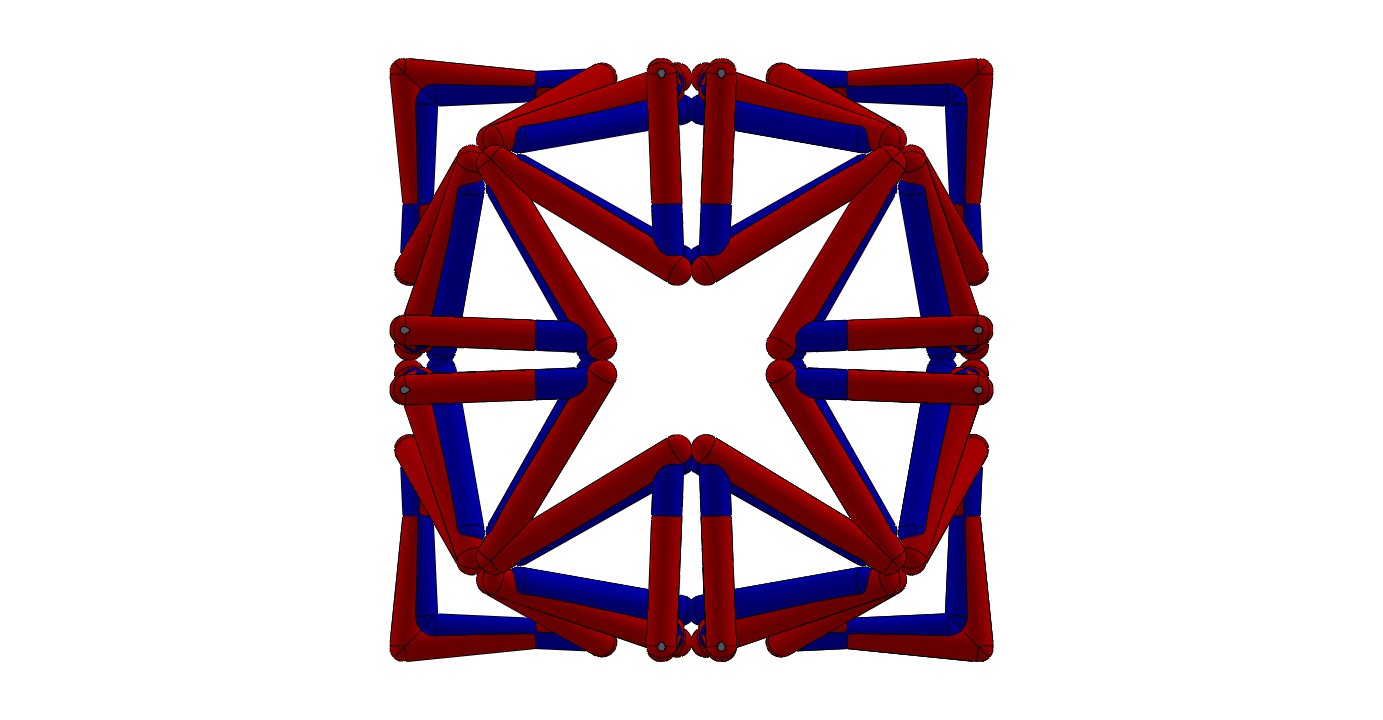} \tabularnewline
$ -0.11753 $ & $ 0.0389$ & \includegraphics[width=1.8cm, bb=250 50 1000 700, clip=true]{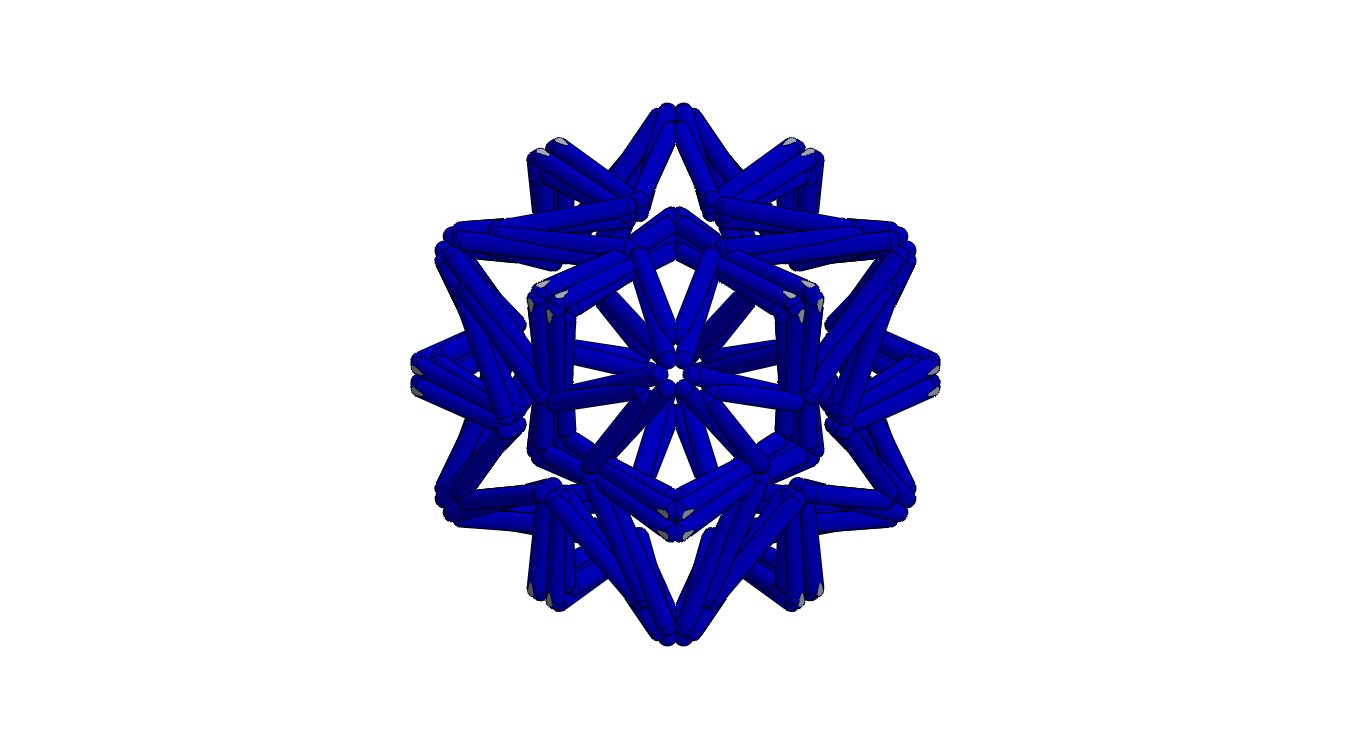} &  \includegraphics[width=1.8cm, bb=250 50 1000 700, clip=true]{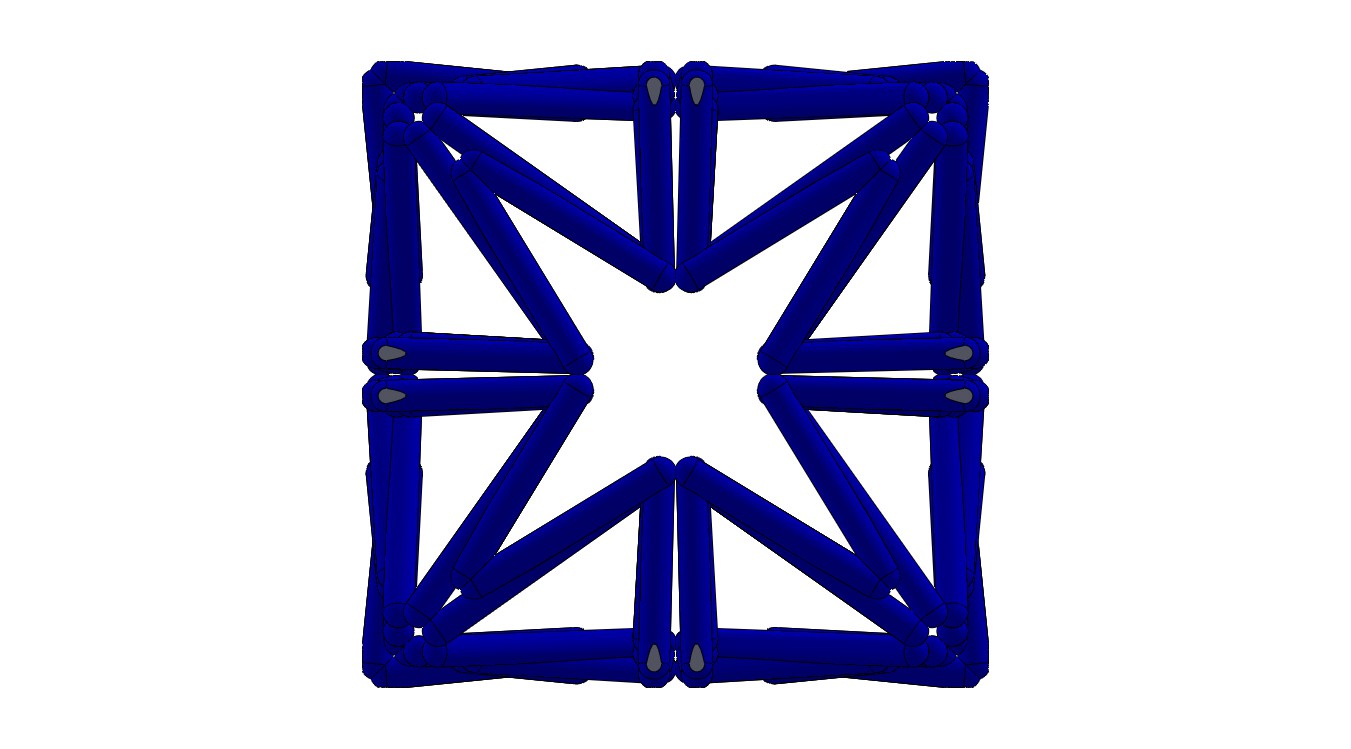} &$ -1.57259 $ & $0.0667$ & \includegraphics[width=1.8cm, bb=250 50 1000 700, clip=true]{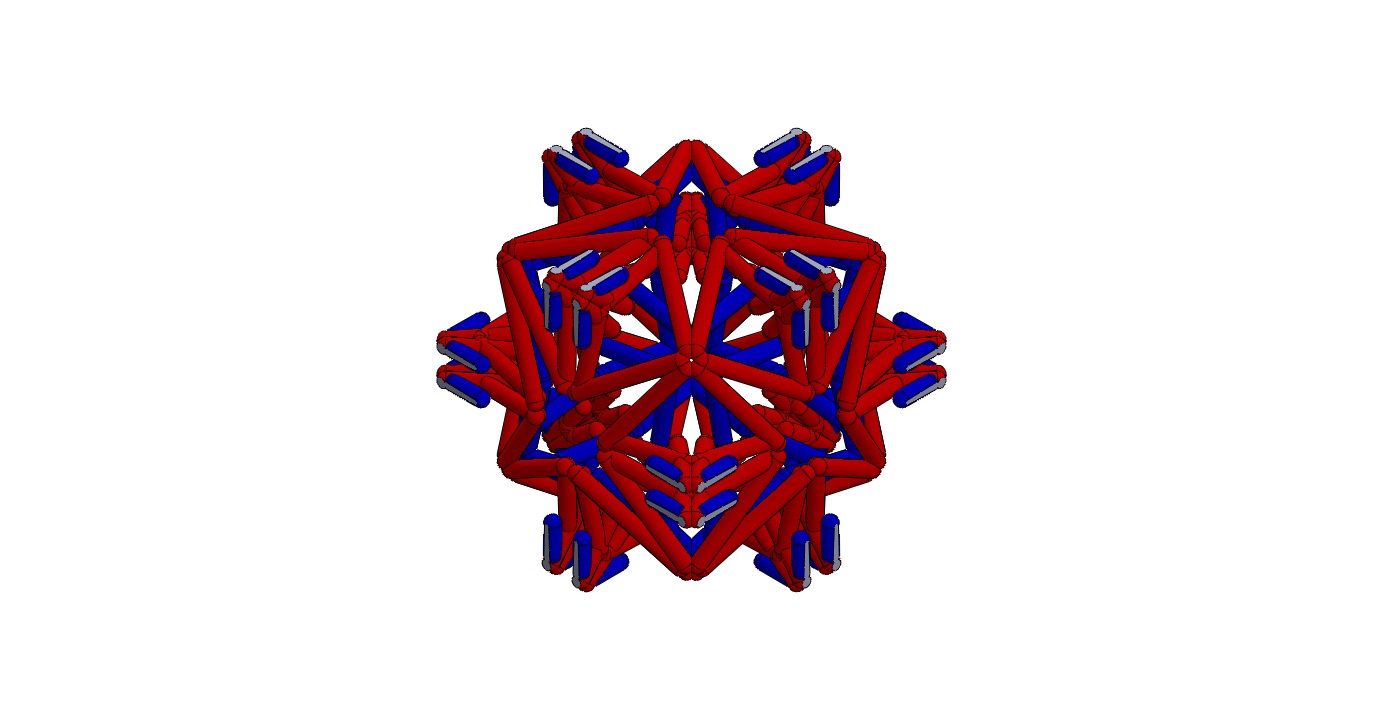} &  \includegraphics[width=1.8cm, bb=250 50 1000 700, clip=true]{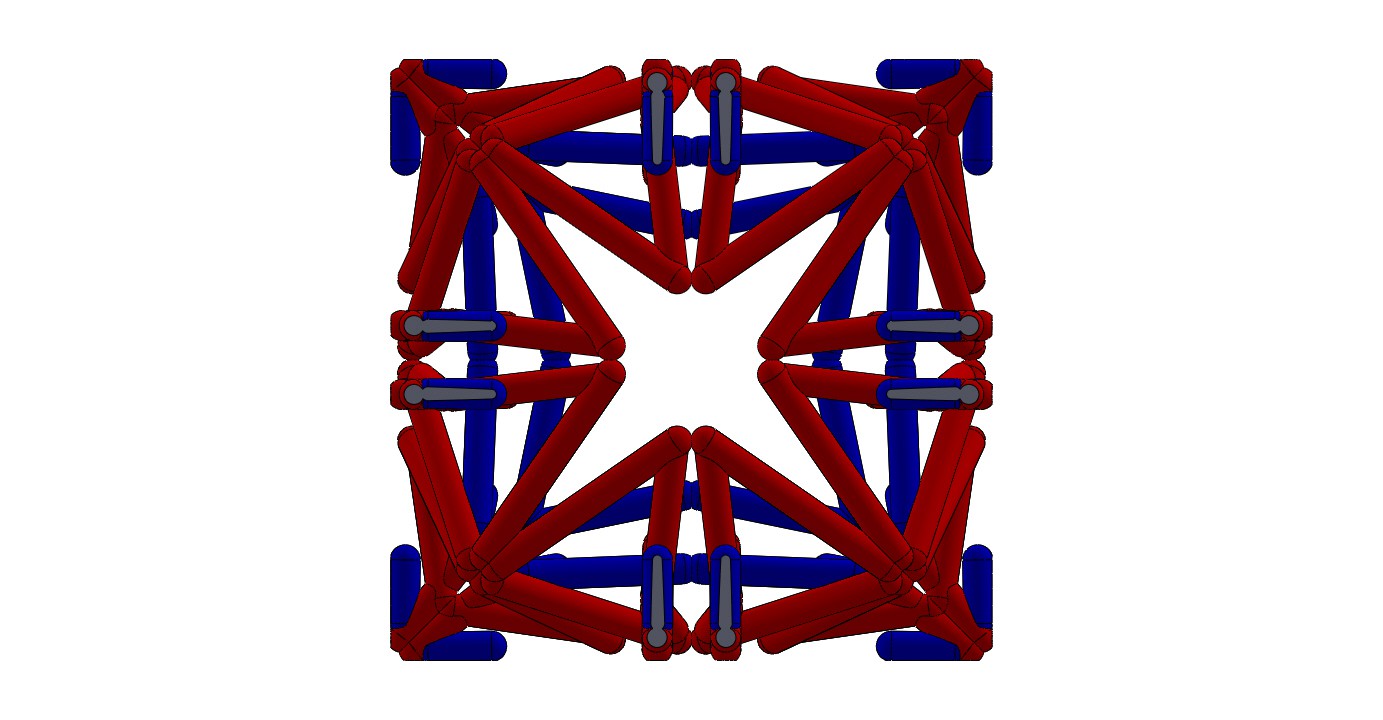} \tabularnewline
$ -0.204 $ & $0.0444$ & \includegraphics[width=1.8cm, bb=250 50 1000 700, clip=true]{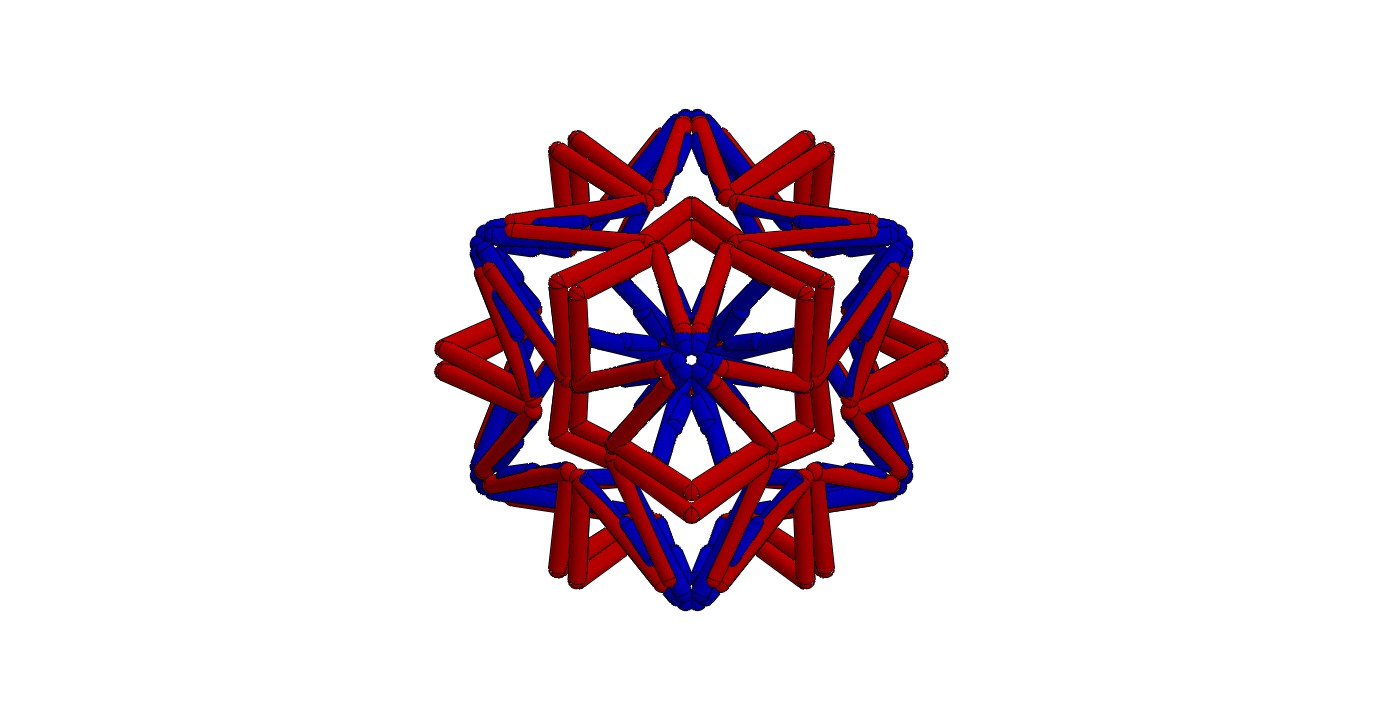} &  \includegraphics[width=1.8cm, bb=250 50 1000 700, clip=true]{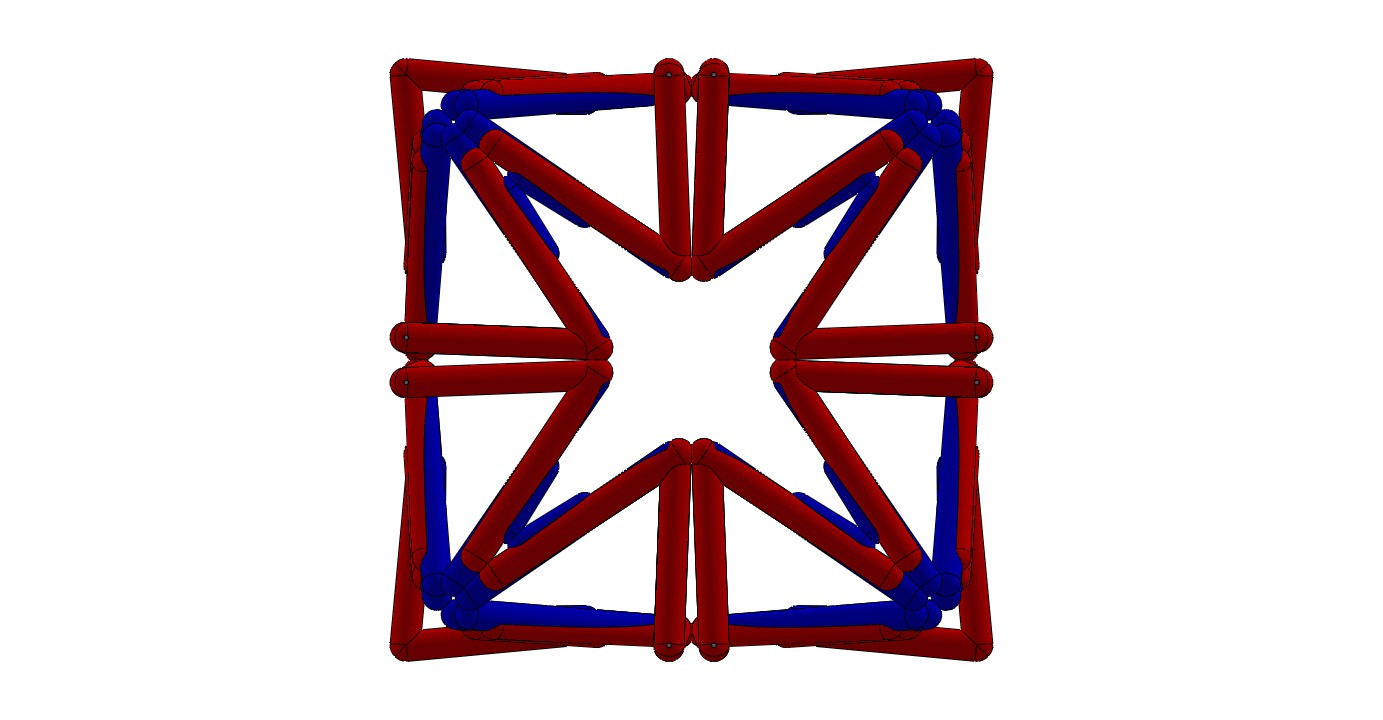}    &$ -3.81031$ & $0.0722$ & \includegraphics[width=1.8cm, bb=250 50 1000 700, clip=true]{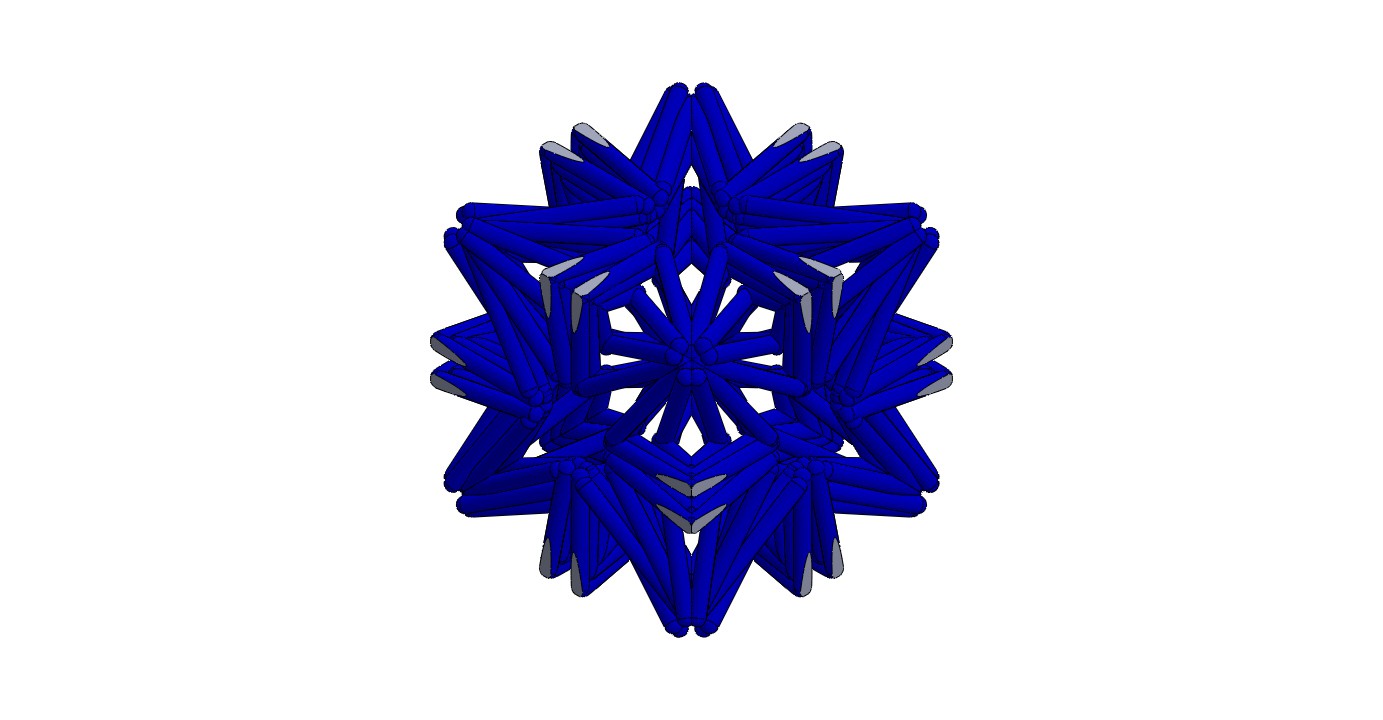} &  \includegraphics[width=1.8cm, bb=250 50 1000 700, clip=true]{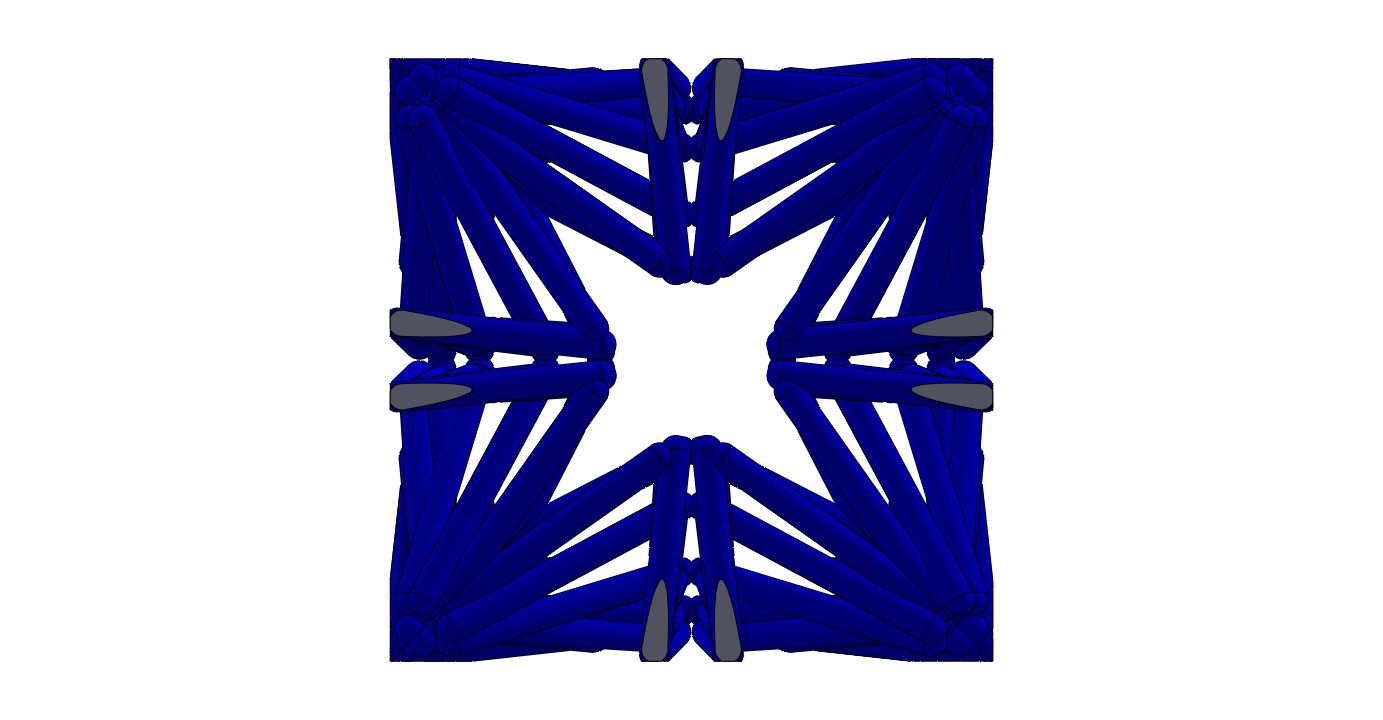}  \tabularnewline

\hline 
\end{tabular}
\par\end{centering}
\caption{Minimal Poisson's ratio designs for cubic two-material lattices for different weight fraction limits $w_f^*$. Red bars are made of material 1, blue bars are made of material 2, and bars that have been removed from the design (i.e., with $\alpha_1^q, \alpha_2^q \approx 0$) are not shown.}
\label{table:poisson}
\end{table*}

\begin{table*}
\begin{centering}
\begin{tabular}{>{\centering}m{1.4cm}>{\centering}m{.6cm}>{\centering}m{1.8cm}>{\centering}m{1.8cm}}
$\nu$ & $w_f^*$ & iso & side
\tabularnewline
\hline 
$  -0.42618 $ & $0.0389$ & \includegraphics[width=1.8cm, bb=250 50 1000 700, clip=true]{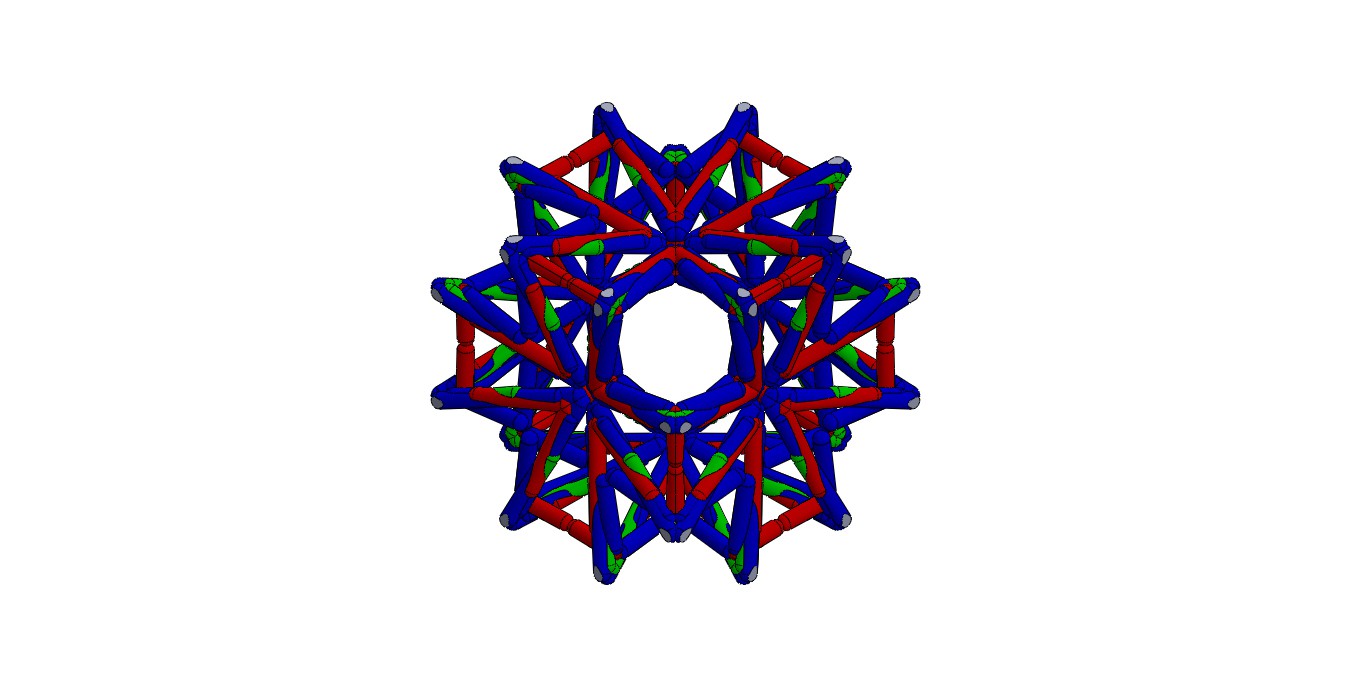} &  \includegraphics[width=1.8cm, bb=250 50 1000 700, clip=true]{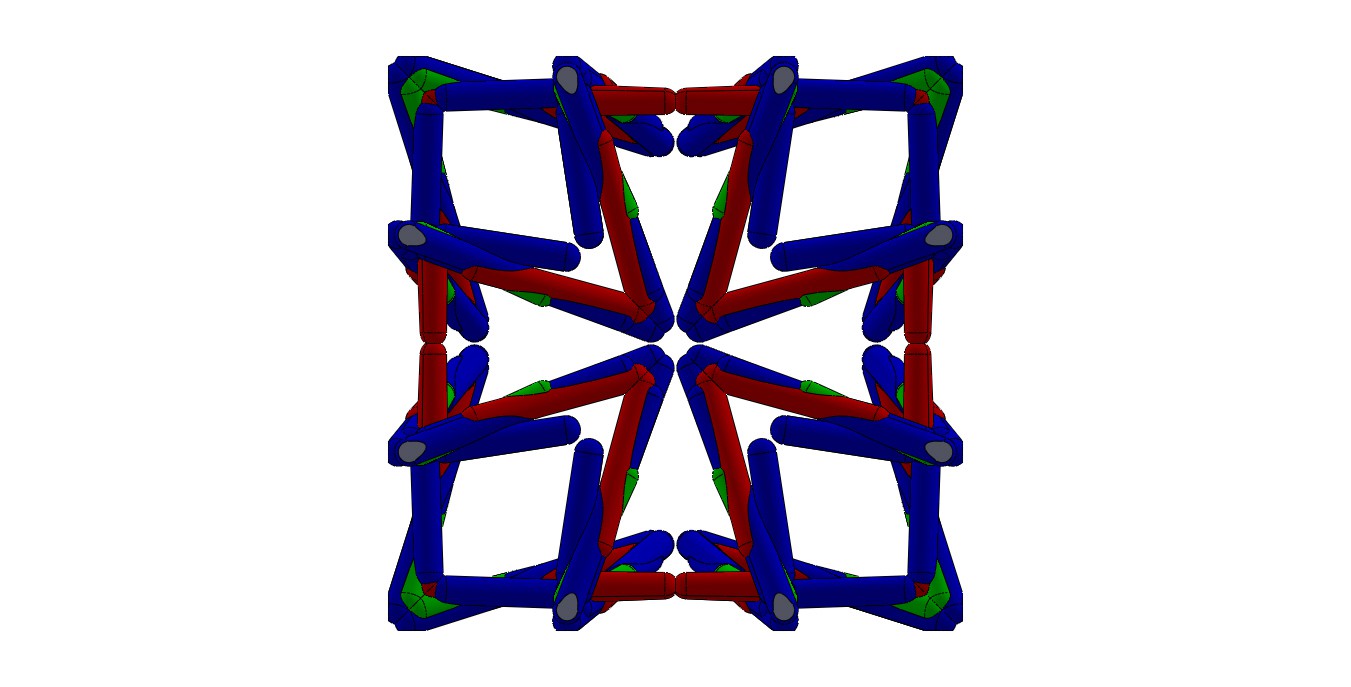}   \tabularnewline
\hline 
\end{tabular}
\par\end{centering}
\caption{Minimal Poisson's ratio design for cubic three-material lattice. Red, blue and green  bars are made of materials 1, 2, and 3, respectively; and bars that have been removed from the design (i.e., with $\alpha_1^q, \alpha_2^q, \alpha_3^q \approx 0$) are not shown.}
\label{table:poisson_3mat}
\end{table*}

\section{Conclusions}
\label{sec:conclusions}

This work presented a topology optimization method for the design of multi-material lattice structures using the geometry projection technique. The numerical examples demonstrate that the proposed method is effective in producing multi-material lattices for the maximization of effective bulk and shear moduli, and for the minimization of effective Poisson's ratio. The proposed formulation effectively imposes any number of symmetry planes to obtain desired material symmetries; in the case of the numerical examples, cubic symmetry is imposed on all lattices.  Moreover, the no-cut constraint is also effective in preventing struts from being cut by the unit cell boundaries or the symmetry planes that would make the manufacturing more difficult, and that may produce closed-cell structures. As expected, the designed lattices satisfy theoretical bounds on effective bulk and shear moduli, and are in fact away from these bounds since they are open-cell structures. The designs produced by the proposed method still pose some fabrication challenges, as some of the struts may have overlaps that are difficult to realize.  Future work will be devoted to minimize strut overlaps, and to impose angle constraints on the struts to remove altogether the need for supports.

 \section*{Acknowledgements}

Support from the National Science Foundation, award CMMI-1634563 to conduct this work is gratefully acknowledged.  This work was also partially supported by a fellowship grant from GE's Industrial Solutions Business Unit under a GE-UConn partnership agreement. The views and conclusions contained in this document are those of the authors and should not be interpreted as necessarily representing the official policies, either expressed or implied, of Industrial Solutions or UConn. We also thank Niels Aage and collaborators for their parallel implementation of MMA \cite{aage2017giga}, which we used for the optimization.

\bibliography{ref}

\end{document}